%% file: d2_maris_08_v10.tex
\documentclass{JINST}

\usepackage{graphicx}
\usepackage{txfonts}
 \usepackage[figuresright]{rotating}

 \input d2_maris_08_symb.tex

\input d2_maris_08_tab.tex

\input d2_maris_08_fig.tex

\def\instOATS{${^1}$}
\def\instUNIMI{$^{2}$}
\def\instLABEN{$^{3}$}
\def\instIAC{$^{4}$}
\def\instISDC{$^{5}$}
\def\instIASFBO{$^{6}$}
\def\instIFPCNR{$^{7}$}
\def\instUNITS{$^{8}$}
\def\instUSA{$^{9}$}
\def\instIASFMI{$^{10}$}

\def\numparts{}
\def\endnumparts{}

 \title{
 Optimization of {\sc Planck}/LFI on--board data handling
   }

 \author{
   M.~Maris{\instOATS} \thanks{Corresponding Author, e--mail: maris@oats.inaf.it}
, 
        M.~Tomasi{\instUNIMI}
, 
        S.~Galeotta{\instOATS}
, 
        M.~Miccolis{\instLABEN}
, 
        S.~Hildebrandt {\instIAC}
, 
        M.~Frailis{\instOATS}
, 
        R.~Rohlfs{\instISDC}
, 
        N.~Morisset{\instISDC}
, 
        A.~Zacchei{\instOATS}
, 
        M.~Bersanelli{\instUNIMI}
, 
        P.~Binko{\instISDC}
, 
        C.~Burigana{\instIASFBO}
, 
        R.C.~Butler{\instIASFBO}
, 
        F.~Cuttaia{\instIASFBO}
,	
        H. Chulani{\instIAC}
,	
        O.~D'Arcangelo{\instIFPCNR}
, 
        S.~Fogliani{\instOATS}
, 
        E.~Franceschi{\instIASFBO}
, 
        F.~Gasparo{\instOATS}
, 
        F.~Gomez {\instIAC}
, 
        A.~Gregorio{\instUNITS}
, 
        J.M.~Herreros {\instIAC}
, 
        R.~Leonardi {\instUSA}
, 
        P.~Leutenegger{\instLABEN}
, 
        G.~Maggio{\instOATS}
, 
        D.~Maino{\instUNIMI}
, 
        M.~Malaspina{\instIASFBO}
, 
        N.~Mandolesi{\instIASFBO}
, 
        P.~Manzato{\instOATS}
, 
        M.~Meharga{\instISDC}
, 
        P.~Meinhold {\instUSA}
, 
        A.~Mennella{\instUNIMI}
, 
        F.~Pasian{\instOATS}
, 
        F.~Perrotta{\instOATS}
, 
        R.~Rebolo{\instIAC}
, 
        M.~T$\ddot{\mathrm{u}}$rler{\instISDC}
, 
        A.~Zonca{\instIASFMI} \\
 \llap{\instOATS} INAF-OATs, 
 Via G.B.~Tiepolo 11, I-34131, Trieste, Italy
    E-mail: first.last@oats.inaf.it\\
 \llap{\instUNIMI}
 Universit\'a di Milano, Dipartimento di Fisica,
 Via G.~Celoria 16, I-20133 Milano, Italy
 E-mail: first.last@unimi.it \\
\llap{\instLABEN}
  Thales Alenia Space Italia S.p.A., 
 S.S. Padana Superiore 290, 20090 Vimodrone (Mi), Italy 
 E-mail: first.last@thalesaleniaspace.com\\
 \llap{\instIAC}
    Instituto de Astrofisica de Canarias (IAC), 
    C/o Via Lactea, s/n E38205 - La Laguna, Tenerife, Espa\~na
    E-mail: first.last@iac.es \\
 \llap{\instISDC}
ISDC Data Centre for Astrophysics, University of Geneva, ch. d'Ecogia 16, 1290 Versoix, Switzerland
  E-mail: first.last@unige.ch \\
\llap{\instIASFBO}
 INAF-IASF Bologna, 
     Via P.~Gobetti, 101, I-40129 Bologna, Italy
     E-mail: first.last@iasfbo.inaf.it \\
 \llap{\instIFPCNR}
        IFP-CNR via Cozzi 53, 20125 Milano
 E-mail: first.last@ifp.cnr.it\\
 \llap{\instUNITS}
    Universit\'a di Trieste, Dipartimento di Fisica,
 Via A.~Valerio 2, I-34127 Trieste, Italy 
 E-mail: first.last@ts.infn.it\\
 \llap{\instUSA}
Department of Physics, University of California, Santa
Barbara, CA 93106, USA.
        E-mail: first.last@deepspace.ucsb.edu\\
 \llap{\instIASFMI}
 INAF-IASF Milano, 
 Via E.~Bassini 15, I-20133 Milano, Italy
 E-mail: first.last@iasfmi.inaf.it
    }

\abstract{
To asses stability against $1/f$ noise, the Low Frequency Instrument (LFI) \onboard\ the {\sc Planck} 
mission will acquire data at a rate much higher than the data rate allowed by the science telemetry 
bandwith of 35.5~kbs. The data are processed by an \onboard\ pipeline, followed \onground\ by a 
decoding and reconstruction step, to reduce the volume of data to a level compatible with the bandwidth 
while minimizing the loss of information. 
This paper illustrates the \onboard\ processing of the scientific data 
used by \Planck/LFI to fit the allowed data--rate, an intrinsecally
lossy process which distorts the signal in a manner
which depends on a set of five free parameters 
($N_{\mathrm{aver}}$, $r_1$, $r_2$, $q$, $\Offset$) for each of the 44 LFI detectors.
The paper quantifies the level of distortion introduced by the \onboard\
processing as a function of these parameters. 
It describes the method of tuning the \onboard\ processing
chain to cope with the limited bandwidth
while keeping to a minimum the signal distortion.
Tuning is sensitive to the statistics of the signal and has to be constantly 
adapted during the flight.
The tuning procedure is based on an optimization algorithm applied to 
unprocessed and uncompressed raw data provided either by simulations, 
pre--launch tests or data taken from LFI set--up to operate in a 
special diagnostic acquisition mode.
All the needed optimization steps are performed by an automated tool, {\tt OCA2}, which 
simulates the \onboard\ processing, explores the space
of possible combinations of parameters, and produces a set of statistical indicators,
among them: the compression rate $C_{\mathrm{r}}$ and the processing noise $\epsilon_Q$.
For \Planck/LFI it is required that $C_{\mathrm{r}} = 2.4$ 
while, as for other systematics, 
$\epsilon_Q$ would have to by less than 10\% of rms of the instrumental white noise.
An analytical model is developed that is able to extract most of the relevant information on the
processing errors and the compression rate as a function of the signal statistics and
the processing parameters to be tuned.
This model will be of interest for the instrument data analysis to asses the level of signal
distortion introduced in the data by the \onboard\ processing.
This method was applied during ground tests when the instrument was operating in 
conditions representative of flight. 
Optimized parameters were obtained and inserted in the \onboard\ processor 
and the performance has been verified against 
and the performance has been verified against the requirements, with the result that the required data rate of 35.5~Kbps has been achieved while keeping the processing error at a level of 3.8\% 
of the instrumental white noise and well below the target 10\% level.
}

\begin{document}

   \CHANGE{
 \section{Introduction}\label{sec:introduction}

One of the most challenging aspects in the design of an astronomy mission in space is the ability to send the collected data 
to the ground for the relevant analysis within the allowable telemetry bandwidth. In fact the increasing capabilities of 
\onboard\ instruments generates ever larger ammounts of data whereas the downlink capability is quite constant being mainly governed 
by the power of the \onboard\ transmitter and and the length of the time window which can be allocated for down linking the data
 \cite{Bertotti:2003}.
In the case of the ESA satellite Planck, which will observe the CMB from the second Lagrangian point (L2) of the Earth -- Sun system, 
$1.5 \times 10^6$~Km far from Earth, the down--link rate is limited to about 1.5~Mbps, and Planck can be in contact with the ground 
station (located at New Norcia, Western Australia) for no more than a couple of hours each day reducing the effective bandwidth by an 
order of magnitude.
In addition, \Planck\ carries two scientific instruments the \Planck\ Low Frequency Instrument (\Planck/LFI), to which this paper is devoted, and the \Planck\ High Frequency Instrument (\Planck/HFI) and both share the bandwidth with the spacecraft to download data 
with other internal spacecraft services and the up--link channel 
The result is that LFI has only about 53.5~Kbps average down link rate while producing a unprocessed data rate of about 5.7~Mbps.
It is evident that some kind of \onboard\ data compression must be applied to fit in to the available telemetry bandwidth.
It is well known that the theoretical maximum compression rate achievable for a given data stream decreases with its increasing 
variance. Thus it is very advantageous before appying any compression algorithm to preprocess the data to reduce its inherent variance.
In the ideal case the preprocessing would not alter the original data, but in practice some information loss can not be avoided 
when the variance is reduced. Thus the \onboard\ preprocessing algorithm should be tunable through some kind of free 
processing--parameters in order to asses at the same time the required compression rate at the cost of a minimal degradation of the 
data.
This paper addresses the problem of the \onboard\ processing and the corresponding ground processing of the scientific data and the 
impact on its quality for the \Planck/LFI mission.
This has also been the topic of two previous papers the first regarding the exploration of possible lossless compression strategies
\cite{maris:2000},  and the second focused to the assessment of the distortions  in data introduced by a simplified model of the \onboard\ plus \onground\ processing \cite{maris:2004}.
Here the work presented by \cite{maris:2004} is completed by introducing in Sect.~\ref{sec:radiometer:model:and:acquisition:chain} a 
brief description of the instrument followed by a quantitative model of the \onboard\ plus \onground\ processing applied in 
\Planck/LFI. The processing can be tuned with the statistical properties of the signal and introduce as small as possible distortion.
to asses the proper compression rate and as small as possible processing distortion. This can be performed by using a set of control 
parameters, as anticipated in \cite{maris:2004}, which are tuned on the real signal. 
The tuning algorithm, which has not been discussed previously is the most important contribution to the \Planck/LFI programme presented in this work and it is discussed in Sect.~\ref{sec:optim}.
The whole procedure has been validated both with simulations and during the pre--flight ground testing. The most signifcative results 
are reported in Sect.~\ref{sec:results}.
Of course, processing has an impact on \Planck/LFI science whose complete analysis is outside the scope of this paper but however is 
briefly analyzed in Sect.~\ref{sec:impact:science}.
Finally Sect.~\ref{sec:final} reports the final remarks and conclusions, while some technical details are presented in appendices 
\ref{app:normalized:entropy}, \ref{app:adc:quantization} and \ref{app:dae:tuning}.
  }

\section{Radiometer model and acquisition chain}\label{sec:radiometer:model:and:acquisition:chain}

\FIGschematic

 %
\Planck/LFI \cite{LFI:general:instrument}
is based on an array of 22 radiometers assembled in 11 {\em Radiometric Chain Assemblies} (RCA) 
in the \Planck\ focal plane.
Each RCA has 4 \radiofrequency\ input lines and 4 \radiofrequency\ output lines, hence the
number of \radiofrequency\ outputs to be measured by the \onboard\ electronics is 44.
Each feed--horn has 2 orthomode transducers extracting the two orthogonal components of linear polarization
in the signal received from the sky and
feeding one of the \radiofrequency\ input lines of a radiometer, the other \radiofrequency\ input line
is connected to a \load\ held at the constant temperature of 4.5~K.

 \CHANGE{
A schematic representation of the flow of information in a single radiometer belonging to an RCA is given in 
Fig.~\ref{fig:schematic}.
 }
Each radiometer acts as a pseudo--correlation receiver
 \cite{LFI:general:rca}
measuring the difference in antenna temperatures, $\DeltaT$, between the sky signal, $\Tsky$, and the \load\, $\Tload$,
\cite{LFI:radiometers:load}.
However, given the sky and the reference--load have different mean temperatures
the reference samples have to be scaled by a {\em Gain Modulation Factor}, $\GMF$,
which balances the difference between $\Tsky$ and $\Tload$
to a $\MEAN{\DeltaT} = 0$ so that

\begin{equation}\label{eq:sky:minus:load}
  \DeltaT=\Tsky -\GMF \Tload.
\end{equation}

 \noindent
A proper choice of $\GMF$ will allow near cancellation out most of the first order 
systematic errors
\cite{R:Factor:Determination,LFI:radiometers:noise:properties},
assuring in this way optimal rejection of systematics, in particular drifts and
the $1/f$~noise
\cite{LFI:general:calibration}. 
As a first approximation it is possible to put

\begin{equation}\label{eq:gmf}
 \GMF \approx \frac{\MEAN{\Tsky} + \Tnoise}{\MEAN{\Tload} + \Tnoise},
\end{equation}

 \noindent
where $\Tnoise$ is the noise temperature.
Eq.~(\ref{eq:gmf}) makes evident how different values of $\GMF$ are needed in the various phases of the mission.
In particular three cases are important: ground tests, in--flight cooling phase and
finally in--flight operations with the instrument in nominal conditions.
As an exemple consider the case of the 30~GHz channel which is the least noisy channel
of \Planck/LFI having an expected $\Tnoise \approx 10$~K.
In on--ground testing the mean $\MEAN{\Tsky} \approx \MEAN{\Tload}$ and so $\GMF \approx 1$
 (\cite{LFI:general:instrument,LFI:general:calibration}).
In flight the $\MEAN{\Tsky} \approx 2.725$~K 
but during the cooling the $\MEAN{\Tload}$ varies from $\approx20$~K down to the 
nominal $\MEAN{\Tload}\approx4.5$~K.
Thus the $\GMF$ varys from $\approx 0.4$ when the instrument starts
to cool--down to $\approx 0.88$ at the end
of the process when its nominal temperature is reached.
 %
With higher values of $\Tnoise$ the other channels will show smaller departures in their $\GMF$ from 1 
as well as a lower sensitivity to the environmental conditions.
 %
%
%
 
To acquire sky and \load\ signals each radiometer has 
two separate \radiofrequency\ inputs, and correspondingly two \radiofrequency\ outputs,
each one connected to a \radiofrequency\ detector and to an acquisition chain 
ending in a 14~bit analog--to--digital converter (ADC) 
housed in the {\em Digital Acquisition Electronics}
box (DAE) \cite{LFI:general:instrument,LFI:general:rca}.
The output of the DAE is sent to the {\em Radiometer Electronics Box Assembly} box
(REBA) 
\footnote{LFI has two redundant REBA units, but since 
they are perfectly equivalent in what regard the \onboard\ data processing,
all over the paper we will consider LFI as having one REBA only.}
which processes the data from the DAE, 
of interpreting and executing telecommands, 
and of interfacing the instrument with the spacecraft Central Data Management Unit.
This unit produces the scientific packets to be sent to the ground 
\cite{LFI:REBA}.

The DAE applies an individually programmable analogue offset to  each input signal prior to applying individual programmablt gains and performing digitization.
The contribution to the read--out noise budget from the ADC quantization is in general
considered marginal. Appendix~\ref{app:adc:quantization} discusses the case in which this
hypothesis is no longer valid.
The offset and the gain are adjustable parameters of the DAE and it is assumed that their calibration
is independent from the REBA calibration \cite{LFI:radiometers:tuning}
with an exception which is discussed in Appendix~\ref{app:dae:tuning}.
The ADCs are fetched in turn and the data are sent to the Science Processing Unit (SPU), 
a {\em Digital Signal Processor} (DSP) based computer
which is part of the REBA  \cite{LFI:REBA} not represented in Fig.~\ref{fig:schematic}.
The SPU stores the data in circular buffers for
subsequent digital processing and 
and then applies the on board software pipeline to the data,
 %
In the process the 14~bit single samples are convert to 16~bits signed integers. 
The content of each ADC buffer is processed separately by the 
\onboard\ processing pipeline and sent to ground.

As usual in these kinds of receivers, the required stability of the radiometers
is assured by switching 
each radiometer between the \sky\ and \load.  Thus each output 
alternatively holds the sky and the \load\ signal (or the \load\
and the \sky) with opposed phases between the two channels.
Hence, each buffer contains 
strings of 
{\em interlaced}
\skyload\ (or \loadsky) samples 
in increasing order of acquisition time, $t$ i.e.

 \begin{equation}
\TskyADCt{0}, \TloadADCt{1}, 
\TskyADCt{2}, \TloadADCt{3}, \dots,
 \end{equation}

 \noindent
or

 \begin{equation}
\TloadADCt{0}, \TskyADCt{1}, 
\TloadADCt{2}, \TskyADCt{3}, \dots.
 \end{equation}

 \noindent
The switching frequency is fixed by the LFI internal clock at
$\SamplingFreqHz$. The switch clock gives also the beat for
the ADCs, which are then synchronized with the switching output, and it is sensed
by the on-board processor, which uses it to reconstruct the ordering of the
signals acquired from the ADCs and to synchronize it with the \onboard\ time.
This frequency also synchronises the ADCs with the input and is used by the SPU 
to reconstruct the ordering of the signals acquired from the ADCs and to synchronise 
them with the on board time.

 %
The data flow of raw data is equivalent to 5.7~Mbps; a large amount of data that  
cannot be fully downloaded to the ground. The allocated bandwidth for the
instrument is equivalent to only 53.5~kbps including all the ancillary data, less 
than 1\% of the overall data generated by LFI. 
The strategy, adopted to fit into the bandwidth, 
relies on three \onboard\ processing steps, downsampling, preprocessing the data to ensure lossless compression, and lossless compression itself. To demonstrate these steps, a model of the input signal shall be used.
 \CHANGE{
It has to be noted that while the compression is lossless, the preprocessing is not, due to
the need to rescale the data and convert them in integers, (a process named data requantizzation). 
However, the whole strategy is designed to asses a strict control of the way in which
lossy operations are done, of the amount of information loss in order to asses optimal compression rate
with minimal information loss.
 }

\FIGprocessing

\subsection{Signal model}\label{sec:OP:signal:model}

We describe quantitatively the kind of signal the pipeline has to process
by modeling the output of the DAE as a function of time, $t$, as 

 \numparts\begin{eqnarray}
 \Tsky(t) & = &  \Tskymean + \Driftsky(t) +\noisesky,\\
 \Tload(t) & = & \Tloadmean + \Driftload(t) +\noiseload.
 \end{eqnarray}\endnumparts

 \noindent
where $\Tskymean$, $\Tloadmean$ are the constant part of the signal. $\Driftsky$, and $\Driftload$  
a possible deterministic time
dependent parts, representing drifts, dipoles, oscillations and so on, $\noisesky$ and $\noiseload$ 
represents the random noise whose moments are 
$\sigmaskynoise^2$, $\sigmaloadnoise^2$, and whose covariance is $\sigmaskyloadnoise$.

The pipeline described in the following sections needs to be tuned to obtain a proper level of data
compression which is largely determined by the covariance matrix of the signal whose components are

 \begin{eqnarray}
  \sigmasky^2  & = &  \VAR{\Driftsky} + \sigmaskynoise^2 \\
  \sigmaload^2  & = & \VAR{\Driftload} + \sigmaloadnoise^2 \\
  \sigmaskyload  & = & \COV{\Driftsky,\Driftload} + \sigmaskyloadnoise 
 \end{eqnarray}

 \noindent
where it has been assumed that the random and deterministic parts are uncorrelated.
It is useful to identify two extreme cases:
the data stream is 
signal dominated, when $\VAR{\Driftsky} + \VAR{\Driftload} \gg \sigmaskynoise^2 + \sigmaloadnoise^2$,
or the data stream is 
noise dominated, when $\VAR{\Driftsky} + \VAR{\Driftload} \ll \sigmaskynoise^2 + \sigmaloadnoise^2$.
In the noise dominated case, the statistics of data will be largely determined by the 
statistics of noise, which in general could be considered normally distributed
and uncorrelated over short time scales,
given the $1/f$--noise
will introduce correlations over long time scales.
In the signal dominated case the statics of data will be instead determined by the kind of time
dependence in the signal. As an example if $|\Tskymean - \Tloadmean|$ is
large compared to the noise, while $\Driftsky$ and $\Driftload$  are negligible, 
the histogram of the signals will resemble the sum of 
two Dirac's delta functions $\delta(x-\Tskymean) + \delta(x-\Tloadmean)$ convolved with 
the distribution of noise. 

If a linear time dependence of the kind
 $\Delta T(t) = \dot{A} t + C$ is present the distribution of the samples will be uniform, and 
bounded between $\overline{T} \pm \dot{A} \tau/2$, where $\tau$ is the time interval 
relevant for the signal sampling. The variance will be $A_\tau^2/12$ where $A_\tau=\dot{A}\tau$
is the drift amplitude over the time scale $\tau$. The signal could be considered
noise dominated if $\tau < \sqrt{12}\sigma/|\dot{A}|$.
From the point of view of data compression, in determining whether a signal is noise
dominated or not, the critical factor is the time scale $\tau$. 
For our coupled signals, denoting with $\Ampdotsky$ and $\Ampdotload$ the drift 
rate in the sky and \load\ signals, and with $\Ampskytau$, 
$\Amploadtau$ the relative amplitudes, 
the relevant components of the covariance 
matrix will be 

 \begin{eqnarray}
  \VAR{\Driftsky}_{\tau} = \frac{\Ampskytau^2}{12} \\
  \VAR{\Driftload}_{\tau} = \frac{\Amploadtau^2}{12}\\
  \COV{\Driftsky,\Driftload}_{\tau} = \frac{\Ampskytau\Amploadtau}{12}
\end{eqnarray}

 \CHANGE{
 \noindent 
In this regard, the most important $\tau$ 
to be considered in this work is the time span for the chunk of data contained in a packet,
which is the minimum unit of formatted data sent by the REBA to the ground. 
Each scientific packet produced by the REBA has a maximum size corresponding to 
1024 octects, part of which has to be allocated for headers carring 
ancillary informations such as the kind of data in the packet or the time stamp.
So, even taking into account data compression, only a small amount of data 
can be stored in a packet corresponding to 
about $6-22$~secs, which depends on details such as the
attained compression rate and the frequency channel involved, as will be shown 
in Sect~\ref{sec:downsampling}.
 }
More complicated distributions may occur for a polynomial time dependence of the kind $\Delta T \propto t^n$,
or for a sinusoidal time dependence of period $P$: $\Delta T \propto sin(2\pi t/P)$, but in most cases a simple linear drift 
$\Delta T \propto t$ could be taken as a reference model given that 
non periodic drifts are bounded in amplitude by corrective actions commanded from the ground station, while
periodic variations have periods much longher than the time span of a packet.
 %
 %
Also in general it is assumed that the mean 
$\MEAN{\Driftsky} = 0$ and the $\MEAN{\Driftload} = 0$ but it is interesting to discuss 
even the case in which this will not be strictly true.

  \subsection{Data compression and \onboard\ processing}\label{sec:OP:scopes}

The strategy adopted to remain inside the downlink bandwidth is based on three processing steps:
 i) signal downsampling, 
 ii) signal conditioning and entropy reduction,
 iii) loss-less compression
 \cite{LFI:general:instrument,Comm:ICD}. 
A schematic representation of the sequence in which these steps are applied \onboard\ and whenever 
possible reversed \onground\ is given in Fig.~\ref{fig:processing}.
The figure refers to a single radiometer chain and it is ideally splitted into two parts: the upper 
part depicts the \onboard\ processing with cyan boxes denoting the main steps. The corresponding \onground\ 
processing is depicted in the lower part with the main steps coloured in yellow. 
 %
Green pads represents the processing parameters.
The first four of them are refered to as REBA parameters, and they are applied both 
\onboard\ and \onground.
The parameters are: the number of ADC raw samples to be coadded to form an
instrumental sample, $\Naver$, the two mixing parameters $\GMFO$, $\GMFT$, the offset $\mathcal{O}$ to be 
added to data after mixing and prior to requantization,  and the requantization step $q$.
The exact meaning of each of these parameters will be explained later in the text, when each step
will be explained in full detail.
 %
The important thing here is to recall that the \onboard\ parameters are imposed by telecommands 
sent from the ground. They are copied in each packet carring scientific data and 
\onground\ they are recovered from the packets to be applied by the \onground\ processing.
The $\GMF$ factor is a parameter of the ground processing and is computed from the total 
power data received on the ground.
The final products in the form of {\em Time Ordered Data} (TOI) either in total power or differentiated
are stored in an archive represented by the light--blue cylinder.

Before entering into the details of the various steps it has to be noted that
in principle a factor of two compression would be immediately gained by 
directly computing the difference between \sky\ and \load\ \onboard, i.e.
sending at Earth differentiated data. 
Although \onboard\ differentiation seems straightforward 
\footnote{This was the baseline of the \onboard\ processing for 
\cite{maris:2000,maris:2004}.}
it implies at least a couple of major disadvantages.
First, once the difference is made, separate information about the sky
and the \load\ is lost,
preventing an efficient detection and removal of other many second order
systematics.
Second a set of 44 $\GMF$\ factors could be in principle easily
uploaded \onboard\ and applied to the data,
but the $\GMF$ for each detector has to be fine--tuned on the real data. 
This would mean that the optimal $\GMF$\ should be continuously 
monitored and adjusted to avoid uncontrolled drifts for each radiometer,
but this is inpractical, having just 3~hours of connection per day.
In addition, an error 
in calibrating the $\GMF$\ will cause an irremediable loss of data.
Therefore, the best solution is to downlink the sky and the \load\ samples 
separately allowing the application on the ground of the optimal $\GMF$.

 \subsection{Downsampling}\label{sec:downsampling}
Each sky sample contains the sky signal integrated over a sky area as wide as the beam,
but given each radiometer is sampled at a frequency of \SamplingFreqHz\ 
the sky is sampled at an apparent resolution of about $1/2$~arcsec. 
On the other hand the beam size for each radiometer goes from 
14~arcmin for the 70~GHz to 33~arcmin for the 30~GHz. 
Consequently it is possible to co--add a number, $\Naver$,
of consecutive samples producing averaged samples whose sampling time correspond 
to a more reasonable resolution without any loss of information.

 \CHANGE{
The downsampling algorithm takes $\Naver$ couples of 
\skyload\ (\loadsky) samples from a given ADC; separates the two subsets
of signals;
computes the sum of sky and load subsets (represented by 32--bits signed integers);
interlaces them; and stores them as \skyload\ (\loadsky) couples
in an circular buffer for subsequent processing.
In normal processing the REBA converts these sums into averages by converting them into
floating--point format and then dividing them by $\Naver$ prior to perfom the 
subsequent steps of mixing, requantization and compression. 
In the case of diagnostic data processing the REBA transfers directly as output
these sums {\em as they are} i.e. without any other processing or compression.
In this case the ground--segment pipeline has the task of converting them 
into averages. This is a trade--off between the need for packets to
carry just data represented by 16 or 32 bits integers, and the need to avoind uncontrolled
round--off errors in the conversion of floating--point averages in integer values.
Note that the diagnostic telemetry is very limited in flight by telemetry bandwith.
}

The value of $\Naver$ depends on the beam--width, $\beamrad$, for the given detector 

 \begin{equation}\label{eq:naver}
 \Naver =  \frac{\omegaSpin \oversampling \sin\boresight }{\beamrad \freqsampling}
 \end{equation}

 \noindent
$\omegaSpin$ [rad/sec] is the rate at which the satellite spins about its spin axis
 \cite{PlanckBlueBook05,Dupac05,ScanningSAIT06},
$\boresight$ is the boresight angle between the telescope line--of--sight 
and the spin axis, and $\oversampling=3$ is the the number of samples per beam. 
Nominal values for the $\Naver$ are 126, 88, 53 respectively for the 30~GHz, 
44~GHz and 70~GHz frequency channels. The corresponding sampling frequencies in
the sky are then 65~Hz, 93.1~Hz and 154.6~Hz, while samples are produced at a rate
twice the sampling frequency.
This drastically reduces the data rate that becomes about 85~kbps without 
introducing an important loss in scientific information. 

The output of the downsampling stage can be seen as a sequence of \skyload\ couples
ordered according to the generation time $t$

$$\left(\begin{array}{c}\Tsky\\ \Tload\end{array}\right)_{t=0}, \left(\begin{array}{c}\Tsky\\ \Tload\end{array}\right)_{t=\delta_t}, \left(\begin{array}{c}\Tsky\\ \Tload\end{array}\right)_{t=2\delta_t},
\dots, \left(\begin{array}{c}\Tsky\\ \Tload\end{array}\right)_{t=n\delta_t}, \dots$$

 \noindent
where $\delta_t = 2\Naver/\freqsampling$
the samples are interlaced to generate a string of time ordered samples as 

$$\Tsky_{,0}, \Tload_{,0}, \Tsky_{,\delta_t}, \Tload_{,\delta_t},  \Tsky_{,2\delta_t}, \Tload_{,2\delta_t},  \dots, \Tsky_{,n\delta_t}, \Tload_{,n\delta_t}, \dots$$

 \noindent
in a manner similar to the output of the ADC.
But, while sky and \load\ samples in each ADC output buffer
are consecutive in time, this is no longer true for the downsampled values.
 %
As an example, assuming a sequence from the ADC where the even samples
are $\TskyADC$ and the odd samples are $\TloadADC$ (i.e., $\Tsky$, $\Tload$ sequences),
then any $\Tsky_{t}$ will be the sum of $\Naver$ samples with times
between $t$ and $t + 2(\Naver-1)/\freqsampling$ while and $\Tload_{t}$ will 
span the time range $t+1/\freqsampling$ and $t + 2\Naver/\freqsampling$. 
While this small time shift is not very important while observing sky sources
it might be thought to be relevant while attempting to correlate the observed 
signal with internal
sensors, such as those used to determine the level of perturbation introduced
by the active cooling. 
However, this problem is probably more theoretical than real as internal temperature variations do not occur on very small timescales.
For simplicity, unless needed, in the remainder of the text we will omit to specify time in our formulas.

 \subsection{Lossless compression, packeting and processing error}\label{sec:lossless:compression:packeting}
 \CHANGE{
To better understand the intermediate step of processing, i.e. mixing and requantization, it is necessary to
introduce here the last step of lossless compression.
The familiar technique of lossless compression is based on the ability of the compression software to recode 
a stream of symbols by using codewords which on average are shorther than those used in input,
and in a way which could be fully recovered 
on ground by a {\em decompression} code.
 }
The coding for the data stream in output to the compressor has to be optimized by taking into account the 
statistical distributions of the symbols in the input data stream. For this reason lossless compressors maintain 
an internal representation of the data distribution, such as the histogram or similar statistical indicators.
In our case the selected compression scheme is based 
on a 16--bit, zero order, adaptive arithmetic entropy encoder 
 \cite{LFI:REBA}.
The compressor assumes the data stream is represented by an uninterrupted
list of couples of 16--bit integers. 
It does not take any particular interpretation of the
content of the samples or of the order in which they are presented. 
Simply it continues to code and store the data in a packet up to when the maximum packet 
length is reached. After that the packet is then closed and a new packet is opened.
In establishing how to code the data, the compressor
uses an adaptive scheme to decide the best coding for the input data
as they are produced by the previous steps of the \onboard\ pipeline.
On ground the decompressor extracts the samples from each packet in the same order in 
which they have been introduced by the compressor. 
In this sense the couple compressor/decompressor act as a First In -- First Out 
device, and becomes nearly transparent in the scientific processing of the data.

The basic requirement for the packets produced by the compression stage 
is that of {\em packet independency} i.e. 
it must be possible to interpret the content of each packet
independently of all the others.
For LFI it means that the pipeline in the ground segment shall be allowed to generate from each single 
packet chunks of differentiated data.
So the compressor has to store consecutive couples of \skyload\ samples within each 
packet together with the information needed by the decompressor to interpret the 
compressed packets.
In addition the compressor has to be able to self--adapt its coding scheme to the 
statistics of the input signal, without the need of any prior information on it.
Finally, the compressor has to be fast enough to allow real--time elaboration of data with limited
memory consumption.
These requirements suggest the use of a compression scheme in which the compressor updates 
its internal statistical table each time it receives a sample. 
An empty statistical table is then imposed at beginning of the compression of a new packet assuring
complete independence.
When a symbol not present in the table is received as input a pseudo--symbol 
corresponding to a ``stop message'' is issued followed by the uncompressed 
new symbol, after that the internal statistical table is updated. 
If the symbol is not new the statistical table is updated and the symbol is compressed accordingly.
On ground the decompressor starts with the same empty internal statistical representation 
assuming the first symbol is a stop followed by a new symbol, and updating accordingly its internal
table as it receives symbols to decode or stop symbols.

The efficiency of a compressor is typically measured by the, so called,
compression $\Cr$ rate: 
the ratio between the length of an output string
$\Lout$ derived from the compression of an input string of length $\Lin$ 

 \begin{equation}
\Cr  = \frac{\Lout}{\Lin}.
\end{equation}

 \CHANGE{
 \noindent
Of course, to accomodate a given data rate $\DataRate$\ inside a given bandwidth $\BandWidth$
a target compression rate has to be obtained leading to the obvious definition
 }

 \begin{equation}
\CrTgt  = \frac{\DataRate}{\BandWidth}.
\end{equation}

 \noindent
It is well known that any lossless compressor based on entropy encoding
has an upper limit for the highest compression rate

 \begin{equation}
\CrTh =\frac{\Nbits}{\entropy},
 \end{equation}

 \noindent
where $\Nbits$ is the number of bits used for coding the samples and 
$\entropy$ is Shannon's entropy for the signal, which in turns depends
on its probability distribution function (PDF).
For an optimal compressor the theoretical $\CrTh$ for a digitized signal 
represented by integers
in the range $\Qmin \le Q \le \Qmax$ is given by 

  \numparts\begin{eqnarray} 
     \label{eq:cr:solita}
   \CrTh & = & \frac{\Nbits}{\entropy}, \\
     \label{eq:entropy:solita}
   \entropy & = &  -\sum_{Q=\Qmin}^{\Qmax} f_Q \log_2f_Q:
  \end{eqnarray}\endnumparts

 \noindent
where $\entropy$ is the Shannon entropy for the data stream,
$f_Q$ is the frequency by which the symbol or value $Q$ occurs in the data stream,
having $\lim_{f_Q\rightarrow0} f_Q \log_2f_Q =0$, and $\sum_{Q=\Qmin}^{\Qmax} f_Q =1$.
Non--idealities in the signal and in the compressor cause the effective $\Cr$ to be different from the 
expected $\CrTh$ having $\CrTh>\Cr$. Usually this is accounted for by scaling $\CrTh$ 
by a multiplicative efficiency factor $\eta$. However its exact determination is a complex task 
described in some detail in Sect.~\ref{sec:optim:oca2k} and for the time being we will neglect it.

From Eq.~(\ref{eq:cr:solita}) and (\ref{eq:entropy:solita}),
to maximize $\Cr$\ we need to minimize $\entropy$ for 
the input signal, forcing the reduction of its variance by requantizing the data. 
I.e. dividing the data by a quantization step, $q$, 
and rounding off the result to the nearest integer

 \begin{equation}
  Q = \round\left(\frac{X+\Offset}{q}\right),
 \end{equation}

 \noindent
where $\Offset$ is an additive constant usually defined by asking 
 \begin{equation}
   \MEAN{X+\Offset} = 0.
 \end{equation}

 \noindent
On ground the data are then decompressed and reconstructed by multiplying them by $q$.

 \begin{equation}
  \tilde{X} = q[Q-\Offset].
 \end{equation}

 \noindent
Some information is lost in the process resulting in a processing distortion, $\Qerr$, which in the simplest case is approximated by

 \begin{equation}\label{eq:Qerr:univariate}
  \Qerr = \RMS{\tilde{X}-X} \approx \frac{q}{\sqrt{12}}.
 \end{equation}

 \noindent
In \cite{maris:2000,maris:2004} we studied the case $X = \Delta T$, there 
it had been shown that for \Planck/LFI the statistics of the differentiated
data stream was approximated by a nearly univariate normal distribution with
$\sigma=\RMS{\DeltaT}$,
and that after re quantization and reconstruction
both $\CrTh$\ and $\Qerr$\ where largely parameterized 
by the $\sigma/q$ ratio with 

 \numparts\begin{eqnarray}
 \label{eq:compression:old}
 \CrTh & \approx &
    \frac{ N_{\mathrm{bits}}
          }{
          \log_2 \left( \sqrt{2\pi e} \frac{\sigma}{q} \right)
            }; \\
 \label{eq:qerror:old}
  \frac{\Qerr}{\sigma} & \approx& \frac{1}{\sqrt{12}} \left(\frac{\sigma}{q}\right)^{-1}; 
 \end{eqnarray}\endnumparts

 \CHANGE{
 \noindent
of course we need to assure that ${\Qerr}/{\sigma}<1$ which is expected to be respected 
``by design'' for \Planck/LFI.
In this regard it has to be recalled how the limit to any instrumental residual error for \Planck/LFI 
was assessed in the context of the overall error budget (including thermal, radiometric, optical 
and data-handling effects), driven by the ultimate requirement of a cumulative systematic error
per pixel smaller than $3\;\mu\mathrm{K}$ (peak--to--peak) at the end of the mission. 
In this regard the 10\% limit for the \onboard\ processing-related errors was set as a reachable 
requirement which should lead to a nearly-negligible impact on science.
 }

 \CHANGE{
We are now in the position of deriving the expected time spans for the compressed 
chunks of data contained in \Planck/LFI packets which have been
reported at the end of Sect.~\ref{sec:OP:signal:model}.
It is sufficient to consider that each packet may carry a maximum number 
$\PacketSamples$ of $\Nbits$ code--words representing scientific data, each 
representing on average $\CrTgt$ samples either \sky\ or \load\ and that
the sampling period after the downsampling is $\freqsampling/\Naver$ to obtain

 \begin{equation}\label{eq:packet:time:span}
   \PacketTimeSpan \approx \frac{2\Naver\PacketSamples\CrTgt }{\freqsampling},
 \end{equation}

 \noindent
where the factor of $2$ 
in front of $\Naver$ 
comes from the fact that the \sky\--\load\ cycle has half the frequency of the ADC sampling.
For \Planck/LFI  $\PacketSamples=490$, while a 
$\CrTgt = 2.4$ would be 
sufficient to allow proper data compression.
Of course, some level of variability among the detectors has to be allowed in order to cope
with non stationarities in the time series, or with the need to share the bandwidth among
different detectors in different manners. 
So a good fiducial range of values for $\CrTgt$ for individual detectors is 
$2 < \CrTgt < 3$, leading the expected values for 
$\PacketTimeSpan$ to vary over $15 -22$~sec, $10 - 16$~sec and $6-9$~sec respectively for the 
30~GHz, 44~GHz and 70~GHz frequency channels.
 }

  \subsection{The mixing algorithm}\label{sec:OP:mixing}

In general a data stream made of alternate sky and \load\ samples can not
be approximated by a normal, univariate distribution. Two different populations of 
samples, with different statistical properties are mixed together. 
In this case the $\Cr$\ could be reduce with respect to the univariate case. 
Furthermore, most of the first order instabilities, such as drifts and $1/f$--noise,
come from the radiometers, producing spurious correlated signals in $\Tsky$ and
$\Tload$. 
So undifferentiated time--lines for $\Tsky$\ and $\Tload$ 
are much more unstable 
than the corresponding $\DeltaT$ timelines further reducing $\Cr$.
In particular, fast drifts may rapidly force the compressor to saturate the packet
filling it with the decoding information, in the worst case resulting in $\Cr < 1$.
According to Eq.~(\ref{eq:compression:old}) It is possible to increase $q$ 
to keep $\Cr$ within safe limits, 
but the $\log_2$ dependence will drive $\Qerr/\sigma$ 
to rapidly grow towards $\Qerr/\sigma\gsim1$.
Alternatively a more complex compression scheme, one taking into account the
sky{--}{--}\load\ correlation could be implemented. But this would be computationally 
demanding and would increase the amount of decoding information to be 
placed in each packet.

One is left with the need to recover the advantage of differentiated
data, i.e. reduced instabilities and more homogeneous statistics, without losing the 
opportunity to have sky and \load\ separately on ground.
The adopted solution is inspired by the principle of the pseudo--correlation receiver.
Instead of sending to ground $(\Tsky, \Tload)$ couples, LFI delivers 
$(\Tone, \Ttwo)$ couples where each $\Tone$, $\Ttwo$ is an independent
linear combination of the corresponding $\Tsky$ and $\Tload$. 
Couples are then quantized and compressed. 
On ground the data are decompressed, dequantized recovering
 the linear combinations which are then reversed, recovering 
the original data \cite{miccolis:2003}. 
The most general formula for the linear combinations is 

 \begin{equation}\label{eq:mixing}
 \left(\begin{array}{c}T_{1} \\ T_{2}\end{array}\right) =
 \left(
 \begin{array}{cc}
  M_{1,\mathrm{sky}} &  M_{1,\mathrm{load}} \\
  M_{2,\mathrm{sky}} &  M_{2,\mathrm{load}} \\
 \end{array}
 \right) 
 \left(\begin{array}{c}\Tsky\\ \Tload\end{array}\right),
 \end{equation}

 \noindent
here the matrix, $\mathbf{M}$, in Eq.~(\ref{eq:mixing}) is named 
{\em mixing matrix} (actually it represents a mixing and a scaling unless 
$|\MixingM|=1$), its inverse 
$\MixingM^{-1}$ is the corresponding 
{\em de-mixing matrix}. 
The demixing matrix is 
applied on ground to recover the string of $(\Tsky, \Tload)$ out of the received 
string of $(\Tone,\Ttwo)$, this imposes $\left|\MixingM\right|\ne 0$.
The structure of $\MixingM$ determines the kind of coding strategy.
A particular structure for $\MixingM$ 
could better fit a given subset of constrains rather than another.
Both the $\Cr$\ and $\Qerr$\ are determined by $q$ as well as $\MixingM$.
In particular it is obvious that the processing distortion
will have the tendency to diverge for a nearly singular $\MixingM$.
A detailed analysis of the whole set of possible structures for $\MixingM$ 
is outside the scope of this paper, but in general $\MixingM$ 
shall be optimized in order to 
i) equalize as much as possible the $\Tone$ and $\Ttwo$ statistics, 
ii) reduce as much as possible the effects of first--order drifts, 
iii) maximize the $\Cr$, 
iv) minimize $\Qerr$.
For \Planck/LFI the following form for $\MixingM$ has been selected, 

 \numparts\begin{eqnarray}\label{eq:baseline:mix}
 \MixingM &=&\left( 
       \begin{array}{cc}
         1, & -\GMFO \\
         1, & -\GMFT \\
       \end{array}
    \right); \\
 |\MixingM| &=& \GMFT - \GMFO; \\
 \DeMixingM &=&\frac{1}{\GMFT - \GMFO} \left( 
       \begin{array}{cc}
         \GMFT , & -\GMFO \\
           1 , & -1 \\
       \end{array}
    \right) \; .
 \end{eqnarray}\endnumparts

 \noindent
which is not completely optimal,
since it allows optimization only on a subset of possible cases,
 but has the advantage of having a reduced amount 
of free parameters to be uploaded for each detector \footnote{Packets independency
imposes that all the free parameters ($\Naver$, $q$, $\Offset$, $\GMFO$ and $\GMFT$) 
have to be stored within each packet.}
and it is directly suggested by
Eq.~(\ref{eq:sky:minus:load}). 
Given in this case $|\mathbf{M}| = \GMFT - \GMFO$ It is evident how for a fixed $q$ the 
distortion will increase decreasing $|\GMFT - \GMFO|$.
In nominal conditions the mean $\MEAN{\Tsky} = 2.735$~K, the $\MEAN{\Tload} = 4~K$, 
and a possible choice for 
$\GMFO$\ and $\GMFT$\ is $\GMFO = 1$, $\GMFT = \GMF = 0.85$. 
But a tuning procedure is required to determine the best parameters
for each radiometer.
 %

\FIGmixingANDinterlacing

 \CHANGE{
The effect of mixing with respect to both signals and distributions is illustrated in 
Fig.~\ref{fig:mixing:interlacing} which
refers to the case of a data--stream which is signal dominated (see page~\pageref{sec:OP:signal:model}).
In the figure the dashed--lines represent the input signals, the full--lines
the corresponding mixed signals. The 
dotted--lines represent instead the limits for the variability induced by the noise. 
So the ramp in frame a) of Fig.~\ref{fig:mixing:interlacing} represents a model signal for 
$\Tsky(t)$ (blue dashed line) and $\Tload(t)$ (red dashed line). 
The corresponding interlaced data are shown in Fig.~\ref{fig:mixing:interlacing}b (blue dashed--line), 
where the limits of variability
induced by noise are not represented to avoid confusion.
The time lines for $\Tone(t)$ and $\Ttwo(t)$ calculated 
for $\GMFO=3/4$, $\GMFT=1/2$, 
are represented in Fig.~\ref{fig:mixing:interlacing}a and Fig.~\ref{fig:mixing:interlacing}b
as full--lines, and they are shifted to avoid
overlapping with the previous plots. It is evident the reduction in the variance which is 
associated with drifts in the mixed data.
}
Mixing transforms the bi--variate PDF which is for $\Tsky$, $\Tload$ signals into that for $\Tone$, $\Ttwo$. 
Fig.~\ref{fig:mixing:interlacing}c represents its effect on the bi--variate PDF for the noise and the drift.
Looking at the normal distributions of randomly variable signals in $\Tsky$ and $\Tload$, $g(\Tsky,\Tload)$, 
the line for which $g(\Tsky,\Tload)/g(\Tskymean,\Tloadmean) = 1/2$ is a dashed line, and the equivalent 
line for or $g(\Tone,\Ttwo)$ is a full-line.
The distribution for the deterministic signal (either a ramp, a drift or a triangular wave) is instead 
represented by a segment, plotted again as a dashed line to denote the $\Tsky$, $\Tload$ signal, and as a 
full line to denote the $\Tone$, $\Ttwo$.
The effect of mixing is a combination of a non--uniform scaling, a rotation and a shift. The circle transforms 
into an ellipse. The line changes its tilt and length. In other terms, the covariance matrix for mixed data will 
be different from the original ones. A very interesting consequence is in the case of a normally distributed noise
a correlated noise will appear in the mixed space even if the input noise is not correlated.
In general after mixing the major axis of the two figures has the tendency to align with the $x=y$ line and 
the center of the two figures shifts. In this case $|\MixingM| = \GMFT-\GMFO = 1$ and the  size of the two figures 
changes proportionally to $|\MixingM|$.
Interlacing transforms the bi--variate PDFs into univariate ones.
Fig.~\ref{fig:mixing:interlacing}d represents the effect of mixing on the PDF of interlaced data. 
Again dashed lines represents the distributions before mixing and the full-lines after mixing.
As in Fig.~\ref{fig:mixing:interlacing}b red indicates the $\Tload$ or $\Tone$ and blue $\Tsky$ or $\Ttwo$.
The bottom part of 
Fig.~\ref{fig:mixing:interlacing}d
represents the resulting distribution of interlaced signals, before (dashed)
and after (full) the mixing.
How these distributions have to be decomposed in terms of the projected distributions 
is shown in the top part of the 
Fig.~\ref{fig:mixing:interlacing}d which 
shows separately the distributions of the random and deterministic
components, respectively a normal and a box distributions
for $\Tsky$ and $\Tload$,
 The resulting distribution will be
the convolution of the two, which for $\Tsky$ and $\Tload$ are very similar to a 
box distribution.
Of course the drift makes the overall signal non Gaussian, in particular for 
$\Tload$.
The central part of 
Fig.~\ref{fig:mixing:interlacing}d
is the equivalent for the mixed signal. Here the drift is reduced and the convoluted signals
are more similar to the original normal distributions of noises.
I.e. mixing not only reduces the distance between the two components but, by reducing the 
drift, make them more normally distributed.

After mixing the $(\Tone, \Ttwo)$ couples are re--quantized giving the 
quantized couples $(Q_1, Q_2)$ which are interlaced and sent to the compressor
 
 \begin{equation}\label{eq:quantization}
  Q_{i}  = \round\left(\frac{T_{i} + \mathcal{O}}{q}\right),
 \; i = 1, 2;
 \end{equation}
 
 \noindent
where $\Offset$ is an offset added to force $(Q_1, Q_2)$ to stay within the range 
$[-2^{15}, +2^{15}]$
\footnote{
As anticipated in Sect.~\ref{sec:downsampling}, to reduce the roundoff error, the
division by $\Naver$ is applied 
generating $(Q_1, Q_2)$, in addition the parameter for digitization is not $q$ but
$\SecondQuant = 1/q$, so that Eq.~(\ref{eq:quantization}) shall be written
 \begin{equation}\label{eq:quantization:true}
  Q_{i}  = \round\left(\frac{\SecondQuant\left(T_{i} + \mathcal{O}\right)}{\Naver}\right),
 \; i = 1, 2;
 \end{equation}

 \noindent
however for consistency with \cite{maris:2000,maris:2004} in the following we will omit
the division by $\Naver$ and we will continue to use $q$ in place of $\SecondQuant$.
}.
On ground packets are entropy decoded, the data streams are de--interlaced
 and the corresponding
$(Q_1, Q_2)$ are used to reconstruct the sky and \load\ samples
 
 \begin{equation}\label{eq:decoding}
  \TalphaR  = \sum_{i=1,2} M^{-1}_{\alpha,i} [q \left( Q_{i} - \Offset \right)], 
 \;\;\; \alpha = \mathrm{sky}, \mathrm{load},
 \end{equation}

 \noindent
where $M^{-1}_{\alpha,i}$ are the components of $\DeMixingM$, and the ``$\xR$''
is used to distinguish a reconstructed quantity out of a processed one $x$.

Mixing will map \skyload\ statistics in the corresponding mixed statistics

  \numparts\begin{eqnarray} 
\label{eq:mixing:mean}
  \Timean & = & \Tskymean - \GMFi \Tloadmean, \; i=1,2; \\
\label{eq:mixing:drift}
  \Drifti   & = & \Driftsky - \GMFi \Driftload , \; i=1,2; \\
\label{eq:mixing:rms}
  \sigma_i^2 & = & \sigmasky^2 + \GMFi^2 \sigmaload^2-2\GMFi\sigmaskyload; \\
\label{eq:mixing:cov}
  \sigmaonetwo^2 & = & \sigmasky^2 + 
                        \GMFO\GMFT \sigmaload^2 - 
\frac{\GMFO+\GMFT}{2} \sigmaskyload .
 \end{eqnarray}\endnumparts

 \noindent
A simplification for the components of the covariance matrix 
can be obtained by assuming $\sigmasky$ as unitary, then defining
$\Talphan = \Talpha/\sigmasky$, $\alpha=$~$\mathrm{sky}$, $\mathrm{load}$;
 and incorporating $\Rsigma=\sigmasky/\sigmaload$ in
the $\GMFi$ factors giving in these normalized parameters

  \numparts\begin{eqnarray} 
\label{eq:mixing:DeltaT:norm}
  \DeltaTn & = & \Tskyn - \GMF\Tloadn \\
\label{eq:mixing:rms:norm}
  \sigmani^2 & = & 1 + \GMFni^2 -2\GMFni\corrsl; \\
\label{eq:mixing:cov:norm}
  \sigmanonetwo^2 & = & 1 + 
                        \GMFOn\GMFTn  - 
\frac{\GMFOn+\GMFTn}{2} \corrsl .
 \end{eqnarray}\endnumparts

 \noindent
where $\corrsl = \sigmaskyload/(\sigmasky\sigmaload)$, $\GMFni=\GMFi/\Rsigma$.
 %
 %
%
The corresponding transforms for the expectations is more complex.
Of course   $\Talphameann = \frac{\Talphamean}{\sigmasky}, \alpha = \mathrm{sky},\;\mathrm{load}$,
but

  \begin{eqnarray}
\label{eq:mixing:mean:normalized}
  \Timeann & = & \Tskymeann - \Rsigma\GMFni \Tloadmeann, \; i=1,2; \\
\label{eq:mixing:drift:normalized}
 \Driftni   & = & \Driftnsky - \Rsigma\GMFni  \Driftnload , \; i=1,2. \\
  \end{eqnarray}

 \noindent
However these normalizations are very useful in discussing the compression rate especially
after defining the obvious $\qnorm = q/\sigmasky$.

  \subsection{Modelling the statistical distribution of processed data}\label{sec:optim:cr}

We want here to define an approximation able to asses $\Cr \ge \CrTh$
in a simple way. 
For this reason we need to model the entropy for the signal entering the compressor.
Of course the accuracy to which it is possible to predict the final $\Cr$ is directly
connected to the accuracy to which the entropy is predicted.
In the following we present two approximations for the entropy of the signal. A lower accuracy approximation
and a high accuracy approximation. 

 \subsubsection{The low accuracy approximation}
Considering the usual reference cases
of a noise dominated
signal and a signal dominated by a linear drift,
in the first case the PDF may be approximated by a normal distribution,
in the second case the PDF may be approximated by a uniform distribution
with $f_Q=q/A$ values and $\sigma = A/\sqrt{12}$, 
but in all the cases for $\sigma/q >> 1$

 \begin{equation} \label{eq:entropy:kpdf}
  H = \log_2 \kpdf \; \frac{\sigma}{q},
\end{equation}

 \noindent
with $\kpdf$ a constant depending on the type of p.d.f. ranging from $\sqrt{12}$ for a uniform distribution
to $\sqrt{2\pi e}$ for a normal distribution.
The difference in $\entropy$ between these two extreme cases is $0.25$ bits.
The argument of the logarithm is the number of symbols in the distribution.
So $\entropy$ may be written also as $\entropy = \log_2 \Nsymbeff$ with 
$\Nsymbeff = \kpdf \sigma/q$. Of course in the case of the uniform distribution
$\Nsymbeff = \Nsymb$.
The PDF for the interlaced signals gives the probability to have a symbol $Q$ either from
processes
$Q_1$ or $Q_2$. Then 

 \begin{equation}
 \mathcal{P}(Q) = \frac{\mathcal{P}_1(Q) + \mathcal{P}_2(Q)}{2}
 \end{equation}

 \noindent
with $\mathcal{P}_i(Q)$, $i=1$,~2 the marginal PDF for the $Q_i$ drawn from the bi--variate PDF
$\mathcal{P}(\Qone,\Qtwo)$.
For our extreme cases both $\mathcal{P}_i(Q)$ are
uniformly distributed or normally distributed according to 
the original PDF from which they are drawn \footnote{In this case the central limit theorem does not apply to the signal
with a uniform PDF given its deterministic nature.}.
This allows one to neglect, in estimating the entropy of the interlace signal, their mutual correlation.
Then the entropy for the interlaced data is just a function of the RMS for the two distributions
$\sigma_1$, $\sigma_2$, and their separation, $\Separationn$, 

\begin{equation}
   \Separationn = \frac{2}{\kpdf} \frac{\EXP{\Ttwon} - \EXP{\Tonen}}{\sigmanone+\sigmantwo}
\end{equation}

 \noindent
which is a normalized measure of the distance between the two peaks.
 %
After some algebra 

\begin{equation}
   \Separationn = 2\frac{{\GMFOn-\GMFTn}}{\kpdf} \frac{\Tloadmeann }{\sigmanone+\sigmantwo}
\end{equation}

Then the entropy will be just a function of 
$\sigmanone$, $\sigmantwo$ and $\Separationn$.
An exact analytical expression for $\entropy$ can not be obtained for this case. 
 %
However, 
it is easy to see that in the limit $|\Separationn| \gg 1$ the entropy takes the limiting value

 \begin{equation}\label{eq:entropyinf:def}
 \entropyinf = \frac{\entropyone + \entropytwo}{2} + 1,
\end{equation}

 \noindent
giving 

 \begin{equation}\label{eq:entropyinf}
  \entropyinf = \log_2(\kpdf) + \log_2(\sqrt{\sigmanone\sigmantwo})-\log_2{\qnorm} + 1.
\end{equation}

 \noindent
On the other side if $\Separationn = 0$ and $\sigmanone=\sigmantwo$ the two PDFs collapse
giving 
$\entropysingle = \entropy_1 = \entropy_2$.
In all the other cases $\entropysingle \le \entropy(\Separation) \le \entropyinf$. 
 %
 %
The important point here is the assumption that

 \begin{equation}\label{eq:entropy:low:approx}
 \entropy \approx \entropyinf,
\end{equation}

 \noindent
would never overestimate the entropy by more than 1~bit or $\approx 30\%$,
so that neglecting the compressor inefficiencies a 
sufficient condition to asses $\Cr\ge\CrTgt$ would be

 \begin{equation}\label{eq:entropy:low:approx:condition}
 \entropyinf < \entropyTgt 
\end{equation}

 \noindent 
with 
$\entropyTgt = \Nbits/\CrTgt$, 
or 

 \begin{equation}\label{eq:sigmaprod:low:approx:condition}
 \sqrt{\sigmanone\sigmantwo} < \qnorm \frac{2^{\Nbits/\CrTgt}}{2\kpdf};
\end{equation}

 \noindent
and so 

 \begin{equation}\label{eq:q:qerr}
 \sqrt[4]{(1 + \GMFOn^2 - 2\GMFOn \corrsl)
       (1 + \GMFTn^2 - 2\GMFTn \corrsl)}
< \qnorm \frac{2^{\Nbits/\CrTgt}}{2\kpdf};
\end{equation}

 \noindent
Eq.~(\ref{eq:entropy:low:approx}),
Eq.~(\ref{eq:entropy:low:approx:condition}),
and the derived Eq.~(\ref{eq:sigmaprod:low:approx:condition}) and 
Eq.~(\ref{eq:q:qerr})
 represent our low--order approximation for the optimization of REBA parameters.
In particular Eq.~(\ref{eq:q:qerr}) puts a lower limit to $\qnorm$ (and $q$)
for a given $\Cr$, in fact for $\GMFOn=\GMFTn=0$, $\qnorm$ must be larger or equal to

 \begin{equation}\label{eq:qmin}
 \qmin(\CrTgt) = \frac{2\kpdf}{2^{\Nbits/\CrTgt}}.
 \end{equation}

\FIGqoptimal

 \CHANGE{
As shown in Fig.~\ref{fig:qoptimal}
for $\corrsl=0$,
the left side of Eq.~(\ref{eq:q:qerr}) in the $(\GMFOn,\GMFTn)$ plane
has a minimum at $(0,0)$. 
Its iso--contour lines
are closed, all centered on the origin, having four axis of symmetry
$\GMFOn=0$, $\GMFTn=\pm\GMFOn$, and $\GMFTn=0$. The maximum distance from the
origin of iso--contour lines occurs for
$\GMFOn=0$ or $\GMFTn=0$, the minimum occurs along the $\GMFTn=\pm\GMFOn$ line.
Changing $\corrsl=0$ toward negative or positive values 
the iso--contour lines are again closed but their symmetry changes taking a more ``cuspidal'' shape,
which is symmetrical about the $\GMFOn=\corrsl$ and the $\GMFTn=\corrsl$ lines. 
In any case the value of the function decreases
going toward the $\GMFTn=\GMFOn=\corrsl$ point where it has a minimum.
From the figure it is evident that
when converting $(\GMFOn, \GMFTn)$ to $(\GMFO, \GMFT)$
a larger $q/\qmin$ ratio or a smaller $\sigmaload/\sigmasky$ increases
the size of the region enclosed by each contour. 
Eq.~(\ref{eq:q:qerr}) and Eq.~(\ref{eq:qmin}) define in this low order approximation the optimal $q$ for which $\Cr=\CrTgt$
 }

 \begin{equation}\label{eq:qoptimal}
  \qnormopt= 
     {\qmin(\CrTgt)} {\sqrt[4]{(1 + \GMFOn^2- 2\GMFOn \corrsl)(1 + \GMFTn^2- 2\GMFTn \corrsl)}}
 \end{equation}

 \noindent
from which $\qopt$ is simply derived as $\qopt=\sigmasky\qnormopt$. 
This equation does not constrain completely $\qopt$ and 
for this reason we have to take into account the processing error
as explained in Sect.~\ref{sec:processing:err}.

 \subsubsection{The high order accuracy approximation}
The accuracy by which $\qopt$ is determined by Eq.~(\ref{eq:qoptimal}) is solely determined by the accuracy
of imposing $\entropy = \entropyinf$. In itself, given the statistics of the input signal it would be not a problem to 
calculate by numerical integration $\entropy$ as a function of $\GMFO$, $\GMFT$ and $q$, but of course this 
would be quite expensive from a computational point of view.
For this reason in 
Appendix~\ref{app:normalized:entropy} is derived a high accuracy 
algorithm to compute $\entropy$ based on simple equations from which $\qopt$ could be readily obtained.
However from the conceptual point of view the high accuracy method
does not introduce any new detail in the discussion, the remaining part 
of this section refers only to the low--accuracy method unless
otherwise stated.

 \subsection{Processing error of the mixing/demixing algorithm}\label{sec:processing:err}

The most important way to quantify the processing error is the measure 
of the distortion in the undifferentiated or differentiated data.
The statistics of such distortions are taken as metrics of the quality of the process. 
For the undifferentiated data 

 \begin{equation}\label{eq:distortion:total}
   \delta_{\alpha} = \TalphaR  - \Talpha,
 \end{equation}

 \noindent
$\alpha = \mathrm{sky}$, $\mathrm{load}$.
By following the methods of \cite{maris:2004} from Eq.~(\ref{eq:decoding}) and 
Eq.~(\ref{eq:distortion:total}) it is easy to derive 
the covariance matrix of the quantization error, 
$\QerrAlphaBeta = \COV{\delta_{\alpha},\delta_{\beta}}$

 \begin{equation}\label{eq:covariance:matrix}
  \mathbf{E}_{\mathrm{q}} = 
   \frac{q^2}{12}
   \frac{1}{(\GMFT - \GMFO)^2}
   \left( 
   \begin{array}{cc}
   \GMFO^2 + \GMFT^2, &  \GMFO + \GMFT \\
       \GMFO + \GMFT, &    2 \\ 
   \end{array}
   \right).
 \end{equation}

 \noindent
The distortion of differentiated data is instead expressed by 

 \begin{equation}\label{eq:distortion:diff}
   \delta_{\mathrm{diff}} = (\tilde{T}_{sky}  -
\GMFtilde \tilde{T}_{load} )  - 
 (\Tsky - \GMF \Tload);
 \end{equation}

 \noindent
where $\GMFtilde$ is the $\GMF$ determined on the processed data, 
which in general will be slightly different 
from the $\GMF$ determined on the original ones. However, assuming 
$\GMFtilde \approx \GMF$ from Eq.~(\ref{eq:covariance:matrix})
the variance of $\delta_{\mathrm{diff}}$ is 

 \begin{equation}\label{eq:differentiated:qerr}
   \Qerrdiff^2 = \frac{q^2}{12} 
   \frac{(\GMFT-\GMF)^2 + (\GMFO-\GMF)^2}{(\GMFT - \GMFO)^2}
 \end{equation}

The first important fact which has to be stressed is that the variances of both errors 
are proportional to $q^2/(\GMFT - \GMFO)^2$.
Of course a nearly singular matrix with $\GMFT \approx \GMFO$ will result in very large
errors.
In addition, Eq.~(\ref{eq:covariance:matrix}) shows that, despite quantization errors
for $Q_1$ and $Q_2$ are uncorrelated, application of demixing causes 
processing errors in $\Tskytilde$ and $\Tloadtilde$ to be correlated
unless

 \begin{equation}\label{eq:decorrelation:condition:LFI}
  \GMFO + \GMFT = 0.
 \end{equation}

 \noindent
However, expanding the numerator of Eq.~(\ref{eq:differentiated:qerr}) produces
$\Qerrdiff^2 \propto \GMFO^2+\GMFT^2 + 2\GMF^2 - 2\GMF(\GMFO+\GMFT)$
suggesting 
the important result that a not null correlation in the quantization errors may lead to a reduction of 
the error distortion in the differentiated data

 \noindent
Another very important case is that in which either $\GMFO = \GMF$ or $\GMFT = \GMF$ 
in this case Eq.~(\ref{eq:differentiated:qerr}) reduces to 

 \begin{equation}\label{eq:differentiated:qerr:reduced}
   \Qerrdiff^2 = \frac{q^2}{12}.
 \end{equation} 

 \noindent
which is the same result that we would have had 
quantizing differentiated data and studied in \cite{maris:2004}.
 \CHANGE{
This fact has been used in the first version of the optimization software, 
designed for the first run of the ground tests (the RAA tests described in
 \cite{LFI:general:instrument})
to increase its speed, together with the fact that 
$\Cr$ and $\Qerr$ are not sensitive to an interchange of $\GMFO$ and $\GMFT$
but rather they are just sensitive to $|\GMFT-\GMFO|$.
However, in the subsequent tests the more general and accurate
procedure described here has been successfully applied.
 }

\FIGquackone
\FIGquacktwo

  \subsection{Saturation}\label{sec:optim:saturation}

A further source of processing error is the saturation of the dynamical range of the data format used
to process the data, or simply {\em Saturation}.

Saturation occurs when the argument of the $y=\round[x]$ function exceeds the maximum
range of values allowed by the computer to represent our results. 
Typically what happens is that $y=\round[x]$ 
returns an $\Nbits$ signed integer. If $|x|> 2^{\Nbits/2}$ 
an overflow or an underflow will occur. Depending on the implementation
of the $y=\round[x]$ function the value of $y$ could be either forced to be $\pm 2^{15}$ with the sign
depending on $x$, or modular arithmetic could be applied so that 
as an example a too large $x>0$ could be mapped into $y<0$. In all cases 
the whole subsequent reconstruction will produce meaningless results.
So it is fundamental to avoid saturation.

For this purpose we define 
as a quantitative index for the saturation the 
instantaneous $\Quack$ ratio \footnote{From QUantization Alarm Check.}

 \begin{equation}\label{eq:quack}
   \Quacki(t) = \frac{T_i(t)}{q 2^{\Nbits-1}}, i=1,2.
 \end{equation}

 \noindent
Saturation occurs if at some time 
$|\Quacki(t)| \ge 1$ and the non--saturation condition is 

 \begin{equation}\label{eq:no:quack}
   |\Quackone(t)| < 1 \wedge |\Quacktwo(t)| < 1; \; \forall t
 \end{equation}

 \noindent
in general this will put limits on 
$\GMFO$, $\GMFT$, $q$ and $\Offset$. 
Assuming to have applied the optimized offset of Eq.~(\ref{eq:offset:optimization})
the linear combinations are 

 \numparts\begin{eqnarray}\label{eq:fluctuations:propagation}
  \Quackone(t) &=& 
   \frac{ 
            \Ampsky(t) - \GMFO\Ampload(t) + 
            \frac{\GMFT-\GMFO}{2} \Tloadmean \pm n\sigmaonenoise 
  }
         {q 2^{\Nbits-1}} \\
  \Quacktwo(t) &=& 
  \frac{
           \Ampsky(t) - \GMFT\Ampload(t) - 
            \frac{\GMFT-\GMFO}{2} \Tloadmean \pm n\sigmatwonoise  
  } 
  {q 2^{\Nbits-1}}
 \end{eqnarray}\endnumparts

 \noindent
where $n\approx5$ is used to assess a safety region against random fluctuations.
In computing $\Quack$ the effect of mutual cancellation of extremal values 
must be considered.
A conservative estimate would be to propagate the modulus of each variation

 \begin{equation}\label{eq:fluctuations:propagation:safe}
  \max(\Delta T_i) = \max{|\Ampsky|} + |\GMFi| \max{|\Ampload|} + 
        \left| \frac{\GMFT-\GMFO}{2} \right| |\Tloadmean| + n \sigma_i,
 \end{equation}

 \noindent
with $\min(\Delta T_i) = -\max(\Delta T_i)$, 
but it is better to explore the various combinations of minima and maxima within
Eq.~(\ref{eq:fluctuations:propagation}), producing a set of partial $\Quack$ indexes
which have to be independently satisfied. 
An example of such method is illustrated in 
Fig.~\ref{fig:quack:one} and
Fig.~\ref{fig:quack:two}.

In general, the separation between $\Tone$ and $\Ttwo$ is a function of time,
whose measure is given by the divergence $\nablaT$, a parameter 
just sensitive to $\Ampload$ and $\Tloadmean$

 \begin{equation}\label{eq:divergence}
  \nablaT(t) = -2\frac{(\GMFT - \GMFO)(\Ampload(t) + \Tloadmean)}
                    {\kpdf (\sigmaone+\sigmatwo)}
 \end{equation}

 \noindent
of course $|\nablaT|$ will be constant when $\Ampload=0$. 

To determine the region of parameter space $\GMFO$, $\GMFT$ which satisfies Eq.~(\ref{eq:no:quack}) it 
is most convenient to work in the $(\GMFO, \GMFT-\GMFO)$ space, there the most general condition is

 \numparts\begin{eqnarray}\label{eq:inclusion:conditions:alternate}
 \label{eq:inclusion:conditions:alternate:a}
   \GMFO (a-1) - b  &< \GMFT - \GMFO <& \GMFO (a-1) - c; \\
      \label{eq:inclusion:conditions:alternate:b} 
  c - (\GMFT-\GMFO) (a-1) &< (a-1) \GMFO <& b - (\GMFT-\GMFO) (a-1);
 \end{eqnarray}\endnumparts

\noindent
with the dimensionless coefficients

 \numparts\begin{eqnarray}\label{eq:inclusion:conditions:coeff}
   \label{eq:inclusion:conditions:coeff:a} 
 a & = & \left(\frac{2\Ampload}{\Tloadmean}+1\right); \\
   \label{eq:inclusion:conditions:coeff:b}
 b_\pm & = & \frac{2}{\Tloadmean}\left(\Ampsky+q2^{\Nbits-1} \pm n\sigma \right);\\
  \label{eq:inclusion:conditions:coeff:c}
 c_\pm & = & \frac{2}{\Tloadmean}\left(\Ampsky-q2^{\Nbits-1} \pm n\sigma \right); \\
   \sigma & = & \max(\sigmaone, \sigmatwo);
 \end{eqnarray}\endnumparts

 \noindent
For $n=0$ those conditions define a diamond--shaped region whose vertices are 

 $$
  \begin{array}{llll}
  A: & \left(\frac{c+b(a-1)}{a(a-1)}, \frac{c-b}{a} \right), & B: & \left(\frac{b}{a-1}, 0 \right), \\
&&&\\
  C: & \left(\frac{b+c(a-1)}{a(a-1)}, -\frac{c-b}{a} \right),& D: & \left(\frac{c}{a-1}, 0\right), \\
  \end{array}
 $$

 \noindent 
Note that $B:$ and $D:$ lie on $\GMFT-\GMFO=0$ line, while $A:$ and $D:$ are above and below it,
the exact ordering depending on the signs. The center of the diamond--shaped region is locate on 
$((b+c)/(a-1),0)$. If $b+c=0$ the region is centered on the origin of the Cartesian system,
$A:$, $C:$ and $B:$, $D:$ are mutually opposed.
In the case 
$a=1$ the region degenerates into a band parallel to $\GMFT-\GMFO=0$, and 
bounded by $-b < \GMFT-\GMFO < -c$.


 %
 \noindent
First considering the stationary case, having $\Ampsky=0$, $\Ampload=0$,
then $a=1$, $b=-c$ the only constraint for $n = 0$ 
in this case $\Quackone$ and $\Quacktwo$ defines the same condition which 
is 

 \begin{equation} \label{eq:saturation:constrain:constant}
     |\GMFT - \GMFO|\frac{\TloadZ}{2} < q\Nsat.
 \end{equation}


\noindent
identifying simply a band in the space center around the $\ForbiddenCase$ line.
But the effect of noise has to be considered i.e. $n>0$ so that now
there is a difference between $a_+$ and $a_-$ so that $\Quackone$ 
defines a vertical band and $\Quacktwo$ a diagonal band whose intersection is 
the diamond--shaped region above, see Fig~\ref{fig:quack:one}a. Changing the ratio
between $\sigmatwonoise/\sigmaonenoise$ will not change the shape but just the size of the region,
Fig~\ref{fig:quack:one}b.  Changing $\Tskymean$ and $\Tloadmean$ instead will
change the shape of the allowed region, Fig~\ref{fig:quack:one}c, Fig~\ref{fig:quack:one}d
and Fig~\ref{fig:quack:two}a.
If $\Tloadmean = 0$
the saturation condition becomes
$|1 - \GMF_i|\cdot|\Amp(t)| < q\Nsat$ i.e.:
$1-q\Nsat/\AmpMax \le \GMF_i \le 1+q\Nsat/\AmpMax$, with $\AmpMax = \max(|\Amp(t)|)$.
This defines a rectangular region with diagonal $\ForbiddenCase$.
In the limiting case for $n=0$, $\AmpMax \rightarrow 0$ the
allowed region becomes the whole plane, while in the opposite case 
$\AmpMax \rightarrow \infty$ the allowed region shrinks toward $\GMFO = \GMFT = 1$.
 \noindent
Perturbations in the sky channel, such as the cosmological dipole,
introduce a fluctuation which affects just the sky,
in this case 
$\Ampload=0$ and $a=1$.
Even here the simplest case is  $\TloadZ = 0$, 
or the forbidden $\ForbiddenCase$ in that case 
$\max(|\AmpD(t)|) < q\Nsat$ is a sufficient condition which puts a limit
just on $q$.
In the most general case from Eq.~(\ref{eq:inclusion:conditions:alternate})
the limits of the allowed region are 
$-2\Ampsky/\TloadZ - 2q\Nsat/\TloadZ < \GMFT-\GMFO<-2\Ampsky/\TloadZ + 2q\Nsat/\TloadZ$
 \noindent
Other effects, such as instabilities of the 4--K \load, could 
affect just $\Tload$ in this case $\Ampsky = 0$, $b=2q\Nsat/\TloadZ=-c$
even here the simplest case is that in which $\Tloadmean = 0$, 
in that case $\max(|\Amp(t)|) < q\Nsat$ is a sufficient condition which 
puts a limit just on $q$.
In general a diamond--shaped region symmetrical around the origin represents the allowed region.
This is represented in Fig~\ref{fig:quack:two}c.

 \noindent
Drifts in the gain of the amplifiers in the radiometers, such as those produced by
thermal effects, add correlated or anticorrelated signals in sky and \load. So the model case 
to be considered is the one in which $\Amp(t) \approx \Ampsky(t) \approx \Ampload(t)$.
In this case with $a=(b+c)/2$ and 
Eq.~(\ref{eq:fluctuations:propagation}) and Eq.~(\ref{eq:divergence})
define the regions in Fig~\ref{fig:quack:two}b and Fig~\ref{fig:quack:two}d.

It is interesting to consider
the range of values assumed by $\Pone$ and $\Ptwo$ with $\Amp$ in the range
$\AmpLow < \Amp < \AmpUp$ then 
$|P_{i,\mathrm{Up}} - P_{i,\mathrm{Low}}| = |\AmpUp - \AmpLow|\cdot|1-\GMF_i|$.
So the condition to have identical ranges, is $\MandatoryCase$ and 
$\GMFT + \GMFO = 2$ or $(\GMFT-\GMFO) = 2 - 2\GMFO $
It has to be noted that in nominal conditions, $\TskyZ\approx2.73$~K, 
$\TloadZ\approx4$~K while $\Tsky$, $\Tload$ fluctuations are expected at the 
level of at most some $10^{-2}$~K giving $\Ampload/\TloadZ \approx \Ampsky/\TloadZ \approx \mathrm{some} \times 10^{-2}$, and hence $a\approx1$, $b\approx-c$. However outside this case, in particular during
testing and the cooling in the transfer phase these conditions could be severely violated. 

 \section{Optimizing the On--board Processing}\label{sec:optim}

The optimization of the algorithm consists in determining the ``best'' combination of
the set of processing parameters i.e. the ``best'' n--tuple
$\Naver$, $\GMFO$, $\GMFT$, $\Offset$, $\SecondQuant$ or $q$.
It is mandatory that the optimization procedure will keep within safe limits 
$\Cr = \CrTgt$. 

The classical approach would require a function of merit
and a searching algorithm through the corresponding 
$N \times R^4$ parameters space to be applied to each of 44 detectors.
However a reduction of the cardinality of space comes from the fact that,
by-requirement, the nominal $\Naver$ is fixed by the oversampling factor for the beam,
so apart from the cases in which a different oversampling is required, the
$\Naver$ in nominal conditions is fixed. The only cases in which $\Naver$ could be 
varied are: i.) sampling of planets for beam reconstruction; ii.) ground testing and 
diagnostics.
The first case occurs when the beam has to be reconstructed with higher detail than the one
reachable with the nominal oversampling factor $\oversampling = 3$. So it is possible 
to ask the \onboard\ processor to decrease $\Naver$ increasing proportionally the
data--rate from the feed--horns which will be affected by a planet. To arrange
the higher--throughput of scientific telemetry, $q$ will have to be increased, increasing
proportionally $\Qerr$. 
In the second case the value of $\Naver$ could be varied either to increase the time
resolution, as an example if sampling of some perturbation characterized by time scales
compatible to $\Naver/\freqsampling$ has to be investigated, or if some temporary 
shortage in the telemetry rate is imposed, asking in this case to increase 
$\Naver$. Also while testing on ground for long term drifts, 
the sky is replaced by a dummy load at constant temperature. In this case 
time resolution is no longer an asset and $\Naver$ could be increased.
A further reduction of the parameter space to be explored 
comes from the fact that usually 
$\Offset$ is optimized in order to have $\MEAN{T_{\mathrm{interlaced}}+\Offset} = 0$ where 
$T_{\mathrm{interlaced}}$ are
the interlaced samples produced after by Eq.~(\ref{eq:baseline:mix}). It is then easy to
derive that 
$\MEAN{T_{\mathrm{interlaced}}+\Offset} = \MEAN{\Tone} + \MEAN{\Ttwo}+2\Offset$ so that
$\Offset_{\mathrm{optimal}} = - (\MEAN{\Tone} + \MEAN{\Ttwo})/{2}$, 
and with some simple algebra

 \begin{equation}\label{eq:offset:optimization}
   \Offset_{\mathrm{optimal}} = - \Tskymean + \frac{\GMFO + \GMFT}{2} \Tloadmean,
 \end{equation}

 \noindent
where the mean has to be computed over a suitable time span.
What remains is a $\Re^3$ parameter space to be explored $(\GMFO, \GMFT, q)$.

\subsection{Target function}\label{sec:target:function}

The target function $\TargetFunction(\GMFO, \GMFT, q)$ 
for the optimization would 
i.) asses $\Cr = \CrTgt$ to be kept within safe limits;
ii.) asses $\Qerr$ to be kept as small as possible;
iii.) asses additive constrains.
These constrains do not allow a unambiguous definition of a 
target function.
As an example, even for a stationary signal dominated by white noise, $\Cr$ computed on each packet
is a random variable. So the question is whether $\Cr = \CrTgt$ has to be interpreted strictly, i.e. 
forcing each packet to have $\Cr = \CrTgt$  or on average leaving space 
for lower and higher $\Cr$?  In general it would not be critical if some fraction of the
packets would be compressed at a rate lower than $\CrTgt$. 
The requirement on $\Qerr$ is even worse defined. Of which $\Qerr$ are we speaking?
As shown in sect.~\ref{sec:OP:mixing} it is evident that 
there is not a general definition for
$\Qerr$. Depending on the scope of the data acquisition it could be more interesting to have
a low $\Qerr$ for $\Tsky$ or $\Tload$ or $\DeltaT$ computed for some reference $\GMF$.
More over neither $\Qerr$ nor $\Qerrdiff$ are functions with a minimum and they vary over the 
full range of positive values. 
In addition within a pointing period $N$ repeated sky samples 
are acquired. In making maps repeated samples are averaged and $\Qerr$ will be reduced by a 
factor $1/\sqrt{N}$
\cite{maris:2004}.
So a relatively high $\Qerr$ could be acceptable at the level of single samples when observing
stationary sources. However the ratio between $\Qerr$ and the noise will not change after averaging.
So a convenient choice would be to consider $\Qerr/\sigma$, where $\sigma$ could be the 
RMS of $\Tsky$, $\Tload$ or $\DeltaT$ depending on the case.
The only hard constraint which has to be considered is that needed to avoid saturation.

The general formula for the target function is 

 \begin{equation}\label{eq:target:function:general}
 \TargetFunction(\mathbf{\Theta}) = \prod_c {\CriterionFunction(\mathbf{\Theta})}^{\Policy}, 
\end{equation}

\FIGgammamincomposition

 \noindent
$\mathbf{\Theta}$ is a vector in the parameter space,
$\CriterionFunction$ is a function varying over the range $[0,1]$ with 
$\CriterionFunction(\Theta)=1$ if $\Theta$ fits the particular criterion $c$ for which the
function is defined, 
$\CriterionFunction(\Theta)=0$ if $\Theta$ does not fit this criterion.
Intermediate values may be also defined in the [0,1] range measuring the
goodness of fits. As an example, a criterion for optimal
$q$ is to have $\gammadiff = \min(\gammadiff)$, the corresponding 
criterion function is $\CriterionFunction(\Theta)=\min(\gammadiff)/\gammadiff$.
The exponents $\Policy\ge0$, with $\sum_c \Policy = 1$
are weights defining the relative importance of each criterion 
within a given policy.
In general is is better to have $\CriterionFunction(\Theta)$ which are derivable.
In some case it is necessary to to deal with poles that have to be avoided.
A solution is to define a metric $\Metric(\Theta)\ge0$ which could have a single pole
for which $\Metric(\Theta)\rightarrow +\infty$
and take $\CriterionFunction(\Theta) = e^{-\Metric(\Theta)}$
or $\CriterionFunction(\Theta) = 1/(1+e^{-\Metric(\Theta)})$.
Typical criteria are shown in Tab.~\ref{tab:typical:criteria}

\TABtypicalcriteria

\FIGequalscrit
\FIGallminimacrit

  \subsection{Analytical Optimization}\label{sec:optim:analytical}

Analytical optimization (AO)
is based on analytical formulas assuming either normally distributed or uniformly distributed signals.
As a starting point for more refined numerical optimization.
At the root of this method of optimization is the requirement of minimizing the processing errors, as an example the $\Qerrdiff$.
However given they diverge at $\ForbiddenCase$ it is necessary to consider the maximization of their inverse normalized to the
minimal value, as an example defining for $\Qerrdiff$ the function 
$\Gammadiff = \min(\Qerrdiff)/\Qerrdiff$. These functions become 0 for $\ForbiddenCasen$, unless either
$\GMFOn=\GMFn$ or $\GMFTn=\GMFn$

Again it is convenient to use the normalized parameters $\GMFni$, 
in this case the covariance matrix of processing errors is 

 \begin{equation}\label{eq:covariance:matrix:normalized}
  \mathbf{\breve{E}}_{\mathrm{q}} = 
   \frac{\breve{q}^2}{12}
   \frac{1}{(\GMFTn - \GMFOn)^2}
   \left( 
   \begin{array}{cc}
   \GMFOn^2 + \GMFTn^2, &  \frac{\GMFOn + \GMFTn}{\Rsigma} \\
       \frac{\GMFOn + \GMFTn}{\Rsigma}, &    \frac{2}{\Rsigma^2} \\ 
   \end{array}
   \right).
 \end{equation}

 \noindent
In the framework of the low--level approximation for $\qopt$ calculation, 
after replacing Eq.~(\ref{eq:qoptimal}) into the $\Qerrdiff$ from Eq.~(\ref{eq:differentiated:qerr}),
substituting 
$\GMFi \rightarrow \GMFni$, $q \rightarrow \breve{q}$ and $\GMF \rightarrow \GMFn = \GMF/\Rsigma$,
and taking its reciprocal one obtain

 \begin{equation}\label{eq:qerr:opt:simple}
  \Gammadiff = 
  \frac{(\GMFTn - \GMFOn)^2}{\left[(\GMFOn-\GMFn)^2+(\GMFTn-\GMFn)^2\right]  
       \sqrt[4]{(1 + \GMFOn^2 - 2\GMFOn\corrsl)(1 + \GMFTn^2 - 2\GMFTn\corrsl)}} 
 \end{equation}

 \noindent
$\Gammadiff$ is symmetrical with respect to the axis $\ForbiddenCasen$ and has a maximum where the processing
error has a minimum.
 %
 %
There is no analytical way to maximize $\Gammadiff$.
However, 
Fig.~(\ref{fig:gamma:min:composition}) shows the contour plot for the various components of this function.
The denominator is the product of a function which is constant over 
circles centered on $\ForbiddenCasen=\GMFn$ (or $\GMFO=\GMFT=\GMF$)
and which increases going far from that point, and $\sqrt{\sigmanone\sigmantwo}$
which has a more or less elliptical form and that for $\corrsl=0$ is centered on 
$\GMFOn=\GMFTn=0$.
The numerator is null for $\ForbiddenCasen$ line of constant
numerator are parallel to $\ForbiddenCasen$  
and increases going far from that line. 
Hence the $\Gammadiff$ maxima must be symmetrically aligned along a line normal to 
$\ForbiddenCasen$. The line has to
cross the $\ForbiddenCasen$ line at $\ForbiddenCasen=\GMFc$,
with $0 \le \GMFc \le \GMFn$, so that
the maxima for  $\Gammadiff$ are located at

  \begin{eqnarray}
    \GMFOn &\simeq& \GMFc \pm \frac{1}{\sqrt{2}} \DistMin, \\
    \GMFTn &\simeq& \GMFc \mp \frac{1}{\sqrt{2}} \DistMin; 
\end{eqnarray}

 \noindent
where $\DistMin$ measures their distance from the $\ForbiddenCasen$ line.
Numerically it is possible to show that in the case $\corrsl = 0$ 
sufficient numerical approximations to 
$\DistMin$ and $\GMFc$ as a functions of $\GMFn$ are,

   \begin{eqnarray}
\GMFc &\approx& 0.6994 \GMFn + 0.2722\\
  \DistMin &\approx& 
   \left\{ 
    \begin{array}{ll}
          &\\
       0, & 0 < \GMFn \le 0.701 \\
       -3.0836 \GMFn^2 + 6.7034 \GMFn +3.0969,  & 0.701 < \GMFn < 1 \\
       -0.3985 \GMFn^2 + 0.3367 \GMFn -0.4369,  & 1 \le \GMFn \le 10 \\
          &\\
   \end{array}
  \right.
  \end{eqnarray}

Fig.~\ref{fig:mosaic:allminima:crit}a represents a typical pattern for $\Gammadiff(\GMFO,  \GMFT)$
(the $\GMFni$ are converted into $\GMFi$). It is assumed $\GMF=1/\Rsigma$. 
The optimization produced by maximizing Eq.~(\ref{eq:qerr:opt:simple}) could be improved by
using the approximation for the entropy in Sect.~\ref{app:normalized:entropy} which takes into account
of possible overlaps between the $\Qone$ and $\Qtwo$ distributions, allowing a better approximation
to $\qopt(\GMFO, \GMFT)$. 
So in the figure $\Gammadiff$ have been computed by using the method in 
Appendix~\ref{app:normalized:entropy} but black contour lines are those obtained
assuming $\entropy = \entropyinf$ at 
the root of Eq.~(\ref{eq:qerr:opt:simple})  it is evident that the two approximations agree quite well.
Crosses mark the position of maxima calculated with the approximated solution described above.
The $\Quack$ factor for this case does not reveal any saturation. So it is possbile to look for
other combinations of optimized parameters. 
As an example Fig.~\ref{fig:mosaic:allminima:crit} b) and c) are the equivalent of $\Gammadiff$ computed for
$\Qerrsky$, and $\Qerrload$. 
Of course while $\Gammasky$ has well defined maxima this is not true for 
$\Gammaload$ given 
$\Qerrload$ has not an upper limit. 
Fig.~\ref{fig:mosaic:allminima:crit} d) represents the product of $\Gammadiff$ and $\Gammasky$.
We may look at combinations of parameters where, $\Qerrsky = \Qerrload$ or 
$\Qerrsky = \Qerrdiff$ or $\Qerrload = \Qerrsky$
as in Fig.~\ref{fig:mosaic:equals:crit}a,b and c or 
$\Qerrdiff \approx \Qerrload \approx \Qerrsky$ and thus use the product of the 
second group of criteria in Tab.~\ref{tab:typical:criteria}, assuming all the 
$\Policy=1/4$. as shown in Fig.~\ref{fig:mosaic:equals:crit}d.

 \subsection{Dealing with saturation}\label{ref:dealing:with:saturation}

A more complex situation could arise if the selected optimal 
$\GMFO$, $\GMFT$ and $\qopt$ lead to saturation.
In this case either a $(\GMFO, \GMFT, \qopt)$
 far from the $\Gammadiff$ peak has to be selected or 
$\qopt$ has to be increased in order to reduce the corresponding $\Quack$ factor.

In the first case the requirement $\Cr = \CrTgt$ will be assessed 
but the quantization error will be larger than the optimal one.
To limit this error the new $(\GMFO, \GMFT, \qopt)$ would have to be selected as much as possible
along the ridge near the $\Gammadiff$ peak ad as much as possible far from the $\GMFO=\GMFT$ line.

In the second case we consider the fact that $\Quack \propto 1/q$ so it is possible 
to take $\GMFO$ and $\GMFT$ at the $\Gammadiff$ peak but to take

$$\qoptquack = \safety \Quackmax \qopt(\GMFOpeak, \GMFTpeak)$$ 

 \noindent
where $\Quackmax = \max(|\Quackonepeak|,|\Quacktwopeak|)$ and $\safety > 1$ is a safety factor,
which typically is $\safety = 2$.
Of course in this case the data are compressed at an higher rate than $\CrTgt$
while the processing error will be increased by a factor 
$\safety \Quackmax$.

  \subsection{OCA2K, non idealities and numerical optimization}\label{sec:optim:oca2k}

Non--idealities in the signal and in the compressor cause the effective $\Cr$ to be different from the 
expected $\CrTh$, and in general $\CrTh>\Cr$.
A formal way to account for this is to define a compression efficiency $\etacr \le 1$ defined as:

 \begin{equation}
  \etacr = \frac{\Cr}{\CrTh},
 \end{equation}

 \noindent
which is the product of the contributions of each non--ideality.
In general it is very difficult to account in a satisfactory way for even the most important non idealities
 as is illustrated by the following examples.

A group of non--idealities comes from the fact that each time a new symbol is discovered in the data stream
the compressor adds at the compressed output a ``stop'' pseudo--symbol followed by the uncompressed symbol.
Then the compressor is coding the symbols in the input data stream
plus the ``stop'' pseudo--symbol and consequently the entropy for the compressed data stream 
to be introduced into Eq.~(\ref{eq:cr:solita}) is changed by a factor $\etastop^{-1}$

 \begin{equation}
  \etastop^{-1} = -\frac{\phistop}{1+\phistop}\log_2 \frac{\phistop}{1+\phistop} + 
                   \frac{ 1 +\log_2(1+\phistop)}{\entropy(1+\phistop)} ,
 \end{equation}

 \noindent
where 
$\phistop = \Nsymb/\Nsamples$, $\Nsymb$ is the number of different symbols in the packet,
$\Nsamples$ the number of samples stored in the packet
 and $\entropy$ comes from Eq.~(\ref{eq:cr:solita}).
Note that for $\phistop \rightarrow 0$, $\etastop^{-1} \rightarrow 1$.
In general for small $\phistop$ the addition of stopping symbols increases the entropy leading to 
$\etastop<1$, but 
when $\phistop$ is sufficiently large the compressed data chunk is diluted in a large number of 
repeated symbols, reducing the entropy of the signals and giving $\etastop > 1$.
However the potential gain in $\CrTh$ is compensated by the need to add uncompressed symbols.
If $\NbitsCode$ is the number of bits needed to store the information used to
decode a symbol, 
 $\Lpck$ is the length of the packet, 
then from the condition 

  $$
    \frac{\Nsamples \Nbits}{\CrTh} + \Nsymb \NbitsCode \le \Lpck;
  $$

 \noindent
and from $\Lin = \Nsamples \Nbits$, assuming the optimal case 
$\Lout = \Lpck$ the dumping factor for the compression efficiency is derived

 \begin{equation}\label{eq:eta:cr:ind}
   \EtaCrInd = \left[ 1 + \frac{\Nsymb}{\Nsamples} \frac{\NbitsCode}{\Nbits} \CrTh \right]^{-1}.
 \end{equation}

 \noindent
In general, for a stationary signal $\Nsymb \ll \Nsamples$ so that $\EtaCrInd$ is a second order correction which
will be neglected in the remaining of the text, but it becomes an important factor for the case 
of non--stationary signals for which $\Nsymb \approx \Nsamples$, which could occur in case of fast drifts.

Two non--idealities very complicated to be analyzed are 
the difference between the expected entropy and the sampling entropy,
and the compressor inertia.

The theoretical estimates of entropy and hence of the expected compression rate,
gives the expected entropy calculated on an ideally infinite number of 
realizations of samples. This means that even very infrequent symbols for the samples are considered
by theory. But the compressor stores a few hundreds of samples for each packet
leading to a truncated distribution of samples and consequently to 
a sampled entropy which in general is smaller than both the theoretical expectation
and the entropy measured on a long data stream.
In theory if $G(Q)$ is the cumulative PDF for the distribution of samples, and if $Q$ is bounded 
between $Q_{\mathrm{inf}}$ and $Q_{\mathrm{sup}}$ it would be sufficient to rescale
the $f_Q$ by $1/(G(Q_{\mathrm{sup}})-G(Q_{\mathrm{inf}}))$ and
redefine accordingly the sum in the definition of the Shannon entropy.
As an example, in the case of a simple normal distribution cutting the distribution 
respectively at $1$, 2, 3 and $4\sigma$ will reduce the entropy as predicted from 
Eq.~(\ref{eq:entropy:kpdf}) respectively by a factor $\etasampling = 0.79$, 0.89, 0.95 and  0.98.
However, the difference between theoretical entropy, or even the entropy 
measured on long data streams, and the sampling entropy measured on short packets 
could be changed 
by the presence of correlations in the signals on scales longer than 
the typical time scale of a packet. 
Last but not least, it is necessary to consider that the compressor takes some time to optimize its coding scheme,
leading to a further loss in compression efficiency. 

The effect of all of these non idealities are
too complicated to be introduced in the theoretical model, so that 
the tuning of REBA parameters based on the theoretical models has to be
refined by numerical optimization.
Numerical optimisation it is important because the handling of difficult
cases in which the hypothesis of the theoretical model completely fails,
it allows experimentation with artificial perturbations introduced in the signal
and it includes higher order effects such as the packet--by--packet variability of
$\Cr$,
In addition numerical simulations must be used to verify the optimized
parameters before uploading them to the instrument.

With these aims the Onboard Computing Analysis (\OCA) software was developed, 
composed of a scanner, able to run the same test on different combinations of REBA parameters;
an analyzer, able to automatically extract relevant statistics on each test;
an optimizer, able to apply different policies defining when a combination of parameters is optimal or not
selecting the best combinations;
a report generator, used to generate automated reports. 
Apart from REBA optimization the development of the \OCA\ libraries has been driven by the need to have a flexible 
environment for testing ground segment operations as explained in \cite{Fraillis:2008b}.
Hence, \OCA\ is able to read, decode, and process small amounts of raw data coming from the
\Planck/LFI scientific pipeline from packets to complete timelines.

At the core of the part of the \OCA\ software dedicated to the REBA optimization there is a {\tt C++} kernel, 
(\OCATK) which processes the input data for each combination of 
REBA parameters performing: 
i. \onboard\ mixing and quantization by using the real algorithm;
ii. \onboard\ compression by using the \onboard\ algorithm;
iii. on ground decompression and reconstruction.

It has to be noted that \OCATK\ uses the same {\tt C} code for compression operated on--board.
So it does not emulate the compressor but uses the real compressor. 
In addition the validation of proper emulation of the \onboard\ and on--ground processing has been provided by
using data generated in the framework of the validation of Level--1. of the \Planck/LFI DPC 
\cite{Fraillis:2008b}. In that way we demonstrated that 
\OCATK\ processes the data in the same manner as the real processing chain.

The input of \OCA\ and \OCATK\ are short data streams of 
raw data downloaded from the instrument just before \onboard\ averaging or 
just after it, see Fig.~\ref{fig:processing}, depending on whether $\Naver$ has to be optimized or not. 
In output \OCATK\ provides measures on a 
packet--by--packet base of $\Cr$ and its 
related quantities such as the estimated packet entropy,
or the measured compression efficiency
$\etaoca$.
It provides also sample--by--sample estimates of critical parameters as $\Qerr$, and $\Quack$.

Despite \OCATK\ is written in {\tt C++},
it remains a heavy, offline tool, which 
can not be directly used for a crude exhaustive 
real--time optimization. This is the reason for which analytical
methods have been developed. 
On the contrary \OCA\ has the ability to use the analytical models to focus on the relevant region of parameters
space.

\OCA\ allows the determination of the optimum parameters according to different optimization strategies
and constraints. 
This is important given the different ways in which REBA parameters are optimized during ground tests
and in flight. 
During ground testing the usual procedure has been to stabilize the instrument and its environment,
calibrate the DAE and then to acquire 
chunks of about 15~minutes of averaged data to be analyzed by \OCA\ to optimize the REBA
parameters
 \cite{LFI:radiometers:tuning}. 
After setting the REBA parameters another 15~minutes of acquisition, this time with the
nominal processing described in Fig.~\ref{fig:processing} is executed as a cross-check. 
 \CHANGE{
In flight the procedure will be to acquire continuously data by using the nominal processing
Short chunks of unprocessed data will be acquired daily in turn from each detector. The comparison
of unprocessed with processed data will allow monitoring of the processing error.
In addition the REBA tuning might be repeated daily in order to test whether some REBA parameters
\onboard\ the satellite should be changed or not.
 }

\OCA\ could be used as a stand--alone application, but 
different interfaces for \OCA\ to other packages have been created for different applications. 
For ground segment testing \OCA\ provided an \IDL\ and {\tt C++} library used in a stand--alone program.
The same occurred for the \Planck/LFI simulation pipeline where 
parts of the \OCATK\ simulating on--board preprocessing and ground processing,
(excluding compression and decompression)
have been included in the \Planck/LFI simulation pipeline.
For the REBA optimization during the ground tests,
\OCA\ has been used within the {\tt LIFE} framework \cite{LFI:dpc:life}.
For routine operations in flight \OCA\ has been included in the 
{\tt PEGASO} \cite{LFI:dpc:life}
software tool designed to monitor the instrument health and performances
at the \Planck/LFI DPC.

 \subsection{The \OCATWO\ optimization algorithm}\label{sec:optimization:algorithm}

As a premise to REBA processing optimization, a value for $\Naver$, 
a $\CrTgt$ and a function of merit $\chi$ 
appropriate to the case under analysis 
have to be fixed. 
As explained, in general $\Naver$ is already fixed by other considerations than 
REBA processing optimization.
A slightly higher than needed $\CrTgt$ is taken
in order to allow some margin. While the $\Gammadiff$ is considered
a sufficient function of merit, but more complex functions, such as those in the family
of functions presented in Eq.~(\ref{eq:target:function:general}) are used as well.

A data chunk long enough to allow the generation of about a hundred compressed packets, is acquired for each radiometer.
In general, the data chunk is \onboard\ processed by allowing coadding for the the given $\Naver$.
For that chunk relevant statistics such as 
$\Tskymean$, $\Tloadmean$, $\sigmasky$, $\sigmaload$, $\sigmaskyload$, 
$\Driftsky$, $\Driftload$ are measured and from that 
$\GMF$ and $\Rsigma$ are evaluated.

The analytical optimization is performed in order to
 determine in an approximate way 
       the region of $\GMFO$, $\GMFT$ where the function of merit could have a peak;
 to grid the region $\GMFO$, $\GMFT$ (typically by regular sampling);
 to determine for each point in the region the function of merit $\chi(\GMFO, \GMFT)$
 and $(\GMFOoptimal, \GMFToptimal)$  as well as the $(\GMFO, \GMFT)$ for which 
       $\chi(\GMFO, \GMFT)$ has its absolute maximum;
 and finally for the previously determined $(\GMFOoptimal, \GMFToptimal)$  the
     $\Offsetoptimal=\Offset(\GMFOoptimal, \GMFToptimal)$ and the
     $\qoptth=\qopt(\GMFOoptimal, \GMFToptimal)$ are determined.
After that $\max(|\QuackUnit(\GMFOoptimal, \GMFToptimal)|)$\ i.e. the maximum value of $|\Quacki(t)|$ among 
the $\Quack$ values determined on the data chunck for $q = 1$ (see Eq.~(\ref{eq:quack})) is measured.
From $\max(|\QuackUnit|)$ $\qoptth$ is could be corrected for saturation. 
In fact, if $\max(|\QuackUnit|) < (1-\safety)\qoptth$ the analytical optimization
returns $\qoptth$ as the best estimate of $q$ otherwise it forces
$\qoptth = \max(|\QuackUnit|) /(1-\safety)$.
In the latter case $\qoptth$ is said to be {\em saturation--limited} and of course
in that case it is expected to have $\Cr > \CrTgt$.

After the analytical optimization the $\qoptth$ has to be numerically refined in order to take 
into account the non--idealities of the compressor. 
If $\qoptth$ is not saturation--limited the \OCATK\ is 
operated to determine, by a polynomial search, the best $\qopt$ allowing 
$\Cr=\CrTgt$ for given $\GMFOoptimal$, $\GMFToptimal$, and $\Offsetoptimal$.
In general the search is performed for $q$ in the range 
$\max(|\QuackUnit(\GMFOoptimal, \GMFOoptimal)|)/(1-\safety)$ and 
$2\qoptth$. 
If $\qoptth$ is saturation--limited the numerical procedure could be 
in principle skipped.
However non--idealities could cause $\Cr<\CrTgt$ even in this case
and to check for this 
a single run of the numerical optimization is performed 
for the selected parameters. 
If $\Cr>\CrTgt$ the procedure is concluded, otherwise the
polynomial search is performed.

When $\qopt$ have been numerically refined a last run of the numerical code for 
$\GMFOoptimal$, $\GMFToptimal$, $\Offsetoptimal$ and $\qoptth$ is performed
to asses the processing error and the histogram of the compression rate.

The typical time to perform the optimization sampling $(\GMFO, \GMFT)$ with
a grid of $25\times25$~samples and a TOI of about 15~minutes of data, is 
about 20~sec, so that the optimization of the whole set of 44 detectors takes
less than 15~minutes including the overheads for data IO.

 \FIGtests

\TABtesttoi

\FIGcsl

 \section{Results}\label{sec:results}

Here are discussed the results of REBA calibration and optimization in the framework of the \Planck/LFI ground tests.
We look at first at a single test to compare the analytical and the numerical optimizations.
We then look at the results of the calibration for the whole set of 44 detectors in a real case.

The results of analytical v.z. numerical optimisation are compared by
using real \Planck/LFI data acquired during
the RAA tests of the instrument performed at \LABEN,
during the summer of 2006. Fig.~\ref{fig:tests}a shows 12~min of data with $\Naver = 52$,
equivalent to about 56715~samples.
while the Tab.~\ref{tab:tests}a gives the relevant statistics of the TOI.
During the test the instrument and its environment was stable, 
no strong drifts are present in the data. 
A clear correlation between \sky\ and \load\ is evident in the plot explaining the $\corrsl \approx 1$.
and the factor of six reduction of the RMS when passing from undifferentiated to differentiated data.
Also the separation between \sky\ and \load\ is not large, being just 18$\sigma$. So after mixing the 
distributions for $\Pone$ and $\Ptwo$ will stay well separated, with $|\Separation|>100$, when 
$|\GMFT-\GMFO| \gtrsim 0.2$. 
In this case it is reasonable to expect that both the low--accuracy and high--accuracy methods to estimate
analytically the entropy will give comparable results.

Indeed, we used both models to optimize the REBA parameters taking $\CrTgt = 2.4$, and both models produced
exactly the same results reported in the second column of Tab.~\ref{tab:tests}b and Tab.~\ref{tab:tests}c.

To test the goodness of the AO, \OCA2K\ was run imposing $q=1$ and taking the same values of $\GMF1$, $\GMF2$ 
used for the AO. 
The predicted entropy of the processed TOI is compared in 
Fig.~\ref{fig:tests}b. There the relative difference between the entropy measured all over the TOI
and the entropy computed analytically by using both methods is reported.
Patches define intervals of accuracy in steps of $0.5\%$ up to $3\%$. Both methods to estimate the entropy
are good predictors of the measured entropy, apart from the region marked with the white boxes where 
the low accuracy method overestimated the entropy.

In a similar manner we compared in Fig.~\ref{fig:tests}c the measured $\Cr$
and the predicted from the high accuracy model. 
Again the model is able to reproduce within $20\%$ or better the measured $\Cr$. 
As discussed before the differences can be ascribed mainly to
the difference between the sampling entropy and the expected entropy (\cite{maris:2000})
and the not--ideal behavior of the compressor  (\cite{maris:2000}).
In general the effect of the sampling entropy would result in a higher $\Cr$ than expected
while non idealities in a lower $\Cr$. 
Different ways can be used to calibrate these effects, however their interplay with the statistic
of the signal is complicated and it is preferable to use OCA2K to fine tune the REBA parameters
optimized by analytical mean,
given that in general the corrections required to properly tune with respect to the analytical
prediction $q$ are at most of about a factor of two.

The numerically refined optimal 
parameters are reported in the third column of Tab.~\ref{tab:tests}c,
as is evident the only variation is just for $q$, the reason is explained by
the contour plots in Fig.~\ref{fig:tests}d and 
Fig.~\ref{fig:tests}e which compares the predicted analytical $\Gammadiff$ and $\Gammasky$
with the $\Gammadiff$ and $\Gammasky$ obtained for the numerically refined parameters.
Very good agreement is obtained in the location of the peaks which determines the optimal
$\GMFO$, $\GMFT$ and in turn the $\Offset$. 
For completeness the last column of Tab.~\ref{tab:tests}d reports the theoretically
estimated $\Quack$, $\Qerrdiff$, $\Qerrsky$ and $\Qerrload$ after replacing the 
analytical optimal $q$ with the numerical one. 
The very good agreement between the theory and the experiment is evident.

As expected the quantization error for the differentiated data is smaller by about a factor of four
than the error for \sky\ or \load\ and anyway the error will be a fraction of ADU, but 
larger or comparable to 
the quantization error introduced by the ADC converter, which for $\Naver=52$ is equivalent to 
$\approx 0.04$~ADU.

Eq.~(\ref{eq:qerror:old}) expresses the processing error for an 
univariate normal distribution as a function of the $\sigma/q$ ratio,
but from Sect.~\ref{sec:OP:mixing} it is evident that in the present case 
the $\sigma/q$ ratio is not a good measure of the processing error.
At the opposite, it is possible to define an effective
$\sigma/q$ ratio in terms of the processing error as

 \begin{equation}\label{eq:sq:fictious}
   \left(\frac{\sigma}{q}\right)_{\mathrm{eff}} = 
     \frac{
         \sigma_{\mathrm{diff}}
        }{
         \sqrt{12} \Qerrdiff
        };
 \end{equation}

 \noindent
which expresses the number of independent quantization levels which could be accommodated within
$1\sigma$. 
So this ratio gives an idea of how well the histogram of the differentiated data is sampled
assuming it could be represented by an univariate normal distribution 
 \footnote{The $\sigmaqeff$ could be used to characterize the processing in the case 
the value of $\Qerrsky$ and $\Qerrload$ is not relevant.}.
A proper sampling would
assure at least $\sigmaqeff > 2$ which is the case for this work
as it is evident from Tab.~\ref{tab:tests}c 
having $\sigmaqeff > 6$.

 %
Before concluding 
 this comparison 
it is worth commenting the way in which the experimental $\Cr$ is reported in 
Tab.~\ref{tab:tests}c.
This is best done by looking at Fig.~\ref{fig:tests}f where the histogram for the $\Cr$ 
of the 111 packets produced in the test is shown. 
Given the true $\Cr$ is a random variable, varying from packet to packet we take the
$5\%$ and $95\%$ percentiles assessing that in less than $5\%$ the $\Cr$ will be respectively smaller
or larger than the quoted $\Cr$, as well as the mean and the median (not quoted in the figure) 
of the measured $\Cr$. In this case it is evident that
the target $\Cr \gsim 2.4$ in something less than half of the packets. 
So it would be better to introduce
some safety factor, requiring for example $\CrTgt \approx 2.6$ or to require that
the median be 2.4 or better
with the 5\% percentile to be $\approx 2.4$.

Fig.~\ref{fig:csl} is representative of the results of the calibration for a whole set of 44
detectors performed during the \Planck/LFI CSL test campaign in 2008
\cite{LFI:radiometers:tuning}. 
Data have been collected over two acquisitions, the first one being  used for the calibration 
itself and the second one to verify the  calibration performances. The environmental set-up and 
the onboard  electronics were kept in a stable state during both acquisitions. 
During the first acquisition, called ``calibration run'', 
the \onboard\ computer was configured to apply
just the downsampling step to the data
but skipping mixing, requantization and compression.
The acquired data have been ingested
into \OCATK\ to generate a list of 
optimized processing parameters for a target $\Cr=2.4$.
Having produced a set of parameters for the REBA, the second  acquisition,
the ``verification run'',
has been run while the instrument was set up to acquire 
data in nominal conditions, by using the same steps that are going to be  used during flight
i.e. mixing, requantization and compression 
followed by ground processing: decompression, demixing and dequantization. 
At the same time data have been also acquired in the
raw format used in the ``calibration run''. 
So for each detector couples of data streams with and without \onboard\ processing were obtained which have been
compared in  order to measure the processing error, following a procedure similar  to the
one described in \cite{Fraillis:2008b}. 

Fig.~\ref{fig:csl} compares the mean $\Cr$, $\Qerrsky/\sigmasky$, $\Qerrload/\sigmaload$, $\Qerrdiff/\sigmadiff$,
where $\sigmadiff$ is the r.m.s. for the differentiated data.
Bars in light colours are for results obtained by processing 
the data taken in the calibration run with the \OCATWO\ simulator.
Bars in dark colours are the results from the data processed by the instrument in the
verification run.  In both cases the same set of optimal REBA parameters have been used.
There is good agreement between the two runs, despite the presence  of a few systematics. 
Such differences are due to slight changes in the environmental
conditions between the two runs \footnote{In the CSL tests the satellite has been kept within
a large cryogenic vacuum chamber which was not as stable as the L2 environment is.}. 
Differences in channels belonging to the same radiometer (00 and 01,  10 and 11) are due to the fact that in the warm back end the two channels go through separate acquisition lines, each of them being  characterized by different noise properties
 \cite{LFI:general:instrument}.
Such differences are usually small, e.g.
detectors (00, 01) and (10, 11) of Feed--Horn \#19.
In a few cases however larger differences occurs, e.g.
detectors (10, 11) of Feed--Horn \#25.
Again, the relative processing errors for \sky\ and \load\ 
are very similar, and in 95\% of the cases they are below 0.4
with some extreme deviations such as detectors 10 of Feed Horn \#25 and 11 of Feed Horn \#26 
for which $\Qerrsky/\sigmasky \approx \Qerrload/\sigmaload \gsim 1$. 
Here the optimal $\qopt$ is not peculiar with respect to the values required for
the other detectors, but optimal $\GMFO$ and $\GMFT$ are very similar having $|\GMFT-\GMFO| = 0.04$
which is the resolution of the search grid in the $(\GMFO, \GMFT)$ space.
Such relativelly ``coarse'' resolution in the search grid for optimal $(\GMFO, \GMFT)$
was imposed by constraining the need to optimize the REBA parameters within a few minutes
after the data acquisition. The coarseness of the $(\GMFO, \GMFT)$ grid 
is also the reason for the apparent coincidence of the mean values of $\GMFO$\ and $\GMFT$\ for different
frequency channels in Tab.~\ref{tab:csl}. In flight such time constraints will be removed allowing
a multi step iteration of the optimization procedure and the use of a thinner grid.
However, even in those extreme cases, the coorrelation between \sky\ and \load\ processing errors 
leads to a much smaller error for the differentiated data.
In such cases the final processing error is always less than
$3.8\%$ of the instrumental white noise..
Scaling those numbers to the calibrated sensitivity per sample and per detector 
the calibrated processing error,
$\Qerrdiffmuk$,
was derived and is reported in units of $\mu\mathrm{K}$ per sample and per detector
in line 10 of Tab.~\ref{tab:csl}. On average $\Qerrdiffmuk$ is 
below the 3~$\mu\mathrm{K}$ level taken as a threshold for
systematics \cite{LFI:general:program} apart from for the case of the detector 01 of the
Feed--Horn \#24 for which $\Qerrdiffmuk \approx 3.2$~$\mu\mathrm{K}$.

The values for the optimal REBA parameters are mainly determined by the frequency of the radiometric channel
with some dispersion from detector to detector.
Tab.~\ref{tab:csl} gives representative median values for $\GMFO$, $\GMFT$, $q=1/\SecondQuant$ from the CSL tests
as well as for the quantities in Fig.~\ref{fig:csl} and the resulting data rate.
$\Offset$ is omitted since it is the most variable parameter and it has no significative impact on
$\Qerr$ and $\Cr$. 
Tab.~\ref{tab:csl} reports also the number of detectors for each frequency channel, the $\Naver$ values
which are kept constant, the compressed data rate per detector, per frequency channel and for the instrument 
as a whole.
Quantities are reported in the form $x \pm \delta x$ where $\delta x$ represents the standard deviation 
taken as a measure of the internal dispersion of $x$ within the given subset of detectors. 
It has not to be interpreted as an error and it must not be propagated.

The total data--rate in Tab.~\ref{tab:csl} is just 7\% higher than the target data--rate 35.5~bits/sec.
Again this departure is mainly due to the limited resolution in the search grid as well as small
changes in the environmental conditions between the two runs. In order to cope with this problem
it is likely that during operations a safer $\CrTgt = 2.5$ target will be set in place of the 
nominal 2.4.

Finally it is worth to consider the gain in the accuracy of the REBA optimization obtained by the complex
procedure described in Sect.~\ref{sec:optim} with respect to the fairly simple scheme used in the
earlier RAA test campaign \cite{LFI:general:instrument}.
During the RAA tests a simplified algorithm had been applied based on the fact that putting
$\GMFO = r$ the processing error for the differentiated data reduces to 
Eq.~(\ref{eq:differentiated:qerr:reduced}) which is independent of $\GMFO$ and $\GMFT$. 
Hence, the only free parameters where $q$ and $\GMFT \ne r$. The
optimization was performed by imposing $\Cr = 2.4$ and selecting those parameters for which
$\Qerrsky \approx \Qerrload \approx \Qerrdiff$. Even in this case the required $Cr=2.4$ 
was achieved but $\Qerrdiff/\sigmadiff$ was between $[0.08, 0.14]$,
when compared to the current $\Qerrdiff/\sigmadiff \le 0.038$ it is evident how this procedure
represents a substantial improvement.
In particular 
LFI has as a target of keeping all of the instrumental systematics and non gaussian noises
in the differentiated data below 10\% of
the instrumental white noise \cite{LFI:general:instrument}.
The optimization scheme described here allows the reduction of the processing error on the differentiated data
by a factor of four pushing it below this ambitious target.

\ATABcsl

 %

\FIGmosaicEntropyApprox

 \section{The impact of the on--board processing noise on the \Planck\ scientific performances}\label{sec:impact:science}

  \CHANGE{
A detailed analysis of the effect of \onboard\ plus on--ground processing on the final scientific 
products of \Planck\ is beyond the scope of this paper. The subject is simply too complex
to be analyzed here, and the analysis has to be specialized to take into account each specific kind of 
product obtained from the \Planck\ data 
i.e.: calibrated time lines, frequency maps, component maps, angular power spectra for those maps and cosmological parameters. 
Neglecting in this discussion the role of $\GMFO$ and $\GMFT$ it is enough to say that
for most of these products the effect of processing 
could be reduced to the effect of processing 
on the signal spectral decomposition in the time or in the spatial domain.
 %
Without to enter in to too many details, some consideration could be derived from the 
analysis reported in \cite{maris:2004} and from a forthcoming work in progress
\cite{maris:2009}.

In the extent in which the processing error for differentiated data
is small when compared to the fluctuations in the signal, 
the noise model for the quantization error data could be applied
to derive the level of degradation in the noise properties of
the instrument.
In the \Planck/LFI case, for differentiated data a typical 
$(\Qerrdiff/\sigmaWN) < 0.1$ 
by using Eq.~(7) of \cite{maris:2004}
the final instrumental performances
will be degraded approximately 
by a factor $\sqrt{1 + (\Qerrdiff/\sigmaWN)^2} < 1.01$.
This immedialty applies to the amplitudes of the spectral decomposition of the signal.
Since processing acts as stationary white noise it will increase by less than $1\%$ the power 
excess introduced by the white noise. 
On the other hand the effect on the phases will be similar to 
to random scrabling over the $[0,2\pi]$ interval,
but even in this case the effect of quantization will be to increase by
few percent the effect of the instrumental whiten noise.
It is immediately posible to extend these results to the \Planck/LFI calibration
which will be based on measurents of the amplitude of the cosmological dipole,
to understand how, when compared to the white noise effect,
the calibration accuracy will be worsened by less than $1\%$ by the processing noise.

It is more complex to consider the case of the \Planck/LFI 
sensitivity to primordial non--gaussianities.
In \cite{maris:2004} the level of perturbation the measurements of primordial 
non--gaussianities introduced by the 
quantization were compared to
the level of perturbation introduced by the residuals 
astrophysical foregrounds concluding
that they are very small.
However,  it is worth analyzing the effect of processing
on non--gaussianities even 
assuming an ideally perfect separation of the foregrounds.
To the extent that the quantizer 
has null expectation, null skewness and in general null central moments of odd order, 
no effect is expected on tests of primordial non--gaussianity 
which are sensitive just to central moments of odd orders. 
Of course a symmetrical quantizer will alter the central moments of
even order, such as the kurtosis.
To estimate this effect it is sufficient to compare the distribution of the 
processing noise from a uniform quantizer with the case of normal white noise
while taking the average over $N$ repeated measures.
In the white noise case the central moments of even order are 
$\mu_n = C_{g}(n)\sigma^{n}/N^{n/2}$, with $C_{g}(n) = n!/(2^{n/2}*(n/2)!)*s^(n)$.
In the case of the quantizer  $\tilde{\mu}_n(N,q) = \tilde{C}(N,n)(q/\sqrt{12})^n/N^{n/2}$,
whatever the form of $\tilde{C}(N,n)$ is the Central Limit Theorem assures that for any $n$ 
$$
  \lim_{N\rightarrow\infty} \tilde{C}(N,n) = C_{g}(n).
$$
	
 \noindent
but given $N$ is finite, a bias in the estimator for each moment of $n$ order will appear.
Expanding about $1/N$ up to the leading term $\tilde{C}(N,n)$ 

$$\tilde{C}(N,n) \approx C_{g}(n)\left(1+ \frac{A_n}{N^{n/2-1}}\right),$$

 \noindent
where $A_n$ is a serie of coefficients having $|A_n| < 1$ whose first three elements are
$A_2=0$, $A_4 = -2/5$, $A_6 = -6/5$, $A_8 = -12/7$.
Since for \Planck/LFI $N \approx 60$ the bias in the expectation of higher order moments will be
very small.
In addition, since $N$ is a characteristic parameter of a mission, any bias due to processing
in the estimators of central moments could be predicted and removed. 

In closing these considerations it is worth noting that for \Planck/LFI the most important limitation 
to apply an hard requantization of data comes from the need to limit the effect on the total power data 
rather than on the differential data. 
Firstly it has to be noted that the cancellation effect of quantizzation errors on \sky\ and \load\ 
from which Eq.~(\ref{eq:differentiated:qerr}) is derived,
 applies only to the $(P_1, P_2)$ 
space of mixed data to the extent in which $q/\sigma_i < \sqrt{12}/3$ for any $i=1$, 2.
Another limit in the maximum amount of processing noise which could be introduced in total power
arises when it is taken in consideration the fact that measurements in 
total power on the \load\ signal have the potential to be a valuable
tool as a source of diagnostics of instrumental systematics.
An example is given by the study of thermal effects induced by the instability of the \load\
which could be detected by cross--correlating $\Tload$ with measurements of temperatures 
acquired by thermometers located in the \Planck\ focal plane,
which limits the quantization on $\Tload$\ to be 
 $\Qerrload/\sigmaload \le 1/2$
\cite{maris:2009}.

 }

 \section{Final Remarks and Conclusions}\label{sec:final}

\CHANGE{

As for many past, present and future scientific space missions, the ESA \Planck\ mission has a 
limited bandwidth to download the scientific data produced by its two instruments.
The bandwidth allocatable for the \Planck/LFI is about 2.4 times lower than 
the raw data flow produced by the 44 detectors comprising the 
instrument which is made of an array of 22 pseudo--correlation receivers, each one
comparing the signal received from the \sky\ with a reference signal.
To fit the allocated bandwidth the data has to be preprocessed on-board and loss--less compressed 
prior to transfer to the ground, where each step of the \onboard\ preprocessing has to be reversed 
to recover the 
original information. Since not all of these steps are completely reversible an overall reduction of the quality of the data 
occurs, which has to be 
quantified and reduced as more as possible not to degrade the instrument performace.

This paper has presented a detailed discussion of the \onboard\ plus on--ground processing 
for \Planck/LFI 
and of its free parameters which can be adjusted in order to fit the compressed data--rate
to the allowed bandwidth. In addition, this paper presented a model to quantify
the level of distortion in the scientific data
as a function of the free parameters for the \onboard\ processing 
and as a function of the attainable compression rate. At last the paper reported on the way
these parameters are optimized to cope with the required bandwidth while limiting
the processing distortion.

Three new results 
about the way in which the output 
of a pseudo--correlation receiver could be handled 
are presented.
First, a new algorithm: mixing followed by requantization and interlacing
to prepare the data stream for compression,
limiting at the same time the amount of 
processing distortion is introduced. 
This method is effective since most of the time variations 
in the \sky\ and \load\ data streams are correlated. Mixing reduces the effective variance
of the signal to be compressed reducing in that way the need for requantizing the data. At the
same time the processing errors in the mixed data are correlated, so that the demixing procedure
operated on the ground introduces cancellation effects which reduce the re--quantization error
in the differentiated signal.
Second, a model which quantifies the level of distortion in the scientific data as a function of 
the free parameters of the \onboard\ processing and as a function of the attainable compression 
rate is given.
Third, it presents a general procedure to search for optimized processing parameters which has to cope 
with the proper use of the allocated bandwidth, requiring a nominal compression factor of 2.4, which 
has to limit as much as possible the processing distortion, and has to be fast given that both in the 
pre--launch tests campaign and in flight only a short time could be allocated for optimization. 

The optimization procedure is based on a combination of the analytical model and of a software simulator,
all integrated into a single applicative program called \OCATWO\
which takes as input a \Planck/LFI data stream applying to it 
the whole \onboard\ and on--ground processing to measure the processing distortion and the data--rate.
The two approaches complement each other: the analytical model is very fast and could be used to rapidly select 
regions of interesting combinations of parameters; the simulator is able to handle conditions which hardly
can be analytically modelled and is used to refine the parameters identified by the analytical model.
The data input to \OCATWO\ can be provided indifferently by the \Planck/LFI flight simulator 
  \cite{PlanckFlightSimulator}, 
or from specific acquisition sessions performed during the test campaign or in flight.
A wide selection of combinations of processing parameters and optimization criteria can be
explored by this code.
In practice this work shows that the analytical model is refined enough to allow 
a full optimization of all of the processing parameters apart from the \onboard\ re--quantization step, $q$,
which has to be tuned numerically to account for a number of non--idealities 
in the data and in the compressor, part of which have been discussed in the paper.
However, it is interesting to see that in no cases is the difference between the analytical and the numerical 
model in the optimized $q$ larger than a factor two.

The last part of this work reports the performances of the optimized \onboard\  algorithms in the framework 
of the pre--launch tests required for instrument qualification. 
In that case it has been demonstrated that the 2.4 compression factor required to operate \Planck/LFI could be 
attained introducing a modest 3.8\% of quantization noise, which is equivalent to an increment in the instrumental 
noise of less than $1\%$, and in particular that processing is not harmful to the scientific exploitation of 
\Planck/LFI data and in particular to the study of primordial non--gaussianities.

In conclusion it is worth noting that the application of mixing follwoed by requantization is of general use and 
could be extended outside the case of the \Planck/LFI since it could be used in any situation in 
which data alternatively taken from a signal source and a reference source are sent to a remote station.

}

\acknowledgments
\Planck\ is a project of the European Space Agency with instruments
funded by ESA member states, and with special contributions from Denmark
and NASA (USA). The Planck-LFI project is developed by an International
Consortium lead by Italy and involving Canada, Finland, Germany, Norway,
Spain, Switzerland, UK, USA.
The Italian contribution to Planck is supported by the Italian Space
Agency (ASI).
We thank the support of the Spanish Ministry of Science and Education.

\appendix

\section{Approximation of the bivariate entropy}\label{app:normalized:entropy}

This section presents the approximation of the entropy for an interlaced bivariate
distribution for the two limiting cases of a uniform distribution or a normal distribution.

 \subsubsection*{The case of a uniform distribution}

For the case of two uniform distributions are taken the intervals where the
distributions are not null as 
$\Qonea \le \Qone \le \Qoneb$ 
and
$\Qtwoa \le \Qtwo \le \Qtwob$.
Also are defined the widths $\None = \Qoneb-\Qonea$ and $\Ntwo = \Qtwob-\Qtwoa$,
and the centers $\Qonec = (\Qonea+\Qoneb)/2$, $\Qtwoc = (\Qtwoa+\Qtwob)/2$.
Without any loss of generality $\None \le \Ntwo$ is assumed.
The entropy is a linear function of $|\Delta| = \left| \Qtwoc - \Qonec \right|$ bounded
between the lower limit 

 \begin{equation}\label{eq:entropy:min}
  \entropymin = 
  \frac{1}{2}\left( 1-\frac{\None}{\Ntwo} \right)  \entropytwo
 -\frac{\None}{2}
    \left(\frac{1}{\None}+\frac{1}{\Ntwo}\right) 
     \log_2 \left(\frac{1}{\None}+\frac{1}{\Ntwo}\right) + 1,
 \end{equation}

 \noindent
for the case the two intervals are completely overlapping, and the upper limit

 \begin{equation}
    \entropyinf = \log_2 \None + \log_2 \Ntwo + 1,
 \end{equation}

 \noindent
for the case of complete separation of the two distributions. So that

 \begin{equation}
    \entropy = \left\{
  \begin{array}{ll}
|\Delta| < \frac{\Ntwo - \None}{2}, & \entropymin \\
\frac{\Ntwo - \None}{2} \le |\Delta| \le 
             \frac{\Ntwo + \None}{2}, & 
                (\entropyinf - \entropymin) \frac{2|\Delta| -\Ntwo+\None}{2\None} + \entropymin \\
\frac{\Ntwo + \None}{2} < |\Delta|, & \entropyinf \\
  \end{array}
    \right. .
 \end{equation}

 \subsubsection*{The case of a normal distribution}

Witout any loss of generality, two sources of normal--distributed interlaced signals having 
respectively variances equal to 1 and $\sigma^2 \ge 1$, quantized with a quantization step $q < 1$
are considered. 

It is necessary to find an approximation for $\entropy(\Separation,q,\sigma)$.
This could not be derived analytically, but
in a manner similar to the case of the uniform distribution
the entropy is bounded between a lower limit, $\entropymin$, and an upper limit $\entropyinf$, while varying 
$\Separation$.
In Fig.~\ref{fig:mosaic:entropy:approx}a shows how the entropy varies as function of $\Separation$ and for
three different values of $\sigma$.
In addition both $\entropy$, $\entropyinf$ and $\entropymin$ are proportional to $-\log_2q$ so that
their differences does not depend on $q$.
For this reason it is convenient to define 
the {\em Normalized Entropy}, $\nentropy$ 

 \begin{equation}
  \nentropy = \frac{\entropy - \entropyinf}{\entropymin - \entropyinf};
 \end{equation}

 \noindent
which is will be just a function of $\Separation$ and $\sigma$
as shown in Fig.~\ref{fig:mosaic:entropy:approx}b as full lines for three values of $\sigma$.

Having $\nentropy$,  $\entropy - \entropyinf$ and $q$ we can readily estimate
$\entropy$ with 

  \begin{equation}
 \entropy =  \entropyinf - \log_2 q + (\entropyinf - \entropymin ) h,
 \end{equation}

 \noindent
note that we do not need to estimate $\entropymin$, and that $\entropyinf$ could be readily
estimated from Eq.~(\ref{eq:approx:hinf-hmin}) by putting $\sigmaone=1$, $\sigma=\sigmatwo/\sigmaone$,
and expressing $q$ in units of $\sigmaone$.

The difference $(\entropyinf - \entropymin)\le 1$~bit is always positive
and just function of $\sigma$. It is null in the limit
$\sigma \rightarrow +\infty$
as shown in Fig.~\ref{fig:mosaic:entropy:approx}c.
The figure shows as dots an approximation obtained 
numerically for $1\le \sigmatwo \le 4000$  for which

  \begin{equation}\label{eq:approx:hinf-hmin}
  \entropyinf  - \entropymin = \exp\left[\sum_{n=1}^5 A_n(\log\sigmatwo)^n + A_0\right]
 \end{equation}

 \noindent
with 
  $A_0 =  1.0893 \times 10^{-2}$, 
  $A_1 = -8.3819 \times 10^{-2}$, 
  $A_2 = -2.3699 \times 10^{-1}$, 
  $A_3 =  4.8141 \times 10^{-2}$, 
  $A_4 = -5.1620 \times 10^{-3}$, 
  $A_5 =  2.1425 \times 10^{-4}$
within an accuracy of $\pm1.1\%$.

Even for $\nentropy$ a numerical approximation is possible within a $\pm1\%$ accuracy 

  \begin{equation}\label{eq:nentropy:gaussian}
  \nentropy \approx e^{-\frac{\Separation^2}{2\sigmastar^2}}.
 \end{equation}

 \noindent
Here $\sigmastar$ is just a function of $\sigma$,
and it is bounded between $0.25558 \le \sigmastar \le 0.30797$ with an 
$\approx21\%$ variation, as shown in Fig.~\ref{fig:mosaic:entropy:approx}d.
Given it is interested to have an upper limit for $\nentropy$ a simple approximation
would be to take $\sigmastar = 0.30797$ overestimating the entropy 
of at most $(\sqrt{\entropyinf-\entropymin}) 0.16 \le 0.16$.
However $\sigmastar$ is a function of $(\entropyinf - \entropymin)$ and numerically we obtained

  \begin{equation}\label{eq:approx:sigmastar}
   \sigmastar = \sqrt{6.8497\times 10^{-2} + 2.5965\times10^{-2} (\entropyinf - \entropymin)} + \epsilon
  \end{equation}

 \noindent
with $|\epsilon| \le 2\times10^{-2}$.

With this approximation the typical accuracy in estimating the $\entropy$ is below $0.01 \div 0.03$~bits
and the optimal $q$ for a given $\entropyTgt$ is promptly derived from

 \begin{equation}
  \log_2 \qopt = (\entropymin - \entropyinf ) \nentropy + \entropyinf - \entropyTgt,
\end{equation}

 \noindent
within a relative numerical accuracy of about $3\%$.

 In short, the algorithm of optimization becomes: 
  i.) given $\GMFO$, $\GMFT$, $q$, $\sigmaone$, $\sigmatwo$, $\Tone$, $\Ttwo$
 ii.) compute $\sigma$, $\Separation$,
 iii.) compute $\sigmastar$, $\nentropy$,
 $\entropyinf$, $\entropyinf - \entropymin$, 
 iv.) compute $\qopt$.

\section{ADC quantization}\label{app:adc:quantization}
Throughout this work it is assumed that the ADC quantization is not relevant for our scopes.
However, it is worth to briefly recall its impact, in particular looking at the conditions
at which the ADC noise could be neglected.

The resolution, or quantization step of the ADC, $\qADC$ is given by $(\Vmax - \Vmin)/2^{14}$~Volts/ADU.
After averaging by $\Naver$ samples $\qADC$ is reduced by a factor $1/\sqrt{\Naver}$.
The effect of ADC resolution is to add in quadrature a non--Gaussian noise to the signals 
 of RMS $1/\sqrt{12}$ before averaging and $1/\sqrt{12\Naver}$ after averaging.
In addition the ADC itself adds a random read--out noise of $\sigmaadc$~ADU which after averaging is
reduced to $\sigmaadc/\sqrt{\Naver}$
When combined these two noises the readout noise whose RMS is
$\sigmarout = \sqrt{1/12 + \sigmaadc}$ before averaging and 
$\sigmarout = \sqrt{1/12 + \sigmaadc}/\sqrt{\Naver}$  after averaging

When a signal of RMS $\sigmazero$ is input to the DAE a gain, $G$, is applied and then the measured RMS is 

 \begin{equation}
  \sigma = \sqrt{\sigmarout^2 + G^2\sigmazero^2},
\end{equation}

 \noindent
depending on the ratio $\sigmarout/G\sigmazero$. The measured RMS will be dominated by the ADC noise or 
by the signal RMS. Signals whose RMS is comparable to the read--out noise are defined as weak signals.

Of course in the case of weak signals the read--out noise is no more negligible when, as an example,
the $\sigmazero$ has to be measured in order to estimate the $\Tsys$. The same is true when the variation 
of the RMS of the signal tacking in account of variations of $G$ has to be estimated
\cite{LFI:radiometers:tuning}.

In addition, given the 1/12 factor in front of the variance induced by the ADC contribution, the 
read--out noise could be dominated by the ADC noise when $\sigmaadc > 0.3$. 

As a practical example if $\sigmaadc \approx 0.5$ and $\sigmazero \approx 1$ the $\sigmarout\approx 0.57$ 
and the bias in estimating $\sigmazero$ will be $\approx 15\%$.

\section{DAE Tuning }\label{app:dae:tuning}
In an ideal scheme of operations, the various stages of a complex instrument such as \Planck/LFI 
would have to be calibrated sequentially, so that 
the calibration of the REBA parameters would be the last step of the calibration procedure
\cite{LFI:radiometers:tuning} and would have no effect on 
the previous stages of calibration.
Practical experience has shown that there is a case in which the tuning of the acquisition
electronics has consequences on the subsequent tuning of the REBA parameters.
Indeed, the hypothesis at the root of the whole compression scheme is that the noise
variance of the input signal is large. This is in general true but this hypothesis could fail
if the variance of the signal after ADC quantization, \onboard\ coadding and mixing becomes too small.
In that case the signal will be over compressed with $\Cr > \CrTgt$ and the quantization error
will be larger or equal to the signal variance. To avoid this case either the DAE gain $G$ and 
$\Naver$ have to be properly tuned, or a set of particular combinations of $\GMFO$, $\GMFT$ values
has to be excluded.

The problem is to ensure $\sigmaone$ and $\sigmatwo$ to be greater than a minimal $\sigmatgt$ typically 
assumed to be at least $2$~$\adu$ in a suitable range of $\GMFO$, $\GMFT$ values.
From Eq.~(\ref{eq:mixing:rms}) it is evident that the $\sigmai^2$ as a function of $\GMFi$ defines two 
identical concave parabolas with a minimum in $\GMFO = \GMFT = \rmin = \sigmaskyload/\sigmaload^2$,
where both $\sigmaone$ and $\sigmatwo$ takes the value 
 \begin{equation}
   \sigma_{\mathrm{min}} = \sigmasky \sqrt{1-\corrsl^2},
\end{equation}

\FIGGainExclusion
 
\noindent
where $\corrsl$ is the correlation coefficient between \sky\ and \load. Note that $\sigma_{\mathrm{min}}=0$ 
just as in the case of a perfect correlation between \sky\ and \load.
So a sufficient condition to asses proper DAE calibration is 
 \begin{equation}
\sigmatgt < \sigmasky \sqrt{1-\corrsl^2},
\end{equation}

\noindent
which puts a constraint on the minimum $G/\sqrt{\Naver}$ which could be accepted. In particular assuming the 
quantization and the readout noise are small with respect to the \sky\ and \load\ RMS, at first order 

 \begin{equation}\label{eq:DAE:calib:condition}
 \frac{G}{\sqrt{\Naver}}> \frac{\sigmatgt}{{\sigmasky}_{,0}\sqrt{1-\corrsl^2}} ,
 \end{equation}

\noindent
where ${\sigmasky}_{,0}$ is the \sky\ RMS with $G=1$ and no averaging.

It could happen that in some cases the condition (\ref{eq:DAE:calib:condition}) can not be full--filled
for any reasonable value of $G$ and $\Naver$. So a {\em forbidden region } in
the $\GMFO$, $\GMFT$ space is defined by the need to have 
$\sigmatgt < \mathrm{min}(\sigmaone,\sigmatwo)$. 
This defines a ``cross'' centered into $\GMFO = \GMFT = \rmin = \sigmaskyload/\sigmaload^2$,
see Fig.~\ref{fig:gain:excl},
with ``harms'' parallel to the two axis of the $\GMFO$, $\GMFT$ space and having for each harm
a width give by 
 
 \begin{equation}\label{eq:DAE:Exclusion}
 \Delta r = 2 \frac{\sigmasky}{\sigmaload} 
           \sqrt{ 
            \left(\frac{\sigmatgt}{\sigmasky}\right)^2 - (1 - \corrsl^2) 
         }
\end{equation}

 \noindent
DAE calibrators could monitor the evolution of $\Delta r$ as $G/\sqrt{\Naver}$ varies.
In general the optimization of $\GMFO$ and $\GMFT$ is performed by scanning a rectangular region in the 
$(\GMFO, \GMFT)$ space of limited width. An informative parameter to avoid to harm a proper REBA calibration
after DAE calibration is to check the fraction of area of the region of interest excluded by 
the DAE calibration $\fDAE$. It is not possible of course to write a general formula for all the
possible cases, but often $\corrsl^2$ is small so that the excluded region has a center near $\GMFO=\GMFT=0$, 
while the optimization region is squared, centered on the origin with $-\GMFlim \le \GMFO, \GMFT \le +\GMFlim$ 
in that case

 \begin{equation}
  \fDAE = \frac
            {(4\GMFlim - \Delta r) \Delta r}
            {4\GMFlim^2}.
\end{equation}


\end{document}

%% file: d2_maris_08_symb.tex
\newcommand{\CHANGE}[1]{ {} }

 \def\Planck{{\sc Planck}}
 \def\sky{sky}
 \def\load{reference--load}
 \def\skyload{sky{--}{--}reference--load}
 \def\loadsky{reference--load{--}{--}sky}
 \def\onboard{on--board}
 \def\onground{on--ground}
 \def\radiofrequency{radio frequency}

 \def\OCA{{\ttfamily OCA}}
 \def\OCATK{{\ttfamily OCA2K}}
 \def\OCATWO{{\ttfamily OCA2}}

 \def\IDL{{\ttfamily IDL}}

 \def\LABEN{Thales Alenia Space (Italy)}

 \def\SamplingFreqHz{8192~\mbox{Hz}}
 \def\freqsampling{f_{\mathrm{sampling}}}     
 
 \def\PacketSamples{\mathcal{N}_{\mathrm{smp}}}
 \def\PacketTimeSpan{\tau_{\mathrm{pck}}}

 \def\Nbits{N_{\mathrm{bits}}}     
 \def\Nsymb{N_{\mathrm{symb}}}     
 \def\Nsymbeff{N_{\mathrm{symb}}^{\mathrm{eff}}}     

 \def\Nsamples{n_{\mathrm{samples}}}
 \def\NbitsCode{N_{\mathrm{bits},\mathrm{code}}}     
 \def\Lpck{{L_{\mathrm{pck}}}}                        

 \def\Lin{L_{\mathrm{in}}}                                               
 \def\Lout{L_{\mathrm{out}}}                    

 \def\Cr{C_{\mathrm{r}}}                        
 \def\CrTh{C_{\mathrm{r}}^{\mathrm{Th}}}        
 \def\EtaCrInd{\eta_{\mathrm{store}}}        

 \def\etacr{\eta_{C_\mathrm{r}}}        
 \def\etasampling{\eta_{C_\mathrm{r}}}        
 \def\etaoca{\eta_{\mathrm{oca2k}}}        

 \def\DataRate{R_{\mathrm{data}}}        
 \def\BandWidth{B_{\mathrm{data}}}

 \def\Amp{A} 
 \def\AmpUp{A_{\mathrm{up}}} 
 \def\AmpLow{A_{\mathrm{low}}} 
 \def\AmpD{D} 
 \def\AmpMax{A_{\mathrm{max}}} 

 \def\Driftsky{\Delta T_{\mathrm{sky}}} 
 \def\Driftload{\Delta T_{\mathrm{load}}} 
 \def\Drifti{\Delta T_{\mathrm{i}}} 

 \def\Driftnsky{\Delta \breve{T}_{\mathrm{sky}}} 
 \def\Driftnload{\Delta \breve{T}_{\mathrm{load}}} 
 \def\Driftni{\Delta \breve{T}_{\mathrm{i}}} 

 \def\nablaT{\nabla T} %
 \def\Qone{Q_{1}} 
 \def\Qtwo{Q_{2}} 

 \def\Qonea{Q_{1,\mathrm{l}}} 
 \def\Qoneb{Q_{1,\mathrm{r}}} 
 \def\Qtwoa{Q_{2,\mathrm{l}}} 
 \def\Qtwob{Q_{2,\mathrm{r}}} 

 \def\Qonec{\bar{Q}_{1}} 
 \def\Qtwoc{\bar{Q}_{2}} 

 \def\None{N_{1}} 
 \def\Ntwo{N_{2}} 

 \def\entropy{H}

 \def\entropysingle{H_{0}}
 \def\entropymin{{H_{\mathrm{min}}}}
 \def\entropyinf{{H_{\infty}}}

 \def\qmin{q_{\mathrm{min}}}

 \def\qnormopt{\breve{q}_{\mathrm{opt}}}
 \def\qopt{q_{\mathrm{opt}}}

 \def\entropyTgt{H_{\mathrm{tgt}}}

 \def\nentropy{h}
 \def\sigmastar{\sigma_{*}}

 \def\entropyone{H_1}
 \def\entropytwo{H_2}

 \def\phistop{\varphi_{\mathrm{stop}}}

 \def\etastop{\eta_{\mathrm{stop}}}

 \def\GMFOoptimal{r_1^{\mathrm{optim}}}      
 \def\GMFToptimal{r_2^{\mathrm{optim}}}      
 \def\Offsetoptimal{\mathcal{O}^{\mathrm{optim}}}
 \def\qoptth{q_{\mathrm{optim}}^{\mathrm{th}}}

 \def\Offset{\mathcal{O}}
 
 \def\CrTgt{C_{\mathrm{r}}^{\mathrm{Tgt}}}

 \def\GMF{r}      
 \def\GMFtilde{\tilde{r}}
 \def\GMFO{r_1}      
 \def\GMFT{r_2}      
       
 \def\GMFi{r_i}      
 \def\GMFc{\breve{r}_{\mathrm{c}}}      

 \def\GMFOpeak{r_1^{\mathrm{peak}}}      
 \def\GMFTpeak{r_2^{\mathrm{peak}}}      

 \def\GMFOn{\breve{r}_1}      
 \def\GMFTn{\breve{r}_2}      
 
 \def\GMFn{\breve{r}}      
 \def\GMFni{\breve{r}_i}      

 \def\GMFlim{r_{\mathrm{lim}}}

 \def\MixingM{\mathbf{M}}
 \def\DeMixingM{\mathbf{M}^{-1}}
 \def\SecondQuant{S_{\mathrm{q}}}

 \def\DistMin{\breve{\ell}_{\mathrm{min}}}      

 \def\TargetFunction{\chi}
 \def\CriterionFunction{\mathcal{Q}_c}
 \def\Policy{\Pi_c}
 \def\Metric{\mu_c}

 \def\kpdf{k_{\mathrm{pdf}}}     
 \def\ForbiddenCase{\GMFO = \GMFT}
 \def\MandatoryCase{\GMFO \ne \GMFT}

\def\ForbiddenCasen{\GMFOn = \GMFTn}

 \def\Separationn{\breve{\Delta}_{\mathrm{distr}}}

 \def\Separation{\Delta_{\mathrm{distr}}}

 \def\omegaSpin{\omega_{\mathrm{spin}}}     
 \def\boresight{\beta}     
 \def\beamrad{b_{\mathrm{rad}}}     
 \def\oversampling{n_{\mathrm{over}}}     

 \def\Naver{N_{\mathrm{aver}}}     
 \def\qnorm{\breve{q}}     

 \def\Quack{\mathcal{Q}_{\mathrm{ack}}} 
 \def\QuackUnit{\mathcal{Q}_{\mathrm{ack}}^1} 
 \def\Quackone{\mathcal{Q}_{\mathrm{ack},1}} 
 \def\Quacktwo{\mathcal{Q}_{\mathrm{ack},2}} 
 \def\Quacki{\mathcal{Q}_{\mathrm{ack},i}} 
 \def\Nsat{N_{\mathrm{sat}}} 

 \def\Quackonepeak{\mathcal{Q}_{\mathrm{ack},1}^{\mathrm{peak}}} 
 \def\Quacktwopeak{\mathcal{Q}_{\mathrm{ack},2}^{\mathrm{peak}}} 

 \def\Quackmax{\mathcal{Q}_{\mathrm{ack},1}^{\mathrm{max}}} 

 \def\safety{\mathcal{S}_{\mathrm{ft}}} 

\def\qoptquack{q_{\mathrm{opt},\mathcal{Q}{\mathrm{ack}}}}

 \def\Qerr{\epsilon_{q}}
 \def\QerrAlphaBeta{E_{q,\alpha,\beta}}
 \def\Qerrdiff{\epsilon_{q,\mathrm{diff}}}
 \def\Qerrdiffmuk{\Delta T_{q}}
 \def\Qerrsky{\epsilon_{q,\mathrm{sky}}}
 \def\Qerrload{\epsilon_{q,\mathrm{load}}}

 \def\gammadiff{\gamma_{\mathrm{diff}}}

\def\Gammadiff{\Gamma_{\mathrm{diff}}}
 
\def\Gammasky{\Gamma_{\mathrm{sky}}}
\def\Gammaload{\Gamma_{\mathrm{load}}}
 
 \def\Tnoise{{T_{\mathrm{noise}}}} 
 \def\Tsky{{T_{\mathrm{sky}}}} 
 \def\Tload{{T_{\mathrm{load}}}} 

 \def\TskyZ{{T_{\mathrm{sky},0}}} 
 \def\TloadZ{{T_{\mathrm{load},0}}} 

 \def\Ampsky{{\Delta T_{\mathrm{sky}}}} 
 \def\Ampload{\Delta T_{\mathrm{load}}} 

 \def\Ampskytau{A_{\mathrm{sky},\tau}} 
 \def\Amploadtau{A_{\mathrm{load},\tau}} 

 \def\Ampdotsky{{\dot{A}_{\mathrm{sky}}}} 
 \def\Ampdotload{\dot{A}_{\mathrm{load}}} 

 \def\corrsl{\varrho_{\mathrm{s}\mathrm{l}}} 

 \def\sigmasky{\sigma_{\mathrm{sky}}} 
 \def\sigmaload{\sigma_{\mathrm{load}}} 
 \def\sigmaskynoise{\sigma_{n,\mathrm{sky}}} 
 \def\sigmaloadnoise{\sigma_{n,\mathrm{load}}} 
 \def\sigmaone{\sigma_{1}} 
 \def\sigmatwo{\sigma_{2}} 

 \def\sigmadiff{\sigma_{\mathrm{diff}}} 

 \def\sigmaonenoise{\sigma_{n,1}} 
 \def\sigmatwonoise{\sigma_{n,2}} 

\def\sigmaskyload{\sigma_{\mathrm{sky},\mathrm{load}}} 
 \def\sigmaskyloadnoise{\sigma_{n,\mathrm{sky},\mathrm{load}}} 
 \def\sigmaonetwo{\sigma_{1,2}} 
 \def\sigmai{\sigma_{i}} 
 \def\noisesky{{n_{\mathrm{sky}}}} 
 \def\noiseload{n_{\mathrm{load}}} 

 \def\sigmani{\breve{\sigma}_{i}} 
 \def\sigmanone{\breve{\sigma}_1} 
 \def\sigmantwo{\breve{\sigma}_2} 
 \def\sigmanonetwo{\breve{\sigma}_{1,2}} 

 \def\Rsigma{R_{\sigma}} 

 \def\Tskytilde{{T_{\mathrm{sky}}}} 
 \def\Tloadtilde{{T_{\mathrm{load}}}} 

 \def\Tskymean{\overline{T}_{\mathrm{sky}}} 
 \def\Tloadmean{\overline{T}_{\mathrm{load}}} 

 \def\TskyADC{T^{\mathrm{ADC}}_{\mathrm{sky}}} 
 \def\TloadADC{T^{\mathrm{ADC}}_{\mathrm{load}}} 

 \newcommand{\TskyADCt}[1]{T^{\mathrm{ADC}}_{\mathrm{sky},t=#1}} 
 \newcommand{\TloadADCt}[1]{T^{\mathrm{ADC}}_{\mathrm{load},t=#1}} 

 \def\Tone{T_{1}} 
 \def\Ttwo{T_{2}} 

 \def\Tonen{\breve{T}_{1}} 
 \def\Ttwon{\breve{T}_{2}} 

 \def\Timean{\overline{T}_{i}} 
 \def\Timeann{\overline{{\breve{T}}}_{i}} 

 \def\Talpha{T_{\alpha}} 
 \def\Talphamean{\overline{T}_\alpha} 

 \def\Talphan{\breve{T}_{\alpha}} 
 \def\Talphameann{\overline{\breve{T}}_{\alpha}} 

 \def\Tskyn{\breve{T}_{\mathrm{sky}}} 
 \def\Tloadn{\breve{T}_{\mathrm{load}}} 

 \def\Tskymeann{\overline{\breve{T}}_{\mathrm{sky}}} 
 \def\Tloadmeann{\overline{\breve{T}}_{\mathrm{load}}} 

 \def\DeltaT{\Delta T} 
 \def\DeltaTn{\Delta \breve{T}} 

 \def\Pone{\mathcal{P}_{1}} 
 \def\Ptwo{\mathcal{P}_{2}} 

 \def\TalphaR{\tilde{T}_{\alpha}} 
 \def\xR{\tilde{x}} 

 \newcommand{\EXP}[1]{\mathrm{E}{\left[#1\right]}}
 \newcommand{\MEAN}[1]{\mathrm{mean}{\left[#1\right]}}
 \newcommand{\RMS}[1]{\mathrm{rms}{\left[#1\right]}}
 \newcommand{\COV}[1]{\mathrm{cov}{\left[#1\right]}}
 \newcommand{\VAR}[1]{\mathrm{var}{\left[#1\right]}}

 \def\Vmin{V_{\mathrm{min}}} 
 \def\Vmax{V_{\mathrm{max}}} 
 \def\sigmaadc{\sigma_{\mathrm{ADC}}}
 \def\sigmarout{\sigma_{\mathrm{ron}}}
 \def\sigmazero{\sigma_0}
 \def\sigmatgt{\sigma_{\mathrm{tgt}}}

\def\fDAE{f_{\mathrm{DAE},\mathrm{excl}}}

\def\rmin{r_{\mathrm{min}}}

 \def\round{\mathrm{round}}

 \def\Qmin{Q_{\mathrm{min}}} 
 \def\Qmax{Q_{\mathrm{max}}} 

 \def\adu{\mathrm{adu}}   

 \def\gsim{{\stackrel{_>}{_\sim}}} 

 \def\Tsys{{T_{\mathrm{sys}}}}

 \def\sigmaWN{{\sigma_{\mathrm{wn}}}}

 \def\qADC{{q_{\mathrm{ADC}}}}

 \def\sigmaqeff{(\sigma/q)_{\mathrm{eff}}}

%% file: d2_maris_08_tab.tex
\def\ATABcsl{
 \begin{table*}[ht]
\begin{center}
\caption{
Representative REBA Parameters, the measured $\Cr$ and relative processing errors from the CSL tests for \Planck/LFI.
  Detectors are grouped by frequency channel,
  for each quantity $x$ the table reports its group median and group standard deviation $\delta x$ 
   as a measure of the group internal dispersion, $\delta x$ must be not considered as an error.
 }
 \begin{tabular}{cccc}
\hline
\hline
  & \multicolumn{3}{c}{Frequency Channel} \\
  & 30 GHz & 44 GHz & 70 GHz \\
\hline
  Detectors & 8   & 12 & 24 \\
  $\Naver$  & 126 & 88 & 53 \\
  $\GMFO$   & $1.042 \pm 0.032$ & $1.042 \pm 0.024$ & $1.042 \pm 0.012$ \\
  $\GMFT$   & $0.917 \pm 0.065$ & $0.917 \pm 0.025$ & $0.958 \pm 0.020$ \\
  $q$ [adu] & $0.297 \pm 0.034$ & $0.198 \pm 0.044$ & $0.279 \pm 0.048$ \\
            &                   &                   &                    \\
  $\Cr$     & $2.400 \pm 0.024 $ & $2.440 \pm 0.019 $ & $2.380 \pm0.023 $ \\
  $\left({\Qerr}/{\sigma}\right)_{\mathrm{sky}} $ 
            & $0.420 \pm 0.278 $ & $0.269 \pm 0.184 $ & $0.177 \pm 0.063 $ \\
  $\left({\Qerr}/{\sigma}\right)_{\mathrm{load}} $ 
            & $0.432 \pm 0.267 $ & $0.271 \pm 0.183$ & $0.178 \pm 0.063$ \\
  $\left({\Qerr}/{\sigma}\right)_{\mathrm{diff}} $ 
            & $0.0341 \pm 0.0016 $ & $0.0369 \pm 0.0010$ & $0.0351 \pm 0.0010$ \\
  ${\Qerrdiffmuk}$~[$\mu\mathrm{K}$] 
            & $1.759 \pm 0.148$ & $2.412 \pm 0.356$ & $1.905 \pm 0.287$ \\
            &                   &                   &                    \\
  Data Rate per Detector [bits/sec] 
            & $454.9 \pm 4.1$ & $640.8 \pm 4.7$ & $1108.2 \pm 9.8$ \\
  Data Rate per Frequency Channel [bits/sec] 
            & $3641.8$ & $7689.9$ & $26600.3$ \\
            &                   &                   &                    \\
  Total Data Rate [bits/sec] 
            & \multicolumn{3}{c}{37932}\\
\hline
\hline
 \end{tabular}
 \label{tab:csl}
\end{center}
\end{table*}
}

\def\TABtypicalcriteria{
 \begin{table}[ht]
 \caption{
 List of possible criteria
 }
 \begin{tabular}{lll}
\hline
\hline
 \multicolumn{1}{c}{Condition} & \multicolumn{1}{c}{Criterium} & \\
\hline
 $\Qerrdiff = \min(\Qerrdiff)$ &  $\CriterionFunction = \min(\Qerrdiff)/\Qerrdiff $ & \\
 $\Qerrsky = \min(\Qerrsky)$   & $\CriterionFunction = \min(\Qerrsky)/\Qerrsky $ \\
 $\Qerrload = \min(\Qerrload)$ & $\CriterionFunction = \min(\Qerrload)/\Qerrload $ \\
 $\Qerrload = \Qerrdiff$ & $\CriterionFunction = 1-|\Qerrload-\Qerrdiff|/\max(\Qerrload+\Qerrsky) $ \\
 $\Qerrload = \Qerrsky$ & $\CriterionFunction = 1-|\Qerrload-\Qerrsky|/\max(\Qerrload+\Qerrsky) $ \\
 $\Qerrdiff = \Qerrsky$ & $\CriterionFunction = 1-|\Qerrdiff-\Qerrsky|/\max(\Qerrdiff+\Qerrsky) $ \\
\hline
\hline
 \end{tabular}
 \label{tab:typical:criteria}
\end{table}
} 

\def\TABtesttoi{
 \begin{table}[t]
 \caption{
Table for the optimization example illustrated in Fig.~10.
 Subtable A) reports \sky\ and \load\
 statistics together with their correlation, and the derived $r$ and $\Rsigma$.
 Subtable B) reports the optimized REBA parameters, obtained respectivelly with the analytical approximation,
 and subsequently with a numerical scan of the parameter space. The processing statistics are reported 
 in the Subtable C) where the processing errors and the $\Cr$\ expected from the analytical model are
 compared with the numerical results. The third column of the subtable gives the theoretical expectations after
 replacing $q$ obtained from theory with $q$ obtained by numerical means.
 }
 \begin{center}
 \begin{tabular}{lccc}
\hline
\hline
&&&\\
                & \multicolumn{3}{c}{\bf A) TOI Statistics}   \\
                & \multicolumn{1}{c}{\bf \sky} & \multicolumn{1}{c}{\bf \load}  & \multicolumn{1}{c}{\bf Combined}\\
\hline
{\bf Mean [adu]}      & 12041.29  & 12313.63 & \\
{\bf RMS  [adu]}      &     9.72  & 10.06     & \\
{\bf Slope [adu/sec]} &    0.026  & 0.027    & \\
{\bf $\corrsl$ }      &           &          & 0.9988 \\
{\bf $r$ }            &           &          &  0.9779\\
{\bf $\Rsigma$ }      &           &          & 0.9659\\
{\bf RMS($\DeltaT)$ [adu]} &      &          & 1.45\\
\hline
&&&\\
                & \multicolumn{2}{c}{\bf B) Optimized Parameters}   & \\
                & \multicolumn{1}{c}{\bf Analytical} & \multicolumn{1}{c}{\bf Numerical}  & \\
\hline
$\GMF1$           & 1.25      & 1.25     & \\
$\GMF2$           & 0.83      & 0.83     & \\
$\Offset$         & 785.41    & 784.39  & \\
$q$               & 0.203     & 0.317    & \\
\hline
&&&\\
                & \multicolumn{2}{c}{\bf C) Processing Statistics}   & \\
                & 
                  \multicolumn{1}{c}{\bf \begin{tabular}{c} \\ \\ Analytical \end{tabular}} & 
                  \multicolumn{1}{c}{\bf \begin{tabular}{c} \\ \\ Numerical \end{tabular}}  &  
             {\bf \begin{tabular}{c} Analytical \\ with \\ numerical $q$ \end{tabular}} \\
\hline
$N_{\mathrm{pck}}$  & 97       & 111 & -- \\
$\Separation$       & 479.1    & 478.9   &  -- \\
$\sigmaone$ [adu]   & 3.291  & 3.292 & -- \\
$\sigmatwo$ [adu]   & 1.885  & 1.884 & -- \\
$\Qerrdiff$ [adu]   &  0.043 & 0.067 & 0.068 \\
$\Qerrsky$  [adu]   &  0.211 & 0.330 & 0.329 \\
$\Qerrload$ [adu]   &  0.199 & 0.310 & 0.311 \\
$\sigmaqeff$        & 9.7 & 6.2 & 6.2\\
$\max(|\Quack|)$    &  0.388 & 0.228 & 0.248 \\
$H_\mathrm{Tot}$ {\bf bits}   &   6.667 & 6.023  & 6.021  \\
{Mean $H$ \bf bits} &   -- & 5.489  & --  \\
{\bf Mean $\etaoca$}&--  & 0.828 &   --  \\
{\bf Min} $\Cr$     &   --   & 2.286 &   --  \\
{\bf 5\%} $\Cr$     &    --  & 2.333 &   --  \\
{\bf Median} $\Cr$  &   --   & 2.413 &   --  \\
{\bf Mean} $\Cr$    &   2.4  & 2.414 & 2.657 \\
{\bf 95\%} $\Cr$    &   --   & 2.469 &   --  \\
{\bf Max} $\Cr$     &   --   & 2.510 &   --  \\
{\bf RMS} $\Cr$     &   --   & 0.045 &   --  \\
\hline
\hline
 \end{tabular}
 \end{center}
 \label{tab:tests}
\end{table}
}

%% file: d2_maris_08_fig.tex
\newcommand{\ROTATEBOX}[1]{#1}

\def\FIGmixingANDinterlacing{
\begin{figure*}[ht]
\begin{center}
\begin{tabular}{ll}
a) & b) \\
\ROTATEBOX{\includegraphics[height=5.5cm,width=5cm]{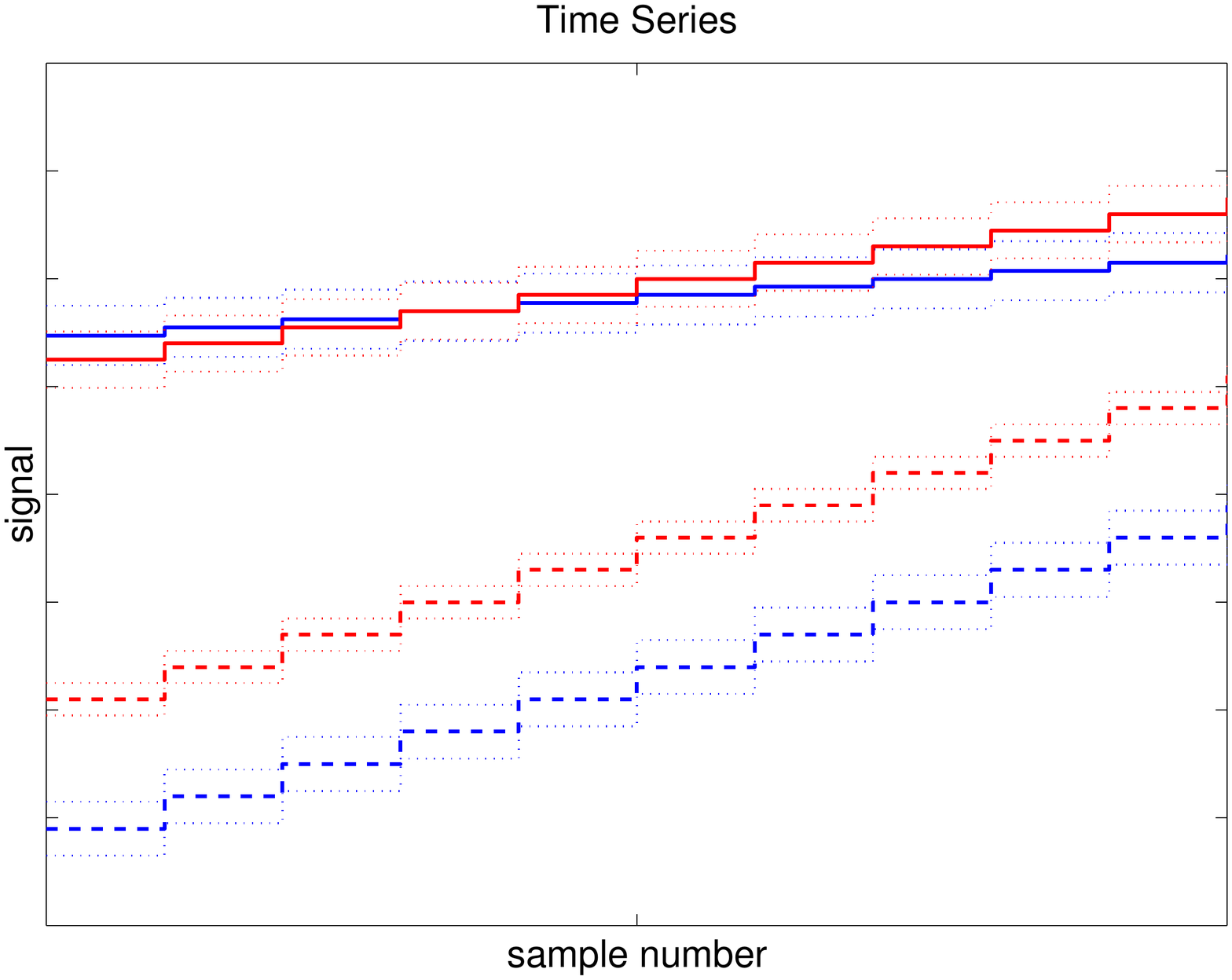}} & 
    \ROTATEBOX{\includegraphics[height=5.5cm,width=5cm]{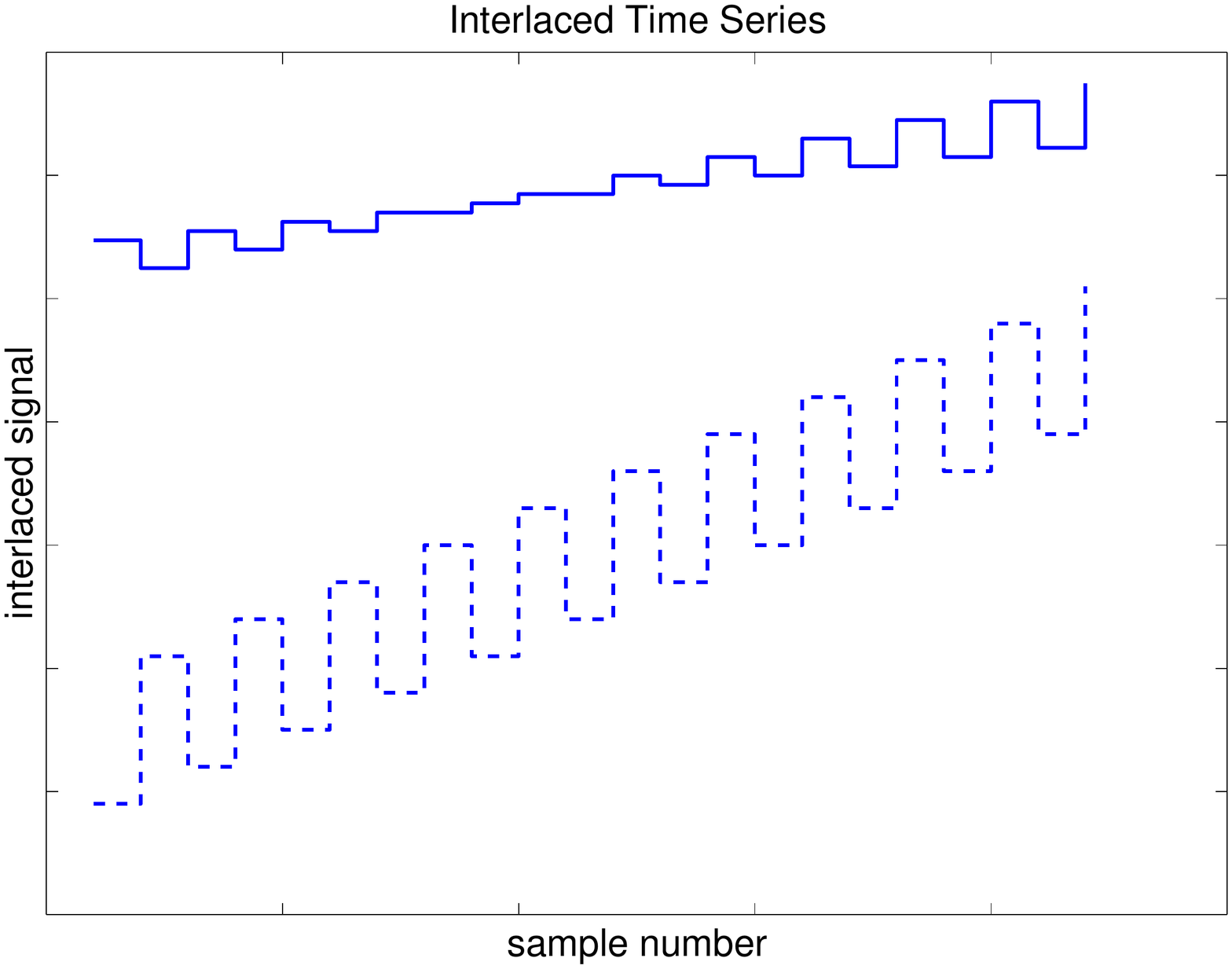}} \\
c) & d) \\
\ROTATEBOX{\includegraphics[height=5.5cm,width=5cm]{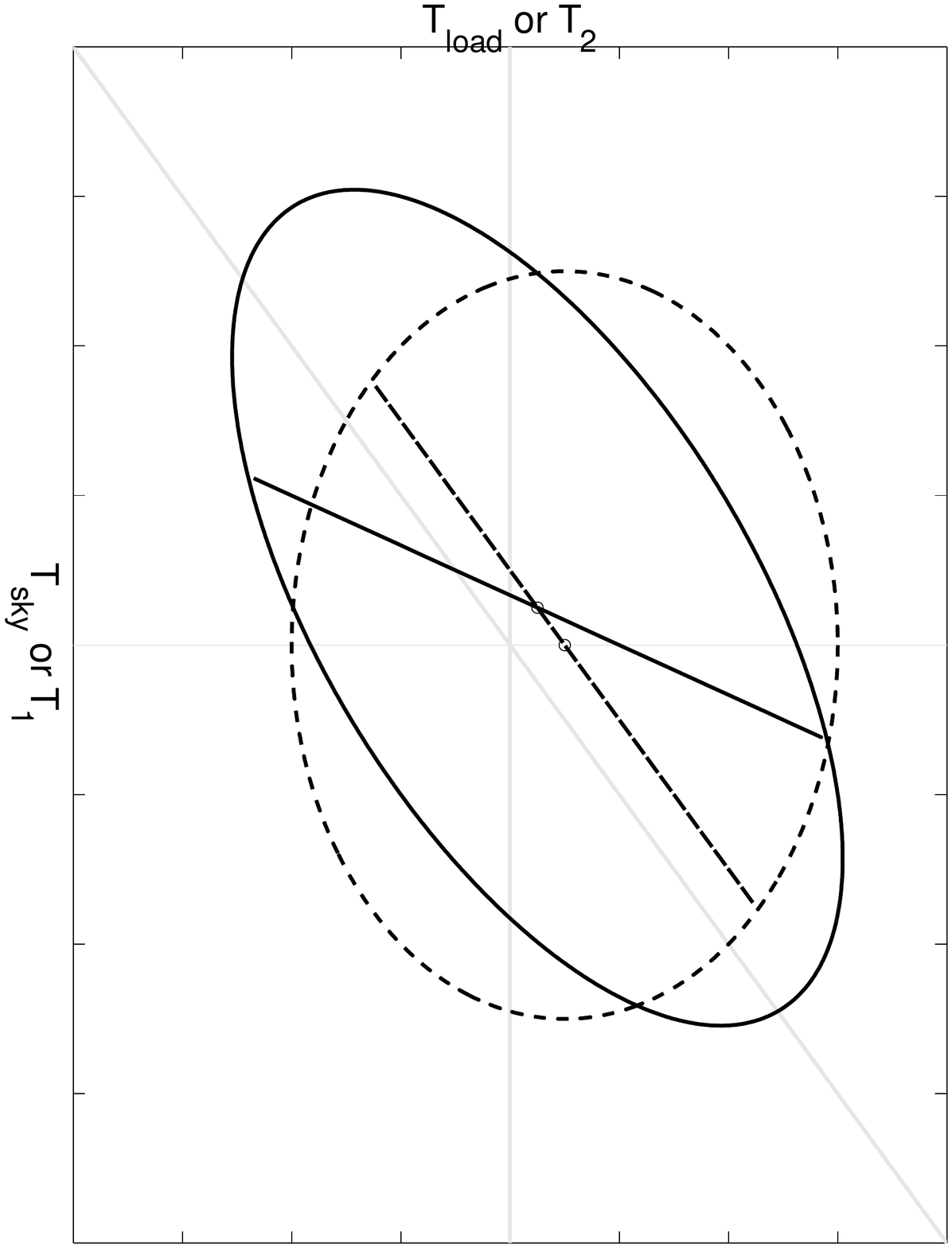} }
  & \ROTATEBOX{\includegraphics[height=5.5cm,width=5cm]{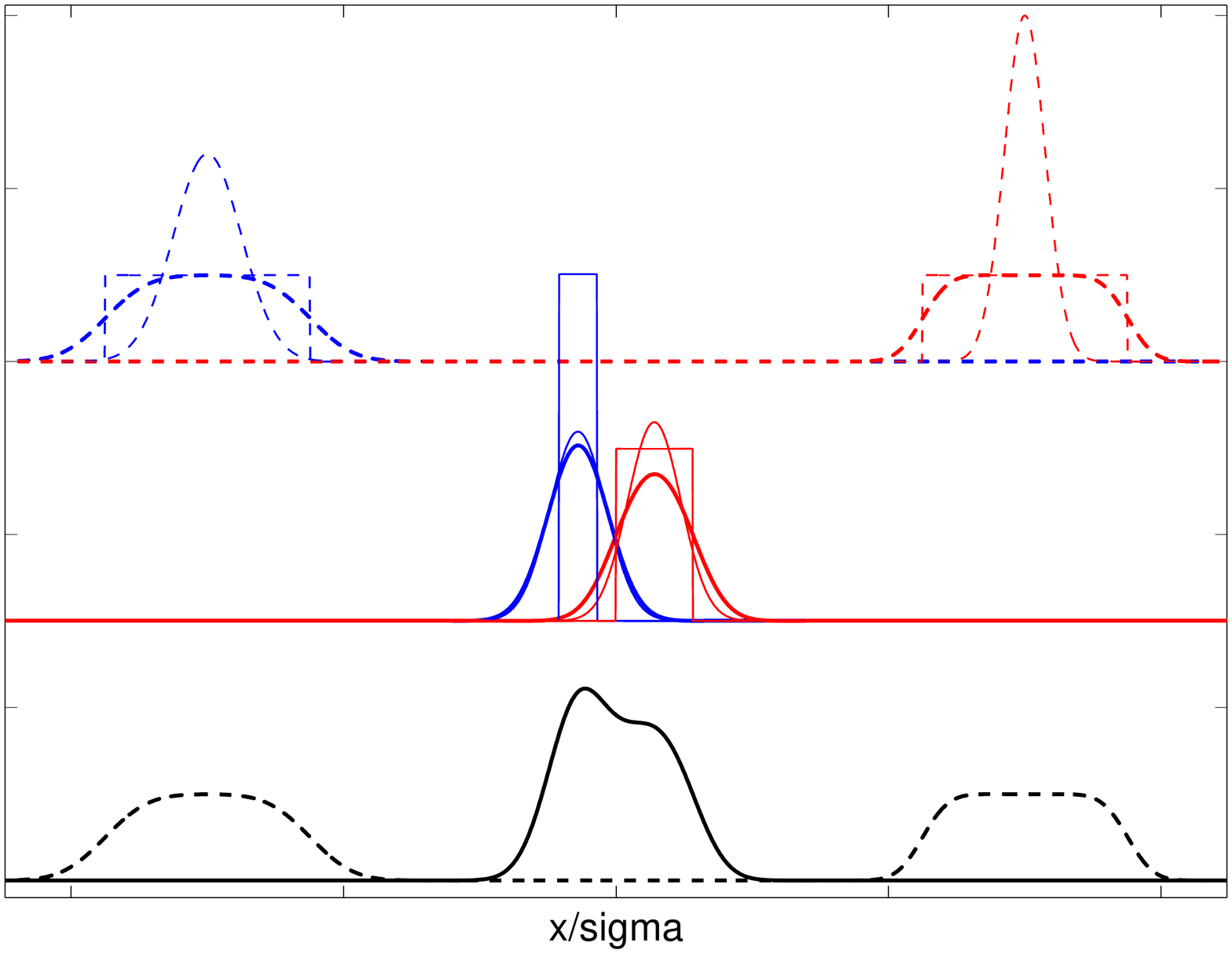}} \\
\end{tabular}
\end{center}
\caption[]{\label{fig:mixing:interlacing}
The effect of mixing and interlacing on time series (top frames) and distributions (low frames).
in the output to the radiometer.
Frame a)
is an input time series, for sky (red dashed--line) and \load\ (blue dashed--line),
and the corresponding mixed quantities $\Qone$ (red full--line), $\Qtwo$ (blu full--line) 
after mixing with $\GMFO = 3/4$, $\GMFT=1/2$. 
The range of values allowed by noise within $\pm 5\sigma$ are represented in both 
mixed and not mixed quantities by the upper and lower dotted lines.
The input time series has identical drifts on sky and \load\ equivalent to several noise $\sigma$s
and corresponds to a signal dominated time serie as defined in Sect.~\ref{sec:OP:signal:model}.
Frame b)
are the signals as seen from the compressor after interlacing, full--line without mixing, 
dashed--line after mixing.
Frame c)
is the effect on a normal distribution and on a ramp, green before mixing and red after mixing.
Frame d)
shows the effect on the projected distributions after interlacing.
}
   \end{figure*}
}

\def\FIGgammamincomposition{
\begin{figure}
\begin{center}
\begin{tabular}{c}
 \ROTATEBOX{
 \includegraphics[height=8cm,width=8cm]{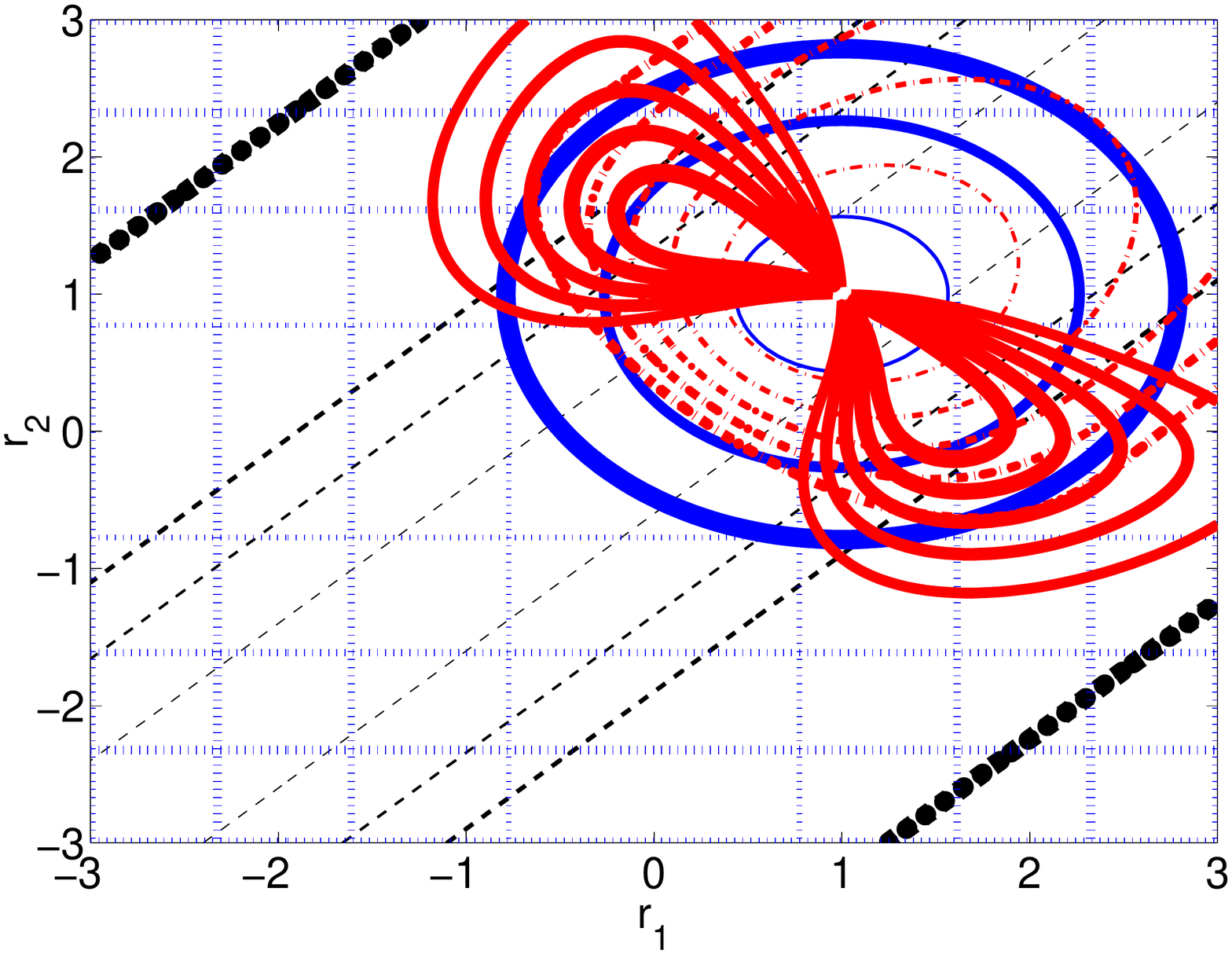}
 }
\end{tabular}
\end{center}
\caption[]{\label{fig:gamma:min:composition}
Iso contour lines for the functions entering $\Qerrdiff$ and $\Qerrdiff$ 
as a function of ($\GMFO$, $\GMFT$).
 Vertical blue dashed lines $\sigmaone$, horizontal blue dashed lines 
$\sigmatwo$, black dashed lines $(\GMFT - \GMFO)^2$, blue contours $(\GMFO-\GMF)^2+(\GMFT-\GMF)^2$,
red dashed contours $\sqrt{\sigmaone\sigmatwo}$, and the red contours are the resulting $\Qerrdiff$.
The width of the lines varies with the value of the function. 
}
   \end{figure}
}

\def\FIGallminimacrit{
\begin{figure*}[ht]
\begin{center}
\begin{tabular}{ll}
a) & b) \\
\ROTATEBOX{\includegraphics[height=5.5cm,width=5cm]{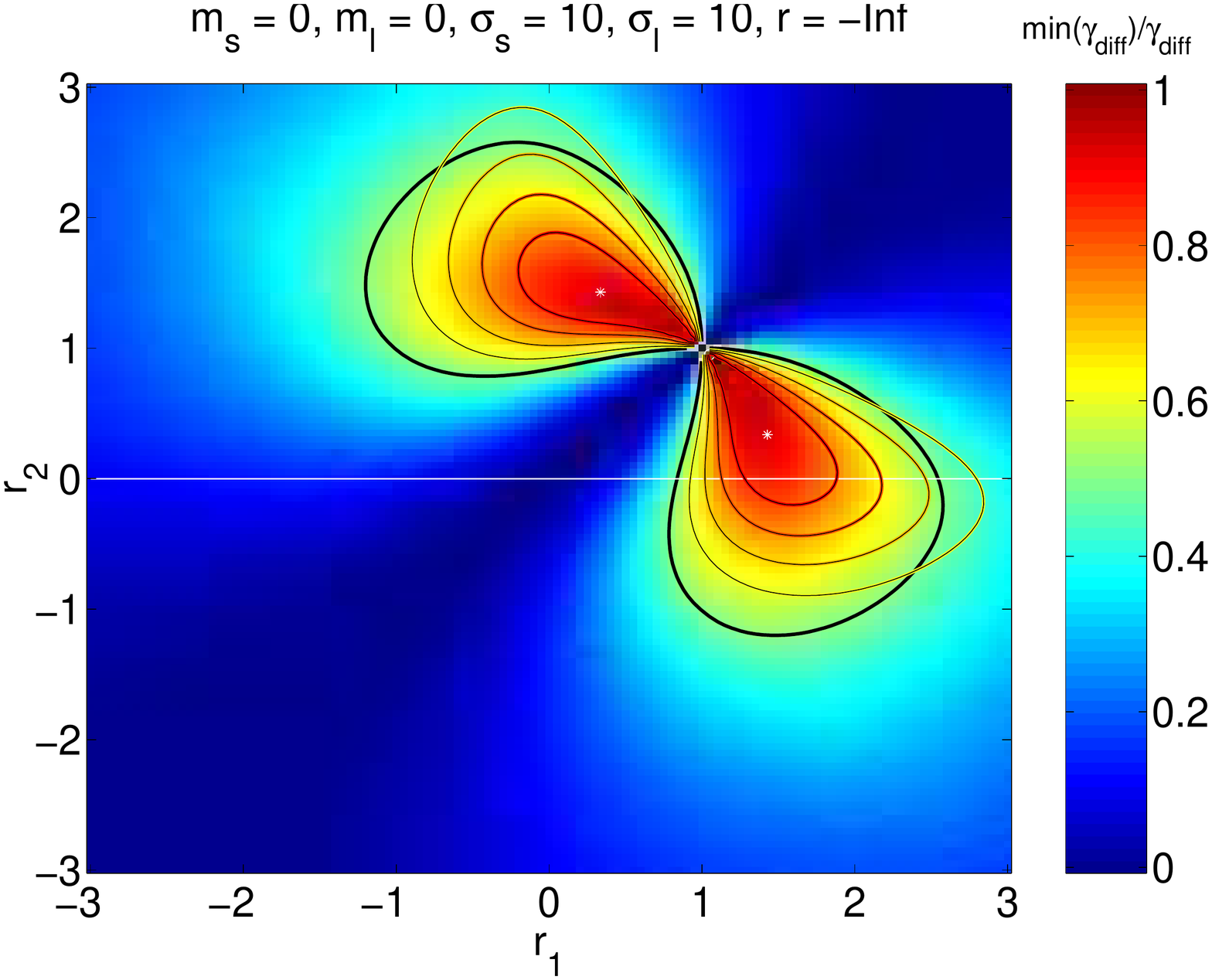}} & 
  \ROTATEBOX{\includegraphics[height=5.5cm,width=5cm]{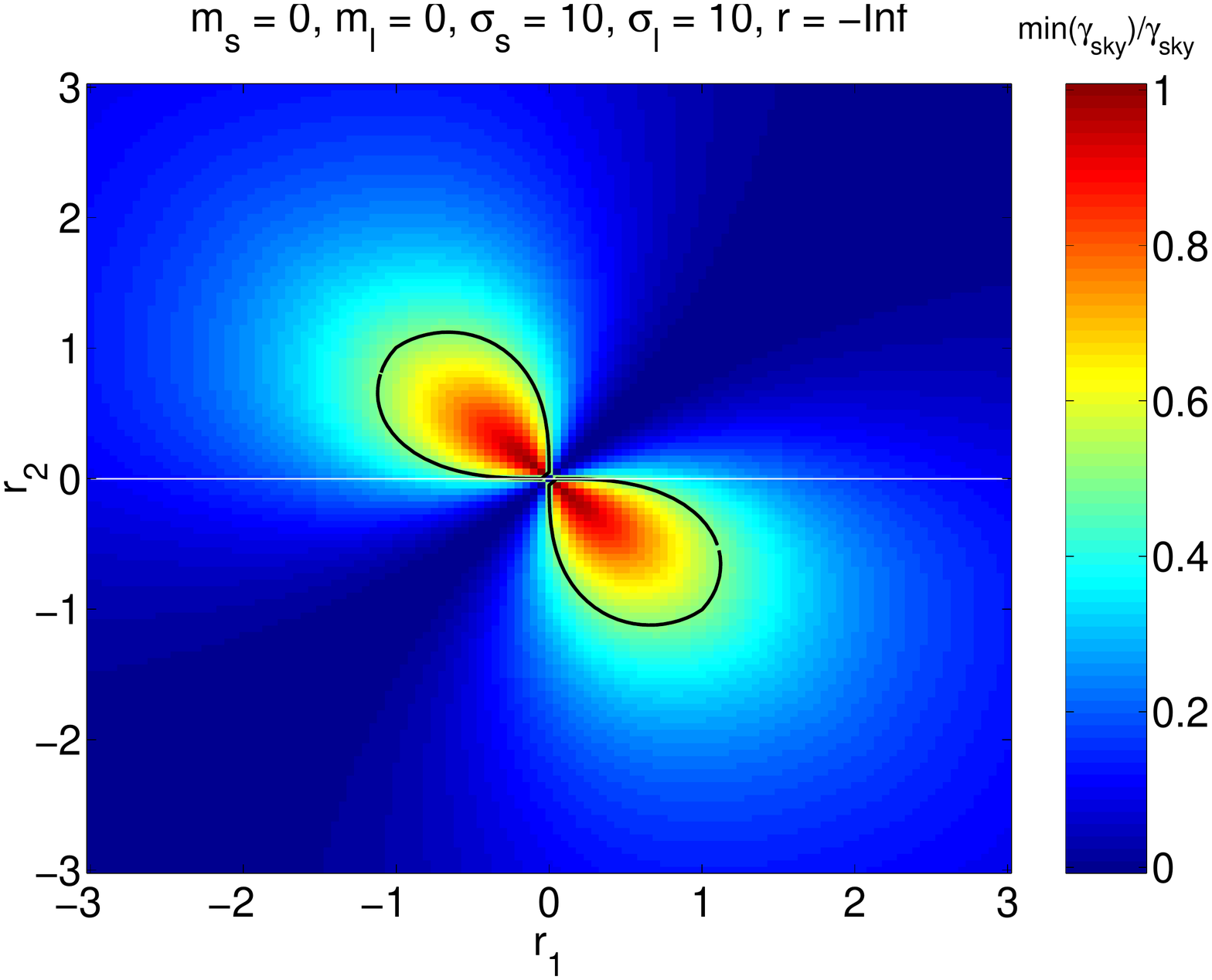}} \\
c) & d) \\
\ROTATEBOX{\includegraphics[height=5.5cm,width=5cm]{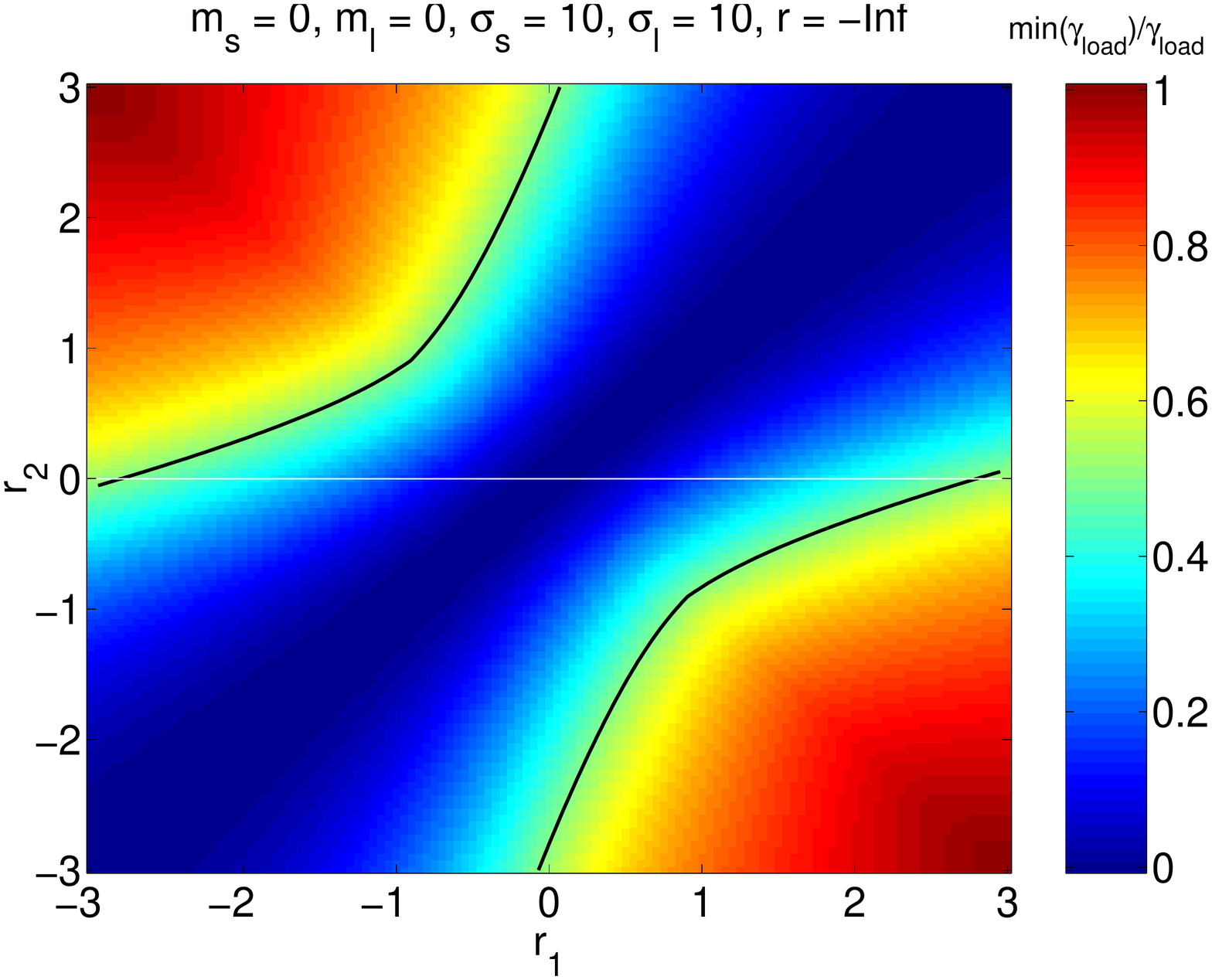}} & 
 \ROTATEBOX{\includegraphics[height=5.5cm,width=5cm]{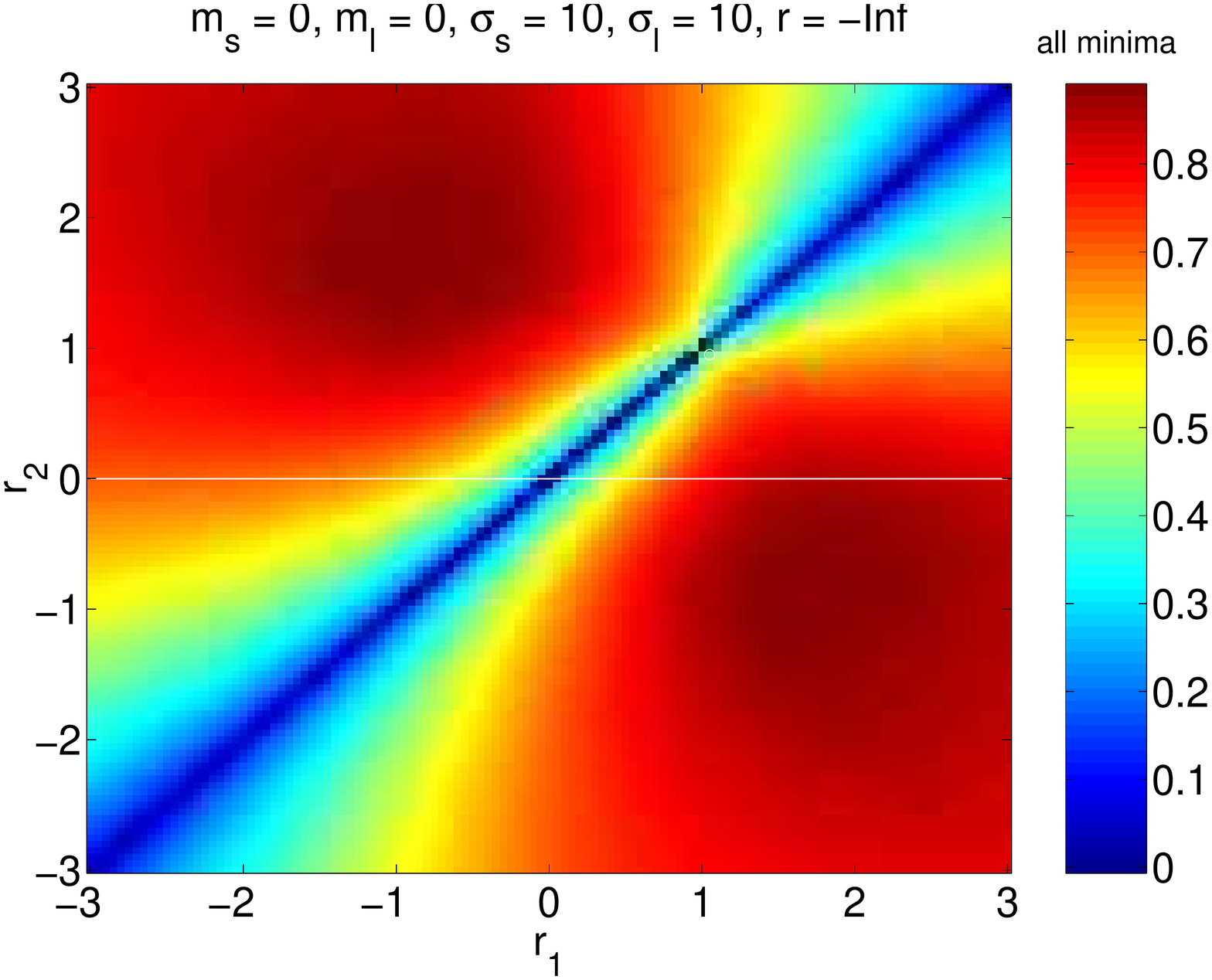}} \\
\end{tabular}
\end{center}
\caption[]{\label{fig:mosaic:allminima:crit}
  Iso contour lines identifiing the regions in the 
($\GMFO$, $\GMFT$) space
where
$\min(\Qerrdiff)/\Qerrdiff$ top--left,
$\min(\Qerrsky)/\Qerrsky$ top--right,
$\min(\Qerrload)/\Qerrload$ bottom--left,
and where all of the $\Qerr$ are as near as possible at their minima,
bottom--right.
 The lines are isocontours of $\min(x)/x$ which is 1 when the minimum is reached.
For the bottom--right frame
the iso--contours are the cubic root of the product of the other frames.
In the top--left frame
white $+$ denotes the positions of the analytical solution, 
the black thick contour is for $\min(\Qerrdiff)/\Qerrdiff = 0.5$,
the black thin contour lines are for $\min(\Qerrdiff)/\Qerrdiff$ assuming $\entropy \equiv \entropyinf$.
The simulation is for $\Tskymean=\Tloadmean=0$, $\sigmasky=\sigmaload=10$~ADU,
$\GMF=1$. No drifts are included.
}
   \end{figure*}
}

\def\FIGequalscrit{
\begin{figure*}[ht]
\begin{center}
\begin{tabular}{ll}
a) & b) \\
 \ROTATEBOX{\includegraphics[height=5.5cm,width=5cm]{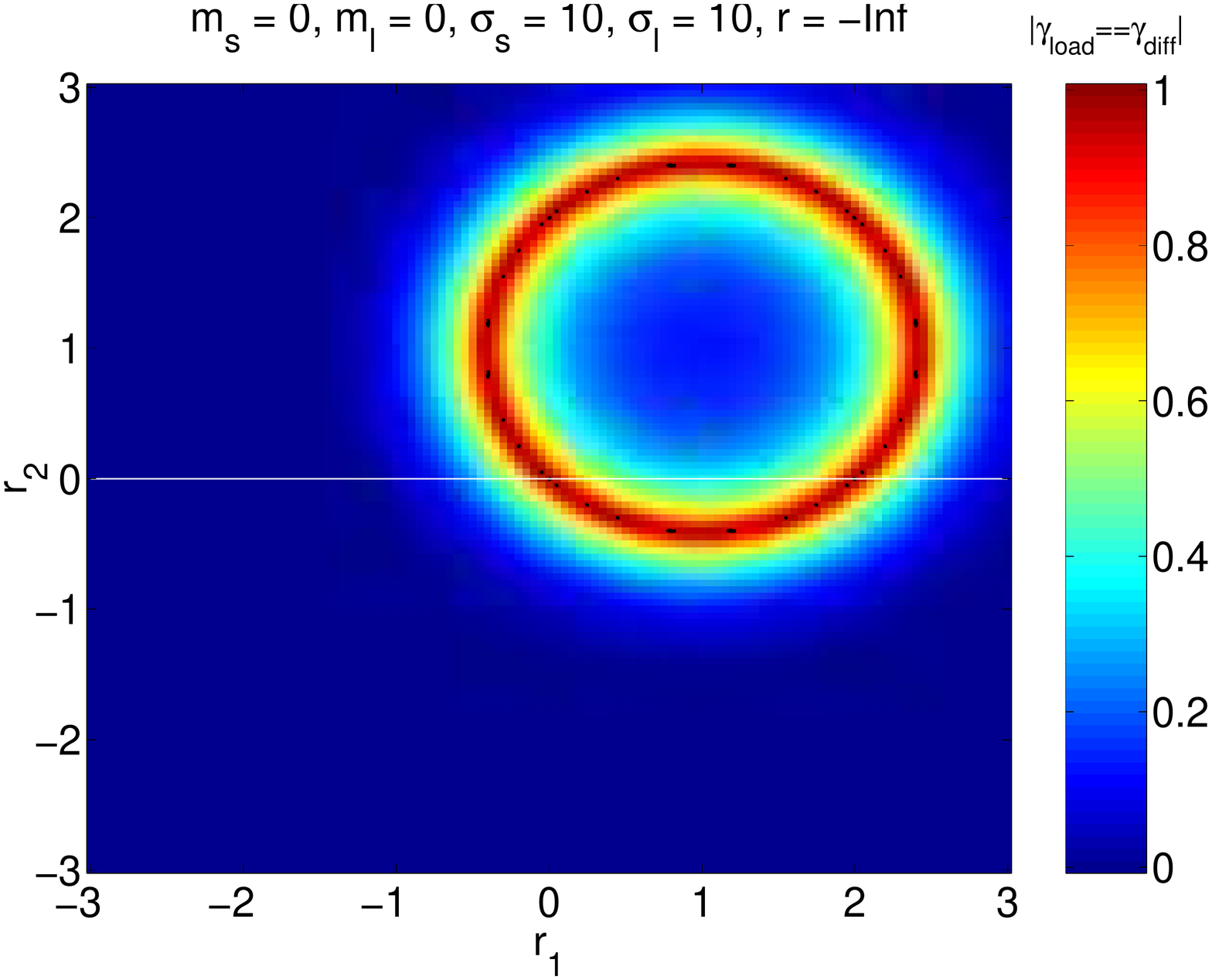}} & 
 \ROTATEBOX{\includegraphics[height=5.5cm,width=5cm]{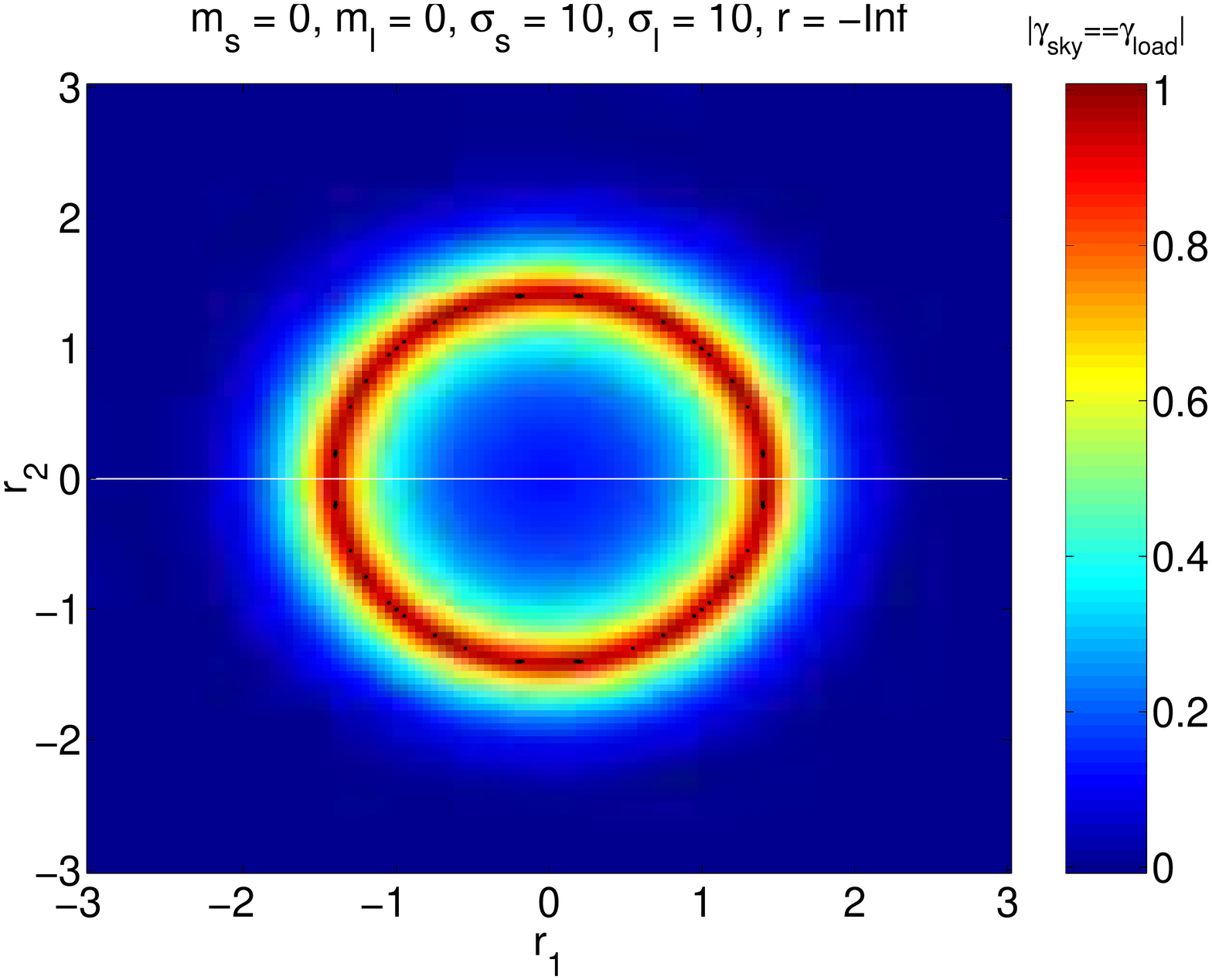}} \\
c) & d) \\
 \ROTATEBOX{\includegraphics[height=5.5cm,width=5cm]{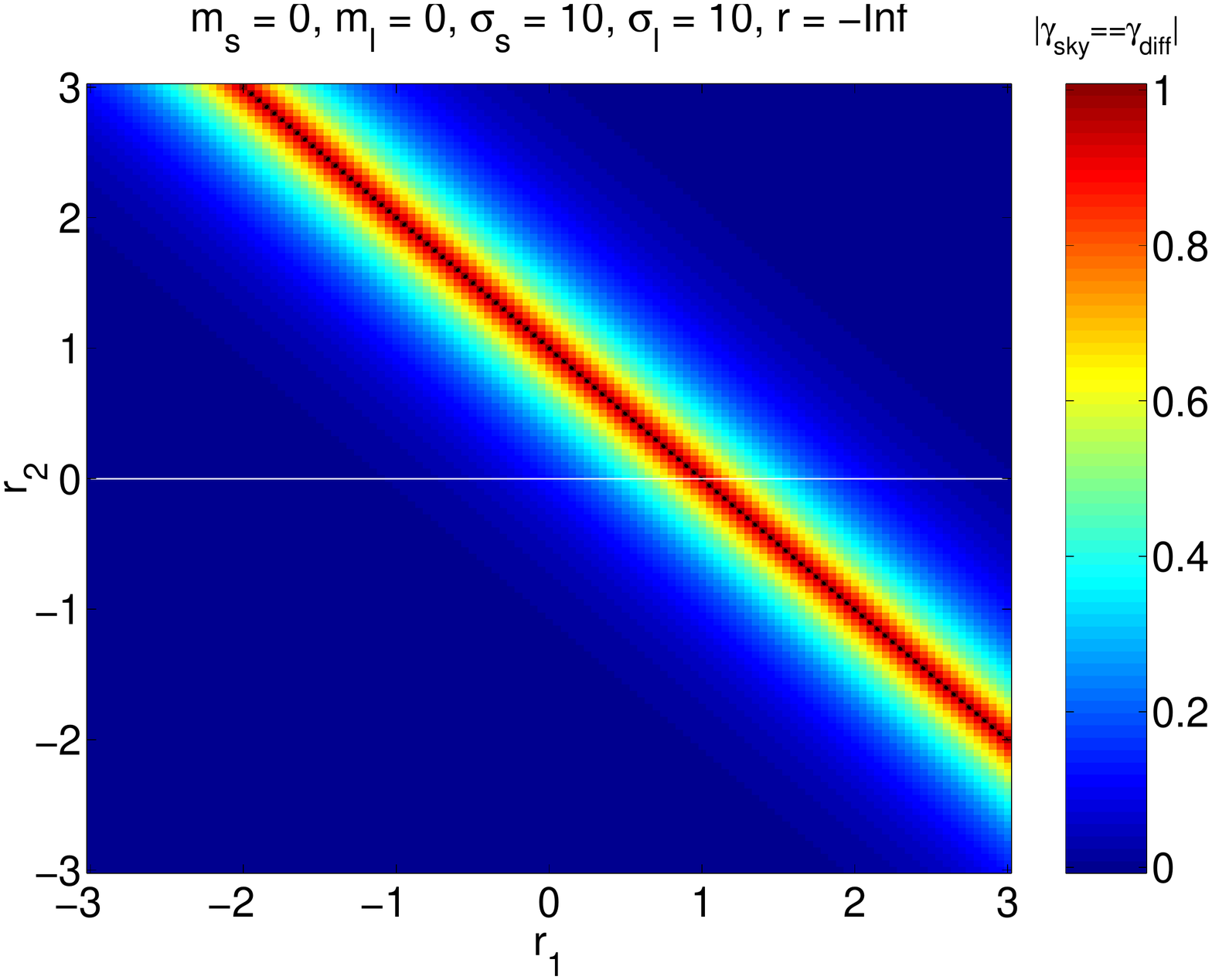}} & 
 \ROTATEBOX{\includegraphics[height=5.5cm,width=5cm]{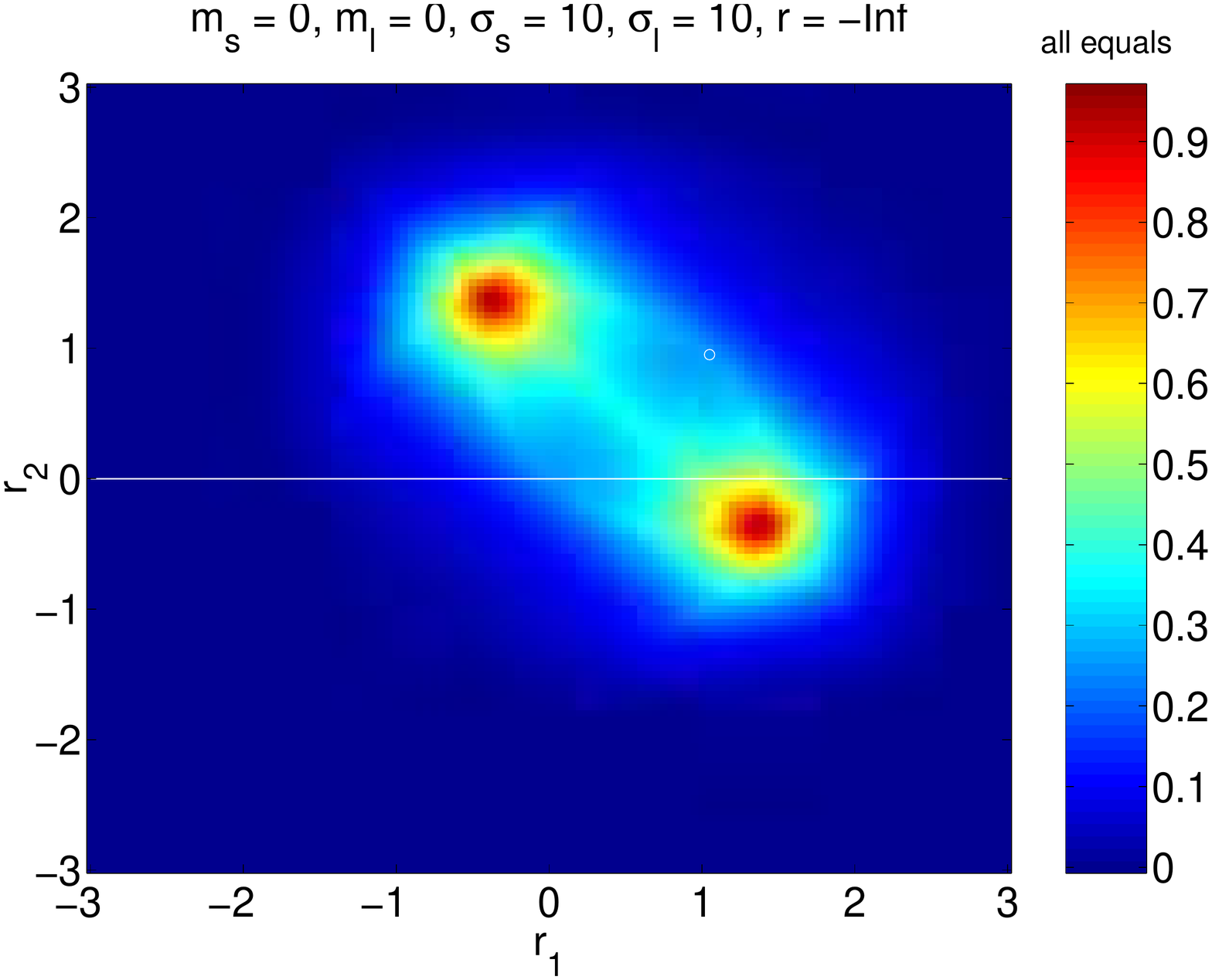}} \\
\end{tabular}
\end{center}
\caption[]{\label{fig:mosaic:equals:crit}
    Comparison  for $\Qerrdiff=\Qerrsky$ top--left, 
$\Qerrdiff=\Qerrload$ top-- right, $\Qerrsky=\Qerrload$ bottom--left and the 
corresponding best fit regions bottom--right;
as a function of ($\GMFO$, $\GMFT$).
 The lines are iso--contours of $\exp(-|x|)$ where $x$ is the difference
between the two arguments to be compared. For the bottom--right frame
the iso--contours are tracked for cubic root of the product of the comparison functions.
The simulation is for $\Tskymean=\Tloadmean=0$, $\sigmasky=\sigmaload=10$~ADU,
$\GMF=1$. No drifts are included.
}
   \end{figure*}
}

\def\FIGmosaicEntropyApprox{
\begin{figure*}[ht]
\begin{center}
\begin{tabular}{ll}
a) & b) \\
 \ROTATEBOX{\includegraphics[height=5.5cm,width=5cm]{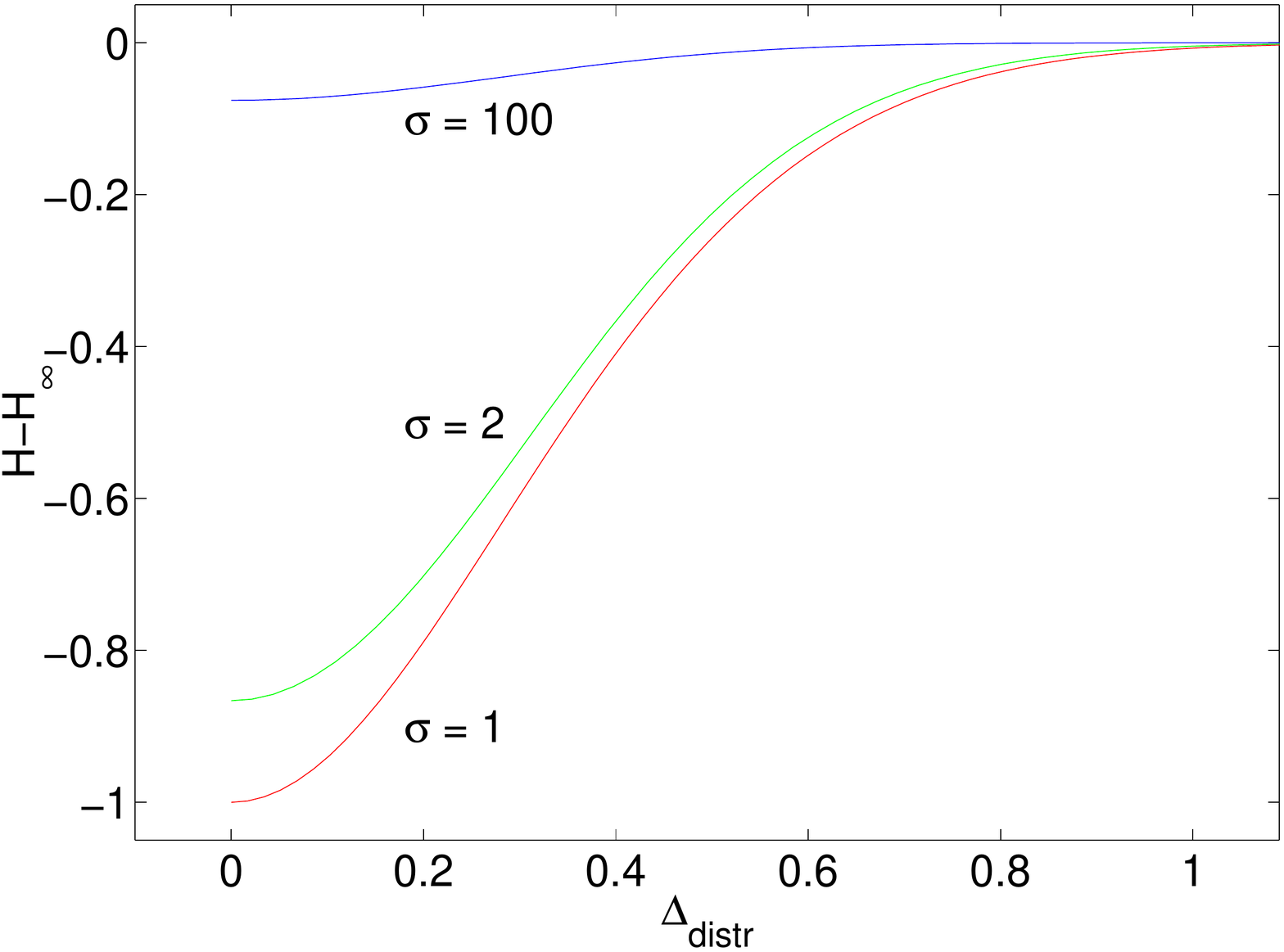}} & 
 \ROTATEBOX{\includegraphics[height=5.5cm,width=5cm]{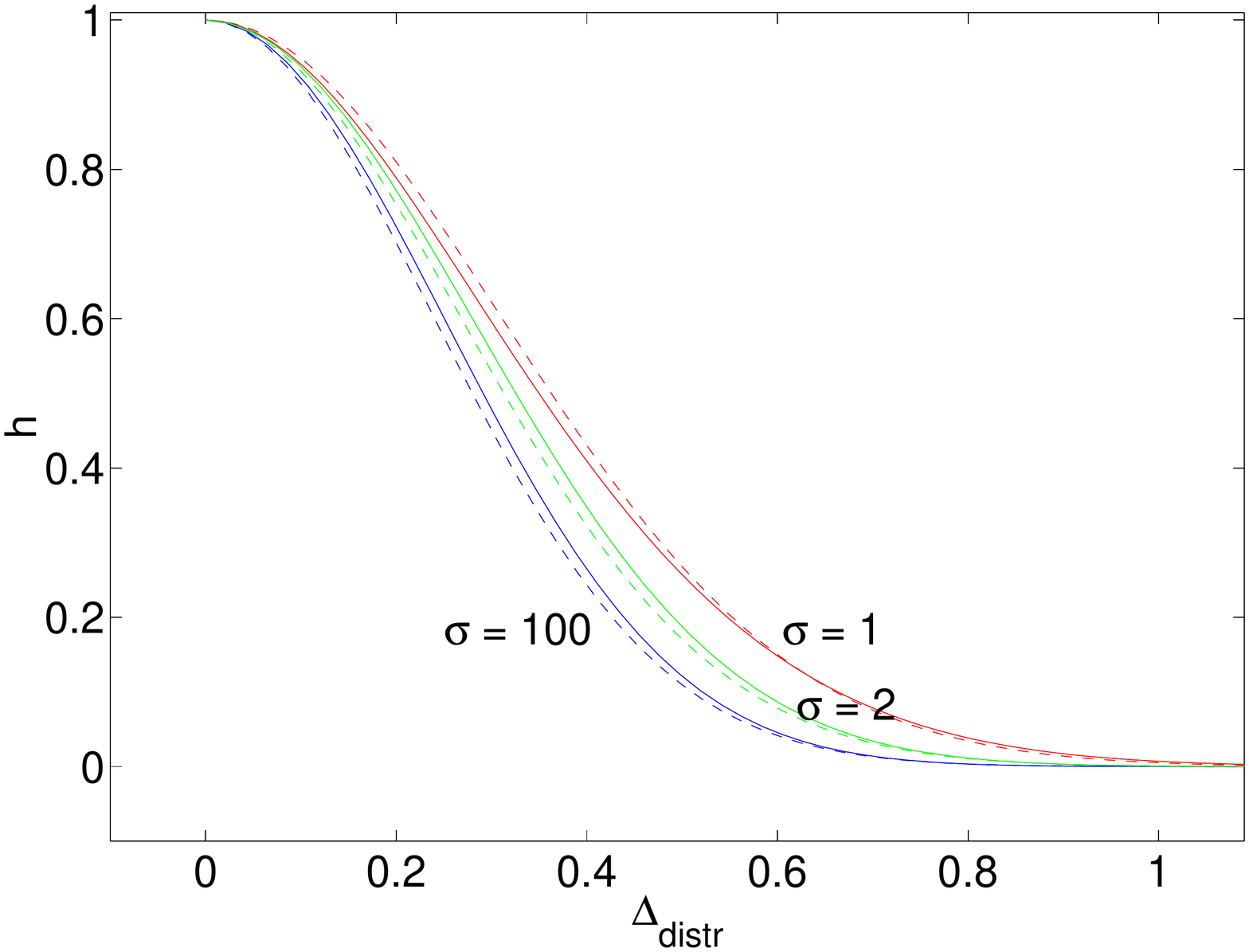}} \\
c) & d) \\
 \ROTATEBOX{\includegraphics[height=5.5cm,width=5cm]{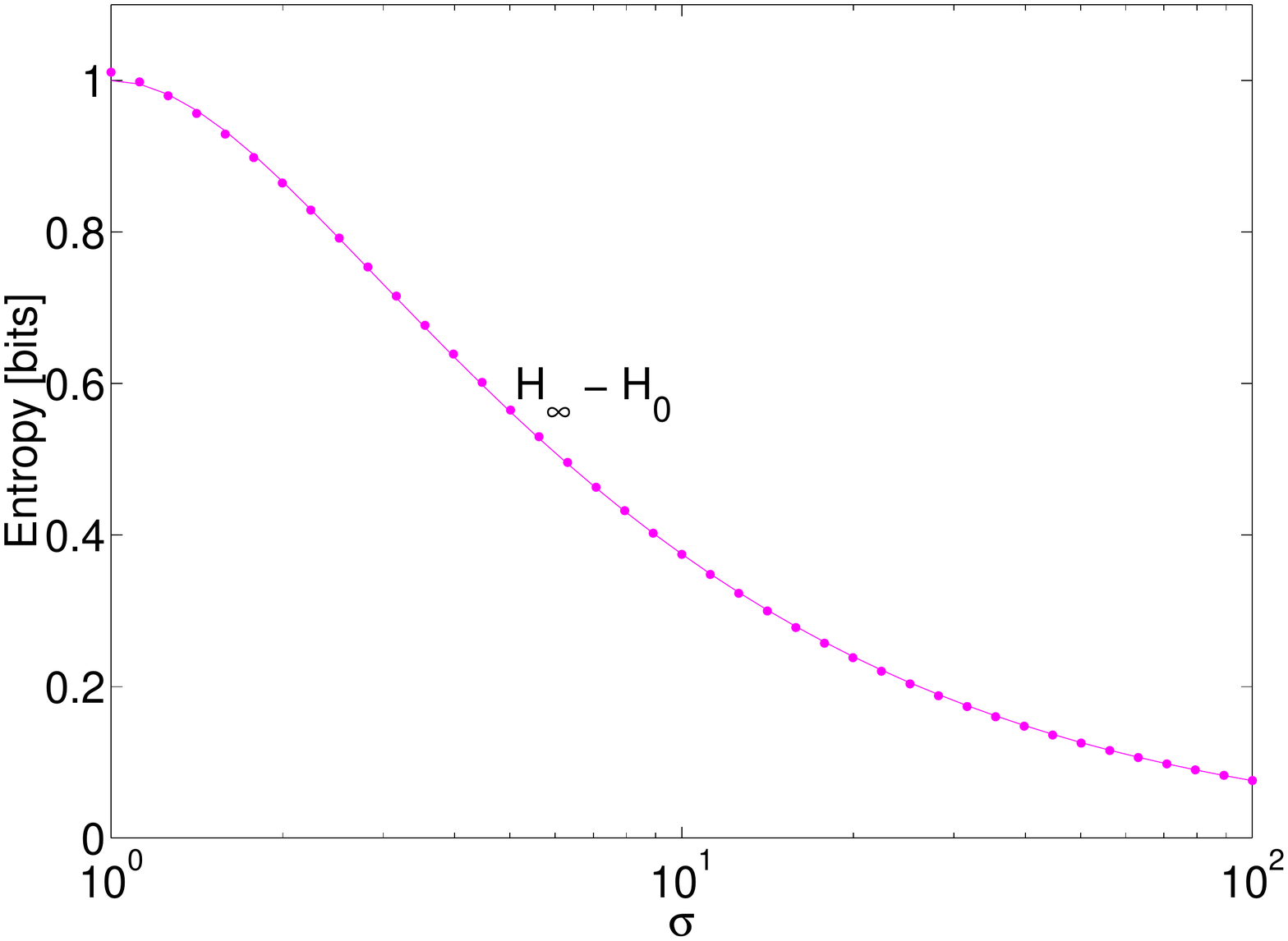}} & 
 \ROTATEBOX{\includegraphics[height=5.5cm,width=5cm]{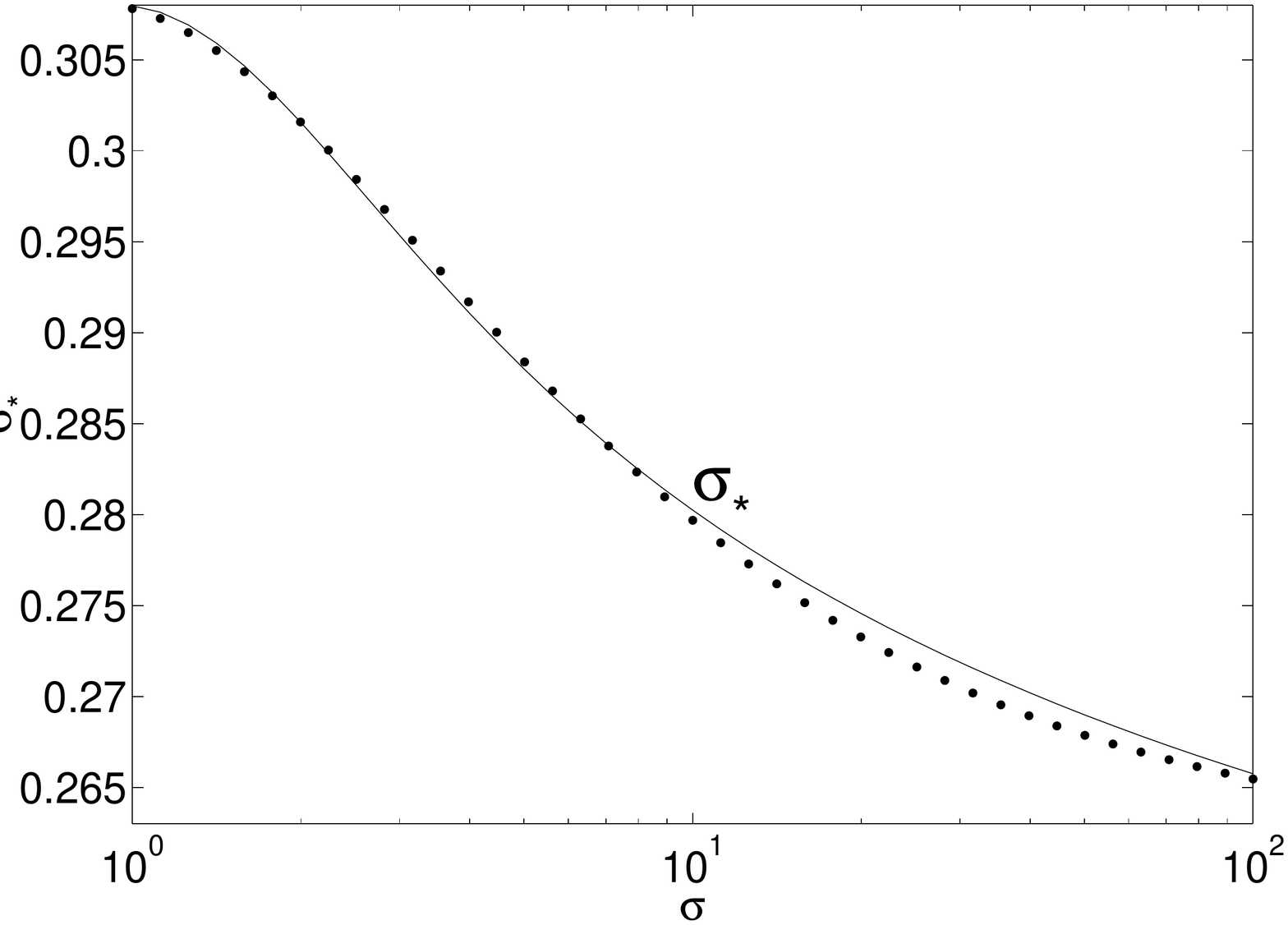}} \\
\end{tabular}
\end{center}
\caption[]{\label{fig:mosaic:entropy:approx}
Frame a: 
Entropy variation for a couple of normally distributed signals as a function  of $\Separation$ 
for three values of $\sigma$. The reference  entropy is 
$\entropyinf$.
Frame b: 
Normalized entropy 
for a couple of normally distributed signals as a function 
of $\Separation$ for three values of $\sigma$. Full lines: numerical integration.
Frame c: 
Normalized entropy difference $\entropyinf - \entropymin$ as a function of $\sigma$
for a couple of normally distributed signals. Full lines: numerical integration.
Dots: approximated formula.
Frame d: 
The $\sigmastar$ parameter as a function of $\sigma$ 
for a couple of normally distributed signals. Full lines: numerical integration.
Dots: approximated formula.
}
   \end{figure*}
}

\def\FIGquackone{ 
\begin{figure*}[ht]
\begin{center}
\begin{tabular}{ll}
a) & b) \\
 \ROTATEBOX{\includegraphics[height=5.5cm,width=5cm]{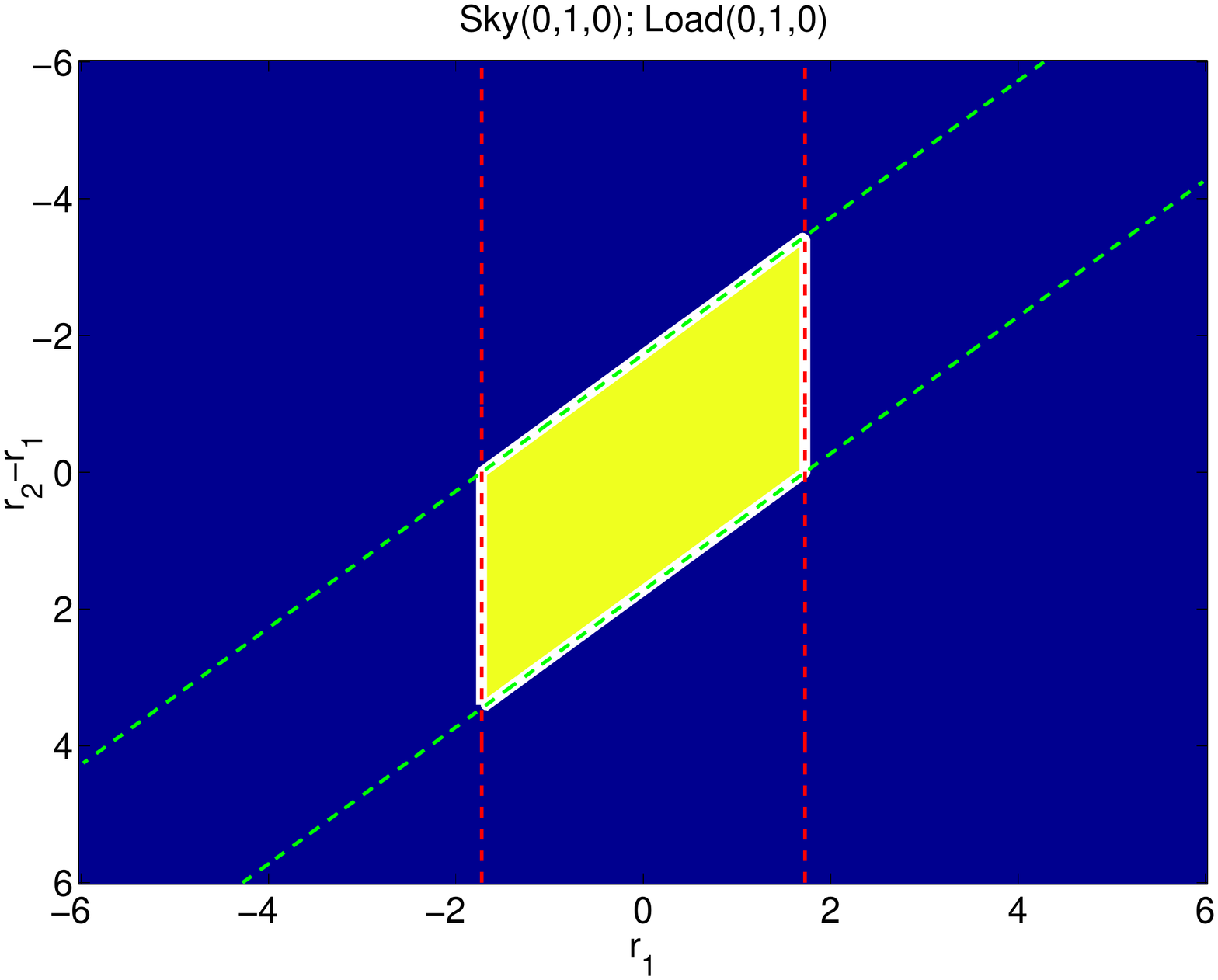}} & 
 \ROTATEBOX{\includegraphics[height=5.5cm,width=5cm]{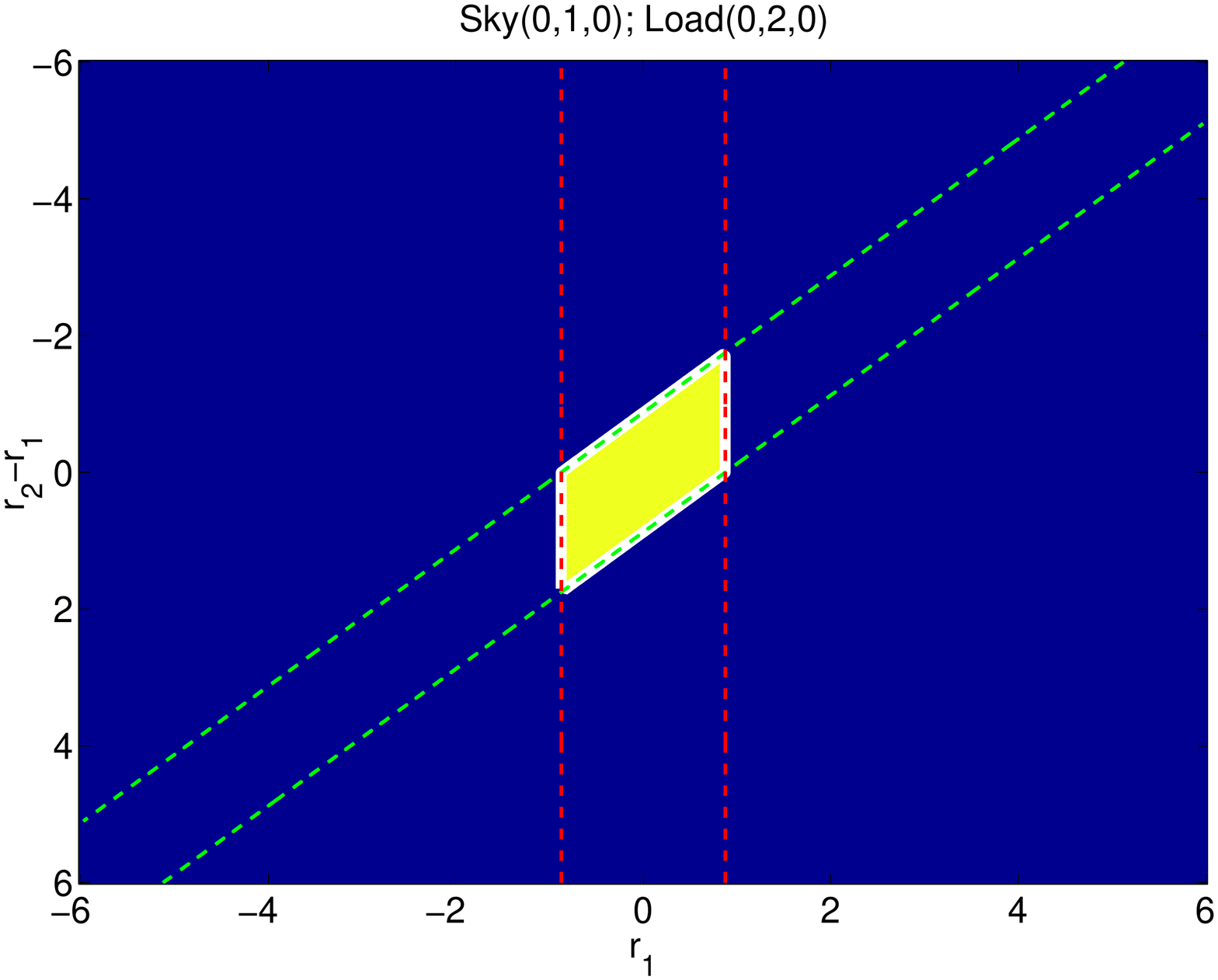}} \\
c) & d) \\
 \ROTATEBOX{\includegraphics[height=5.5cm,width=5cm]{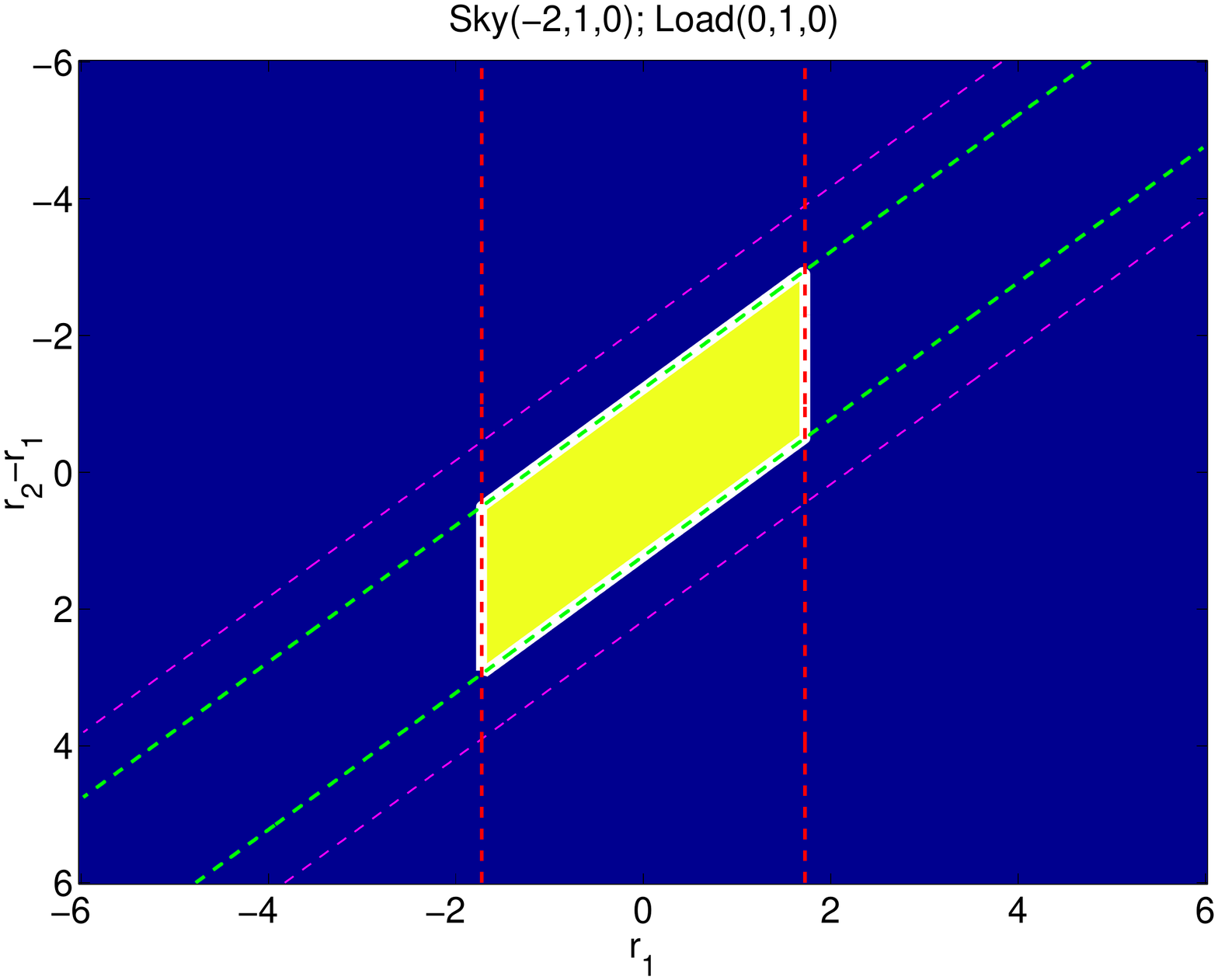}} & 
 \ROTATEBOX{\includegraphics[height=5.5cm,width=5cm]{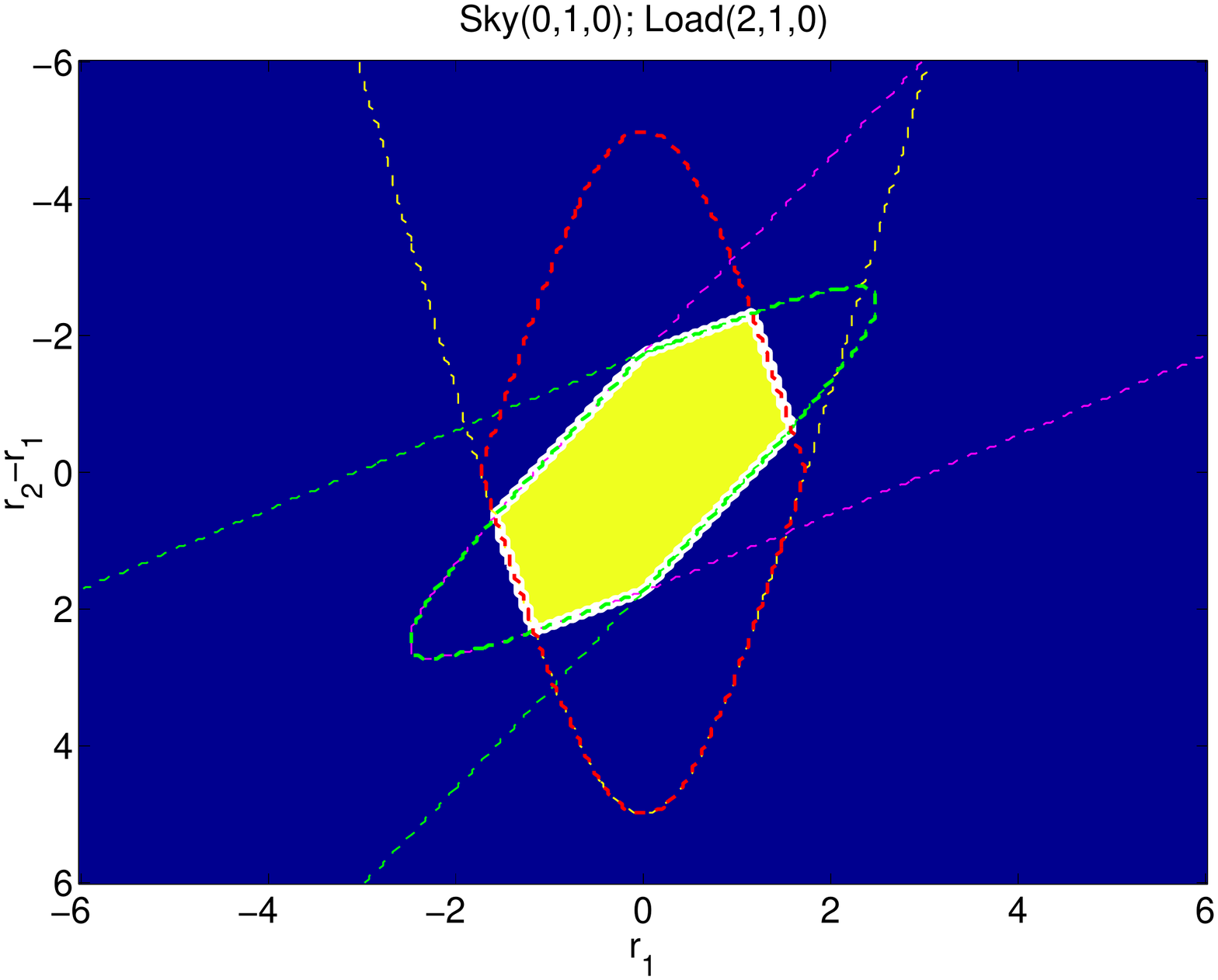}} \\
\end{tabular}
\end{center}
\caption[]{\label{fig:quack:one}
Example of analysis of $\Quack$ factors in the $(\GMFO, \GMFT-\GMFO)$ space.
Various cases for different values of $\Tsky$ and $\Tload$ are considered.
The yellow region is the allowed region when all the $\Quacki<1$, the blue region
is the forbidden one. Thin dashed lines are the limits of allowed
regions $\Quackone$ with positive (red) or negative (yellow) drifts and $\Quacktwo$ for 
positive (green) or negative (violet) drifts.
Thin full lines are the allowed regions for $\Quackone$ and $\Quacktwo$ whose intersection
is marked with a white thick line.
Values of $\Tskymean$, $\sigmasky$, $\Driftsky$ and the corresponding for $\Tload$ are
between parentesis in the title of each frame in the order (mean, sigma, drift).
}
   \end{figure*}
}

\def\FIGquacktwo{ 
\begin{figure*}[ht]
\begin{center}
\begin{tabular}{ll}
a) & b) \\
 \ROTATEBOX{\includegraphics[height=5.5cm,width=5cm]{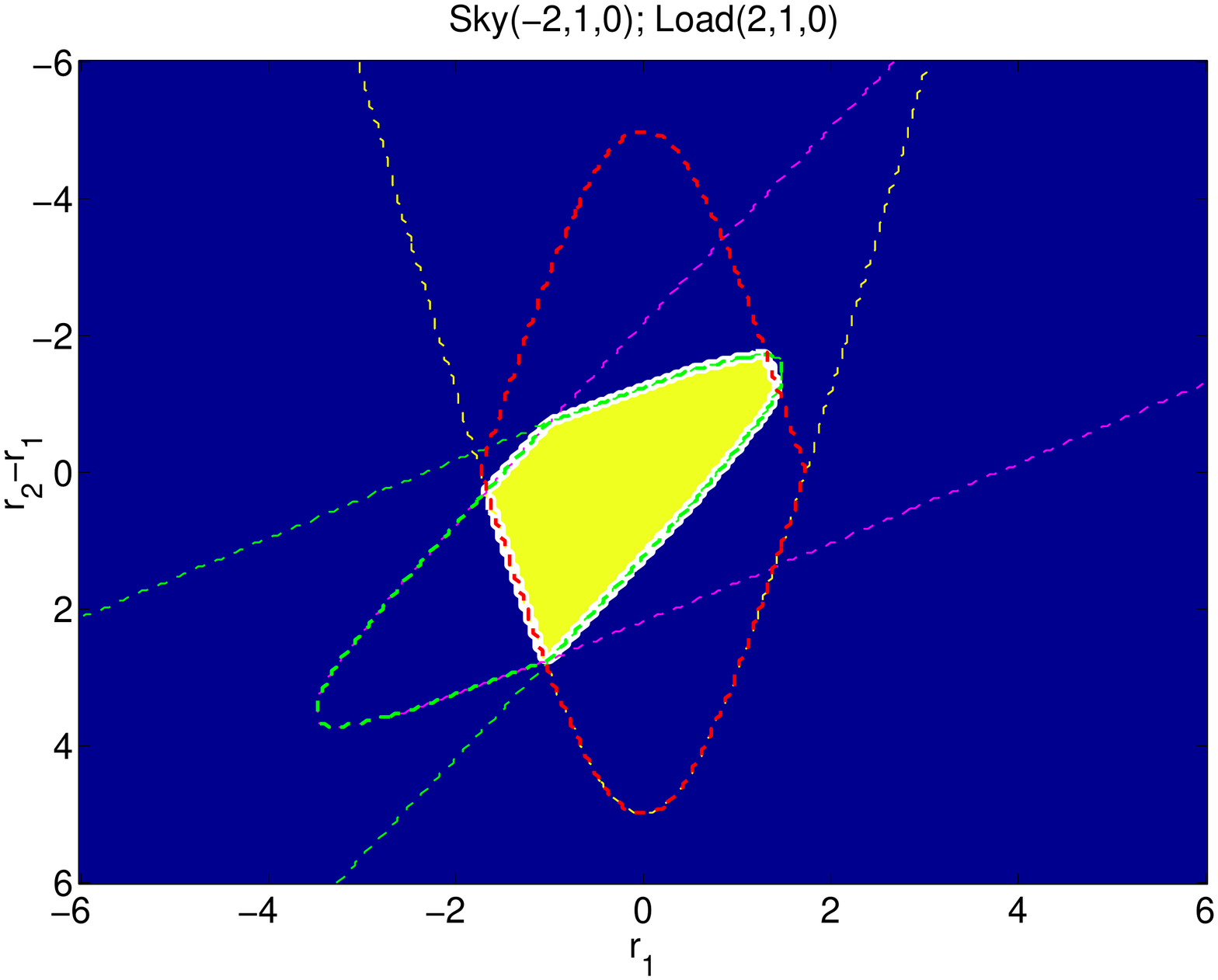}} & 
 \ROTATEBOX{\includegraphics[height=5.5cm,width=5cm]{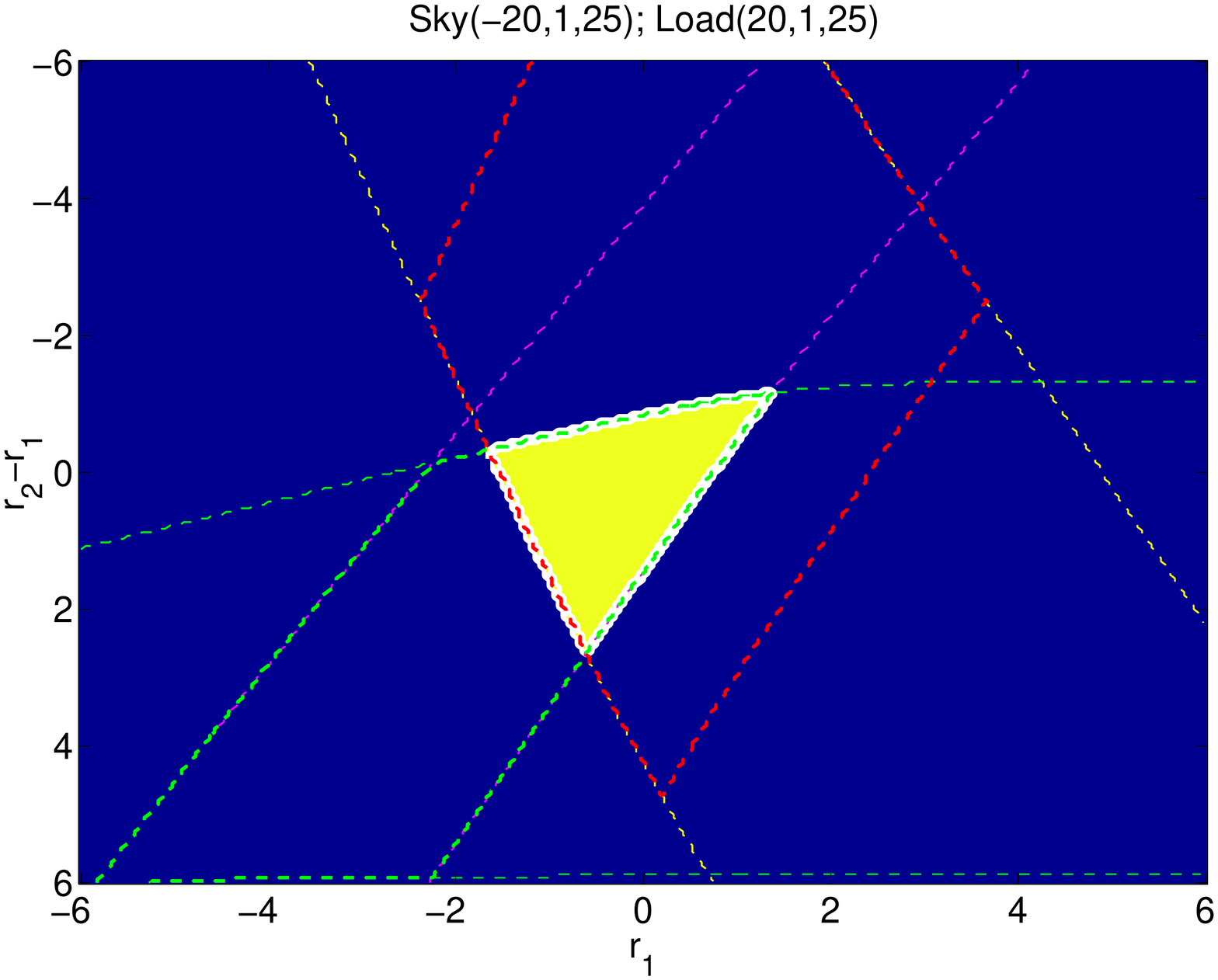}} \\
c) & d) \\
 \ROTATEBOX{\includegraphics[height=5.5cm,width=5cm]{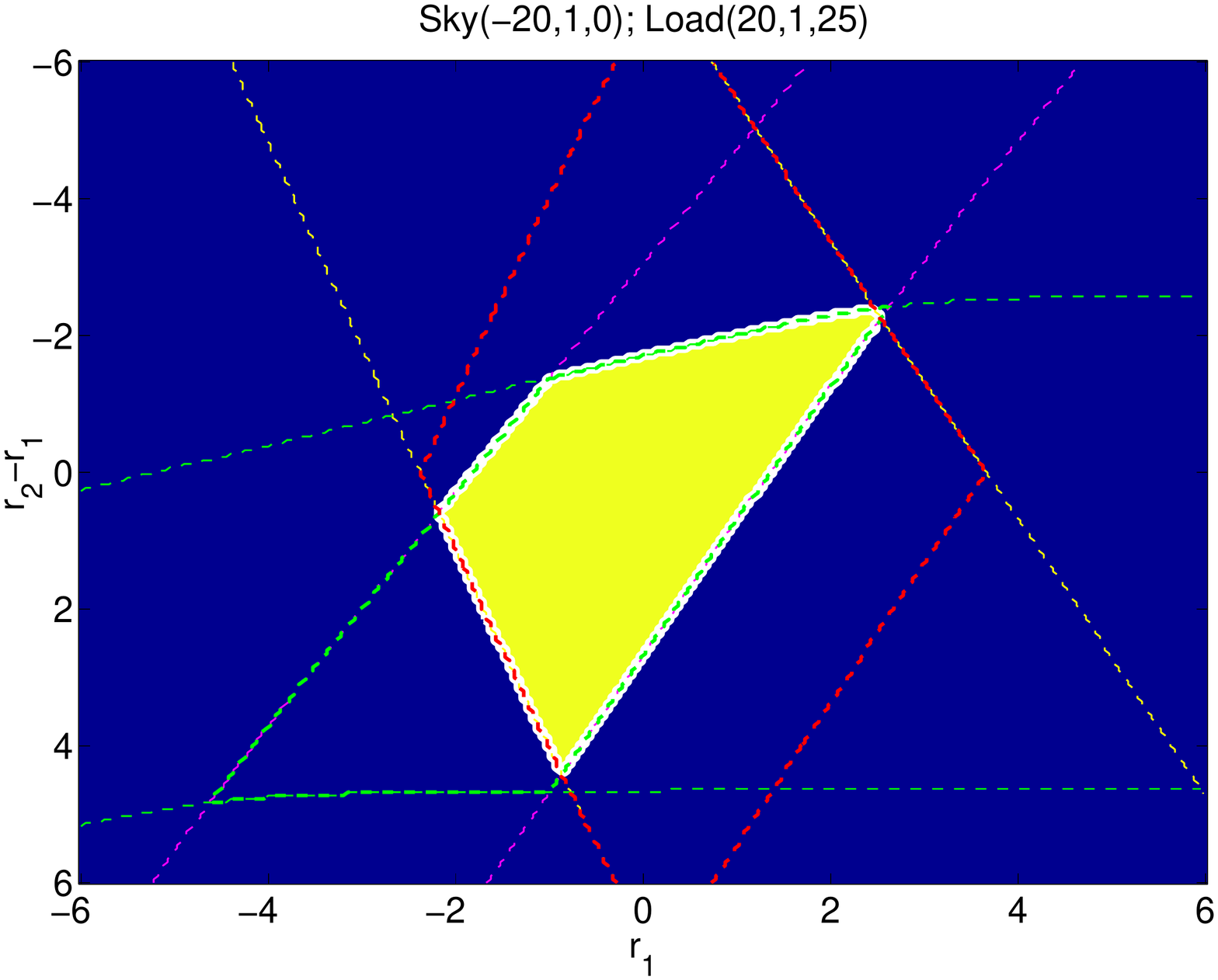}} & 
 \ROTATEBOX{\includegraphics[height=5.5cm,width=5cm]{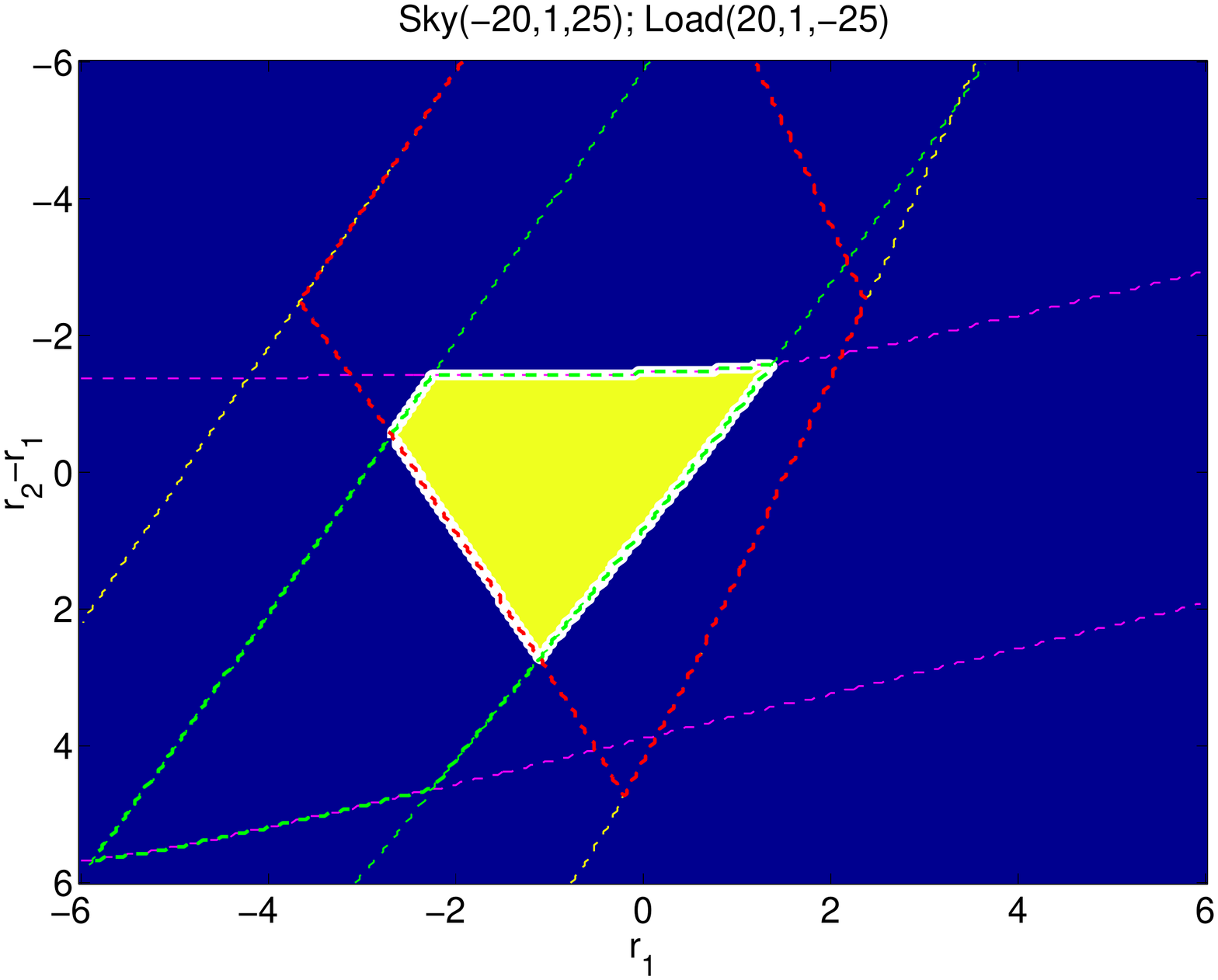}} \\
\end{tabular}
\end{center}
\caption[]{\label{fig:quack:two}
Example of analysis of $\Quack$ factors. 
See caption of Fig.~\ref{fig:quack:one} for explanation of symbols.
}
   \end{figure*}
}

\def\FIGGainExclusion{
\begin{figure}[t]
\begin{center}
\begin{tabular}{c}
 \ROTATEBOX{
 \includegraphics[height=8cm,width=8cm]{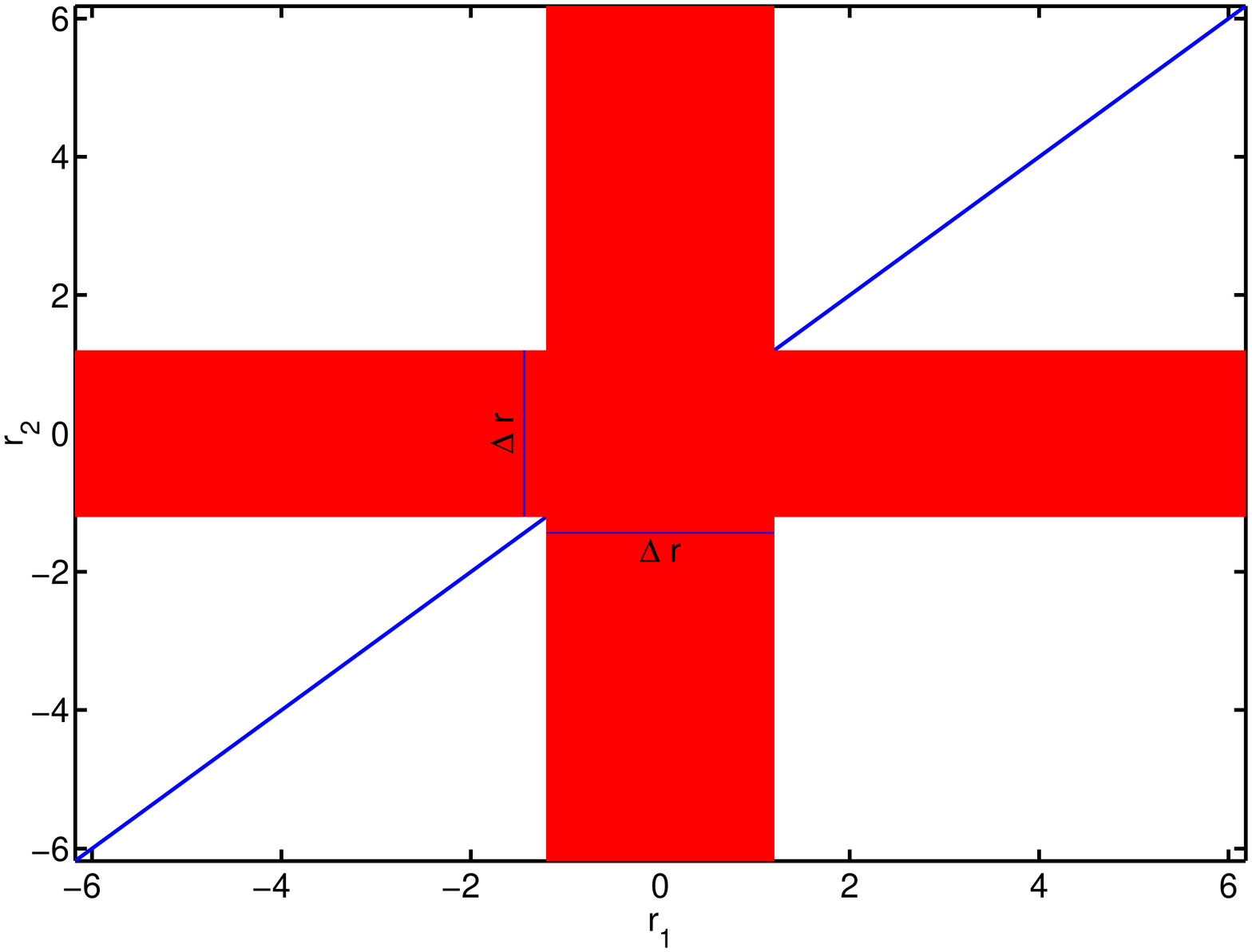} 
 }
\end{tabular}
\end{center}
\caption[]{\label{fig:gain:excl}
The region of the $\GMFO$, $\GMFT$ space excluded by the condition
$\sigmaone, \sigmatwo > \sigmatgt$, red.
The width of the two crossing bands, $\Delta r$, is given by 
Eq.~(\ref{eq:DAE:Exclusion}). In this case $\fDAE \approx 0.35$.
}
   \end{figure}
}

\def\FIGschematic{ 
\begin{figure}[t]
\begin{center}
\begin{tabular}{c}
 \ROTATEBOX{
 \includegraphics[height=10cm]{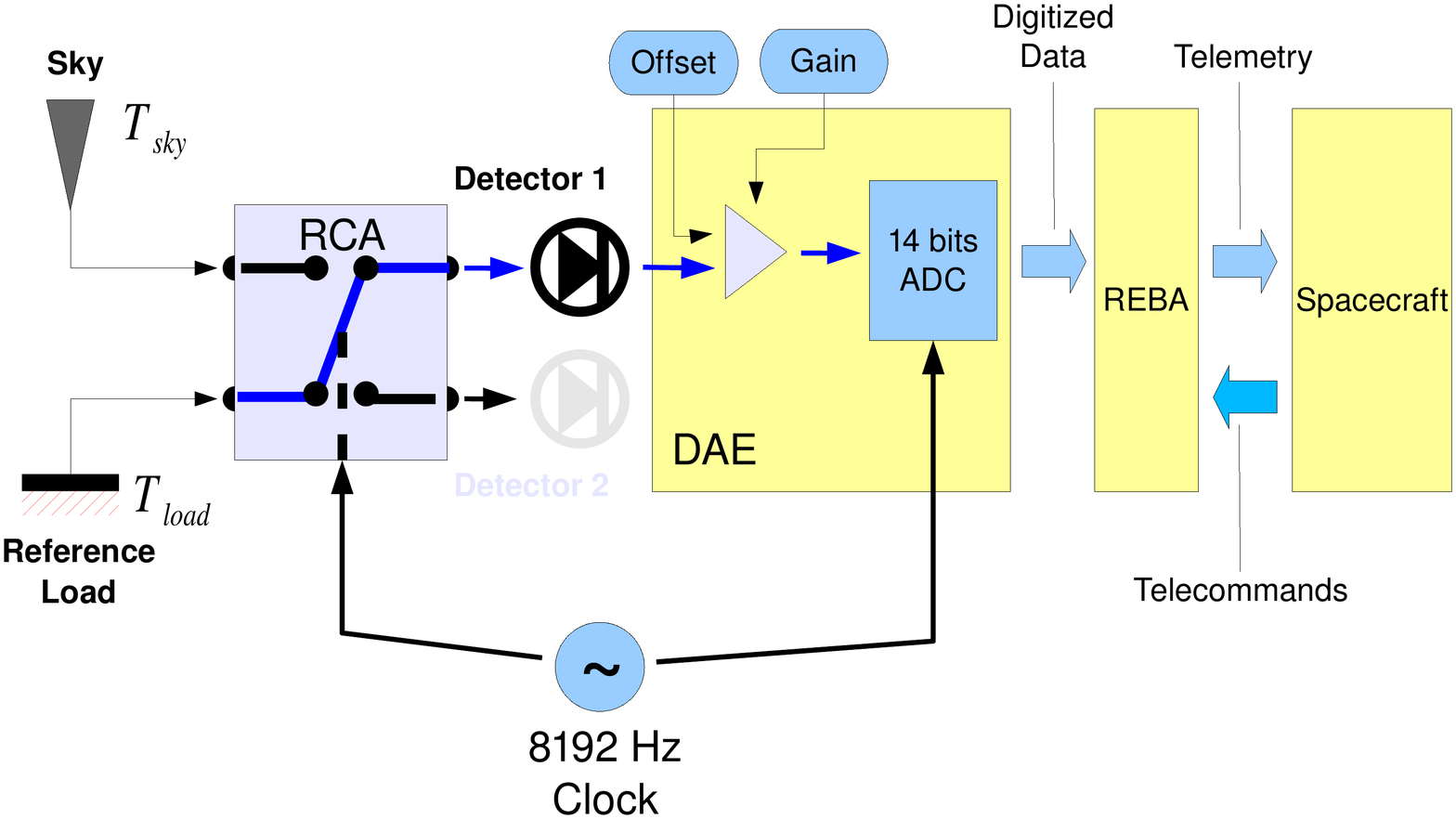}
 }
 \vspace{-2cm}
\end{tabular}
\end{center}
\caption{
\label{fig:schematic}
A schematic view of the main flow of scientific data for a single RCA of \Planck/LFI.
Each RCA has two detectors, but in this scheme only the first is represented
and schematized. For graphical purposes the scheme represents just 
the first detector while connected to the \load, while
Detector 2 would be connected to \sky. At a change of the Clock phase the two detectors will switch
their connections. The block arrows represents the flow of digitized data and telemetry toward the spacecraft and 
the flow of telecommands from the spacecraft.
}
\end{figure}
}

\def\FIGprocessing{ 
\begin{figure}[t]
\begin{center}
\begin{tabular}{c}
 \ROTATEBOX{
 \includegraphics[height=10cm]{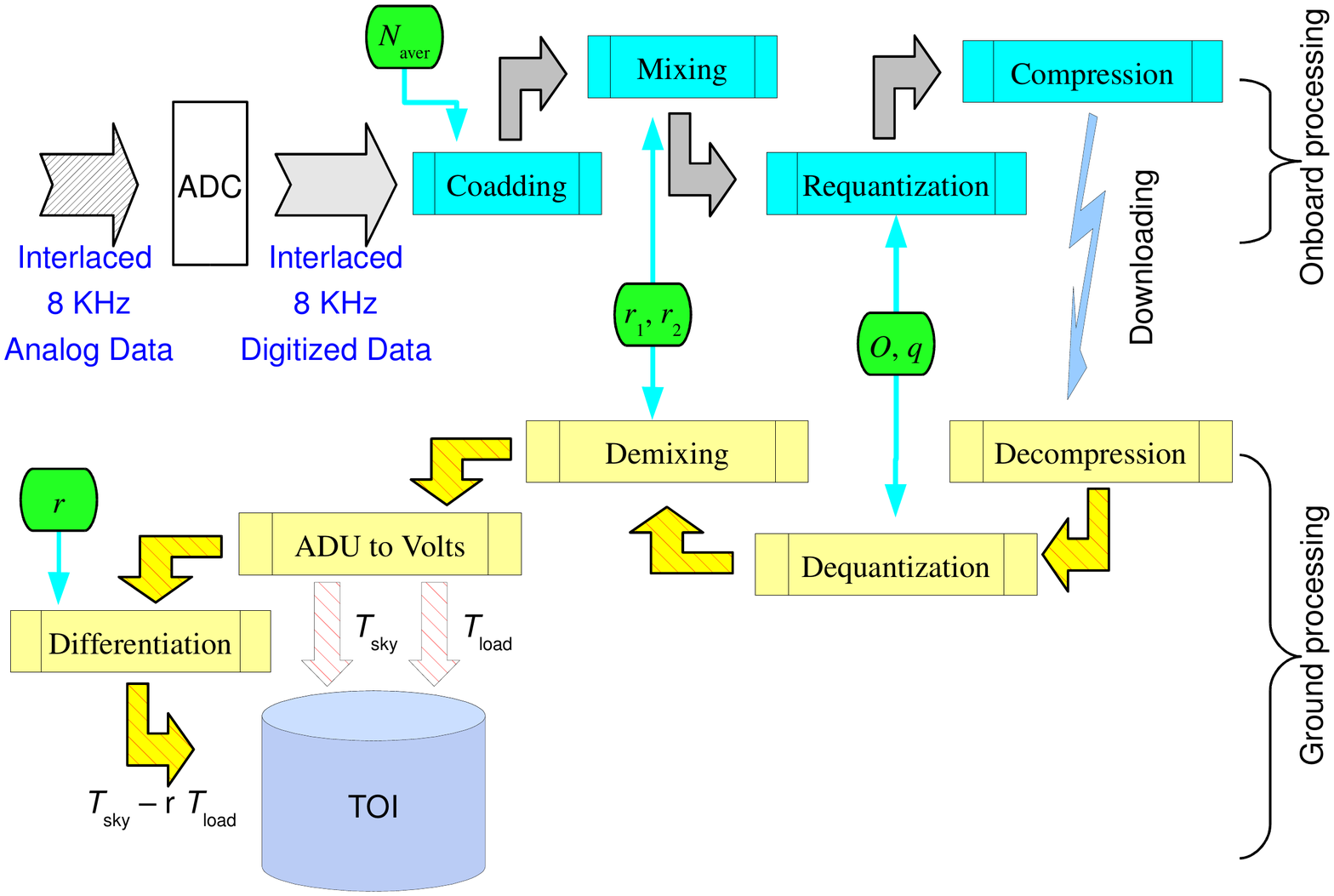}
 }
\end{tabular}
\end{center}
\caption{
\label{fig:processing}
Schematic representation of the scientific onboard and ground processing for the \Planck/LFI.
Cyan boxes represent REBA operations, yellow boxes ground operations. Green pads specify the 
parameters needed by each operation. TOI could be produced both in undifferentiated form 
($\Tsky$, $\Tload$ stored separately) or in differentiated form.
}
\end{figure}
}

\def\FIGtests{
\begin{figure*}[t]
\begin{center}
\begin{tabular}{lll}
a) & b) & c) \\
\rotatebox{90}{\includegraphics[height=5.5cm,width=5cm]{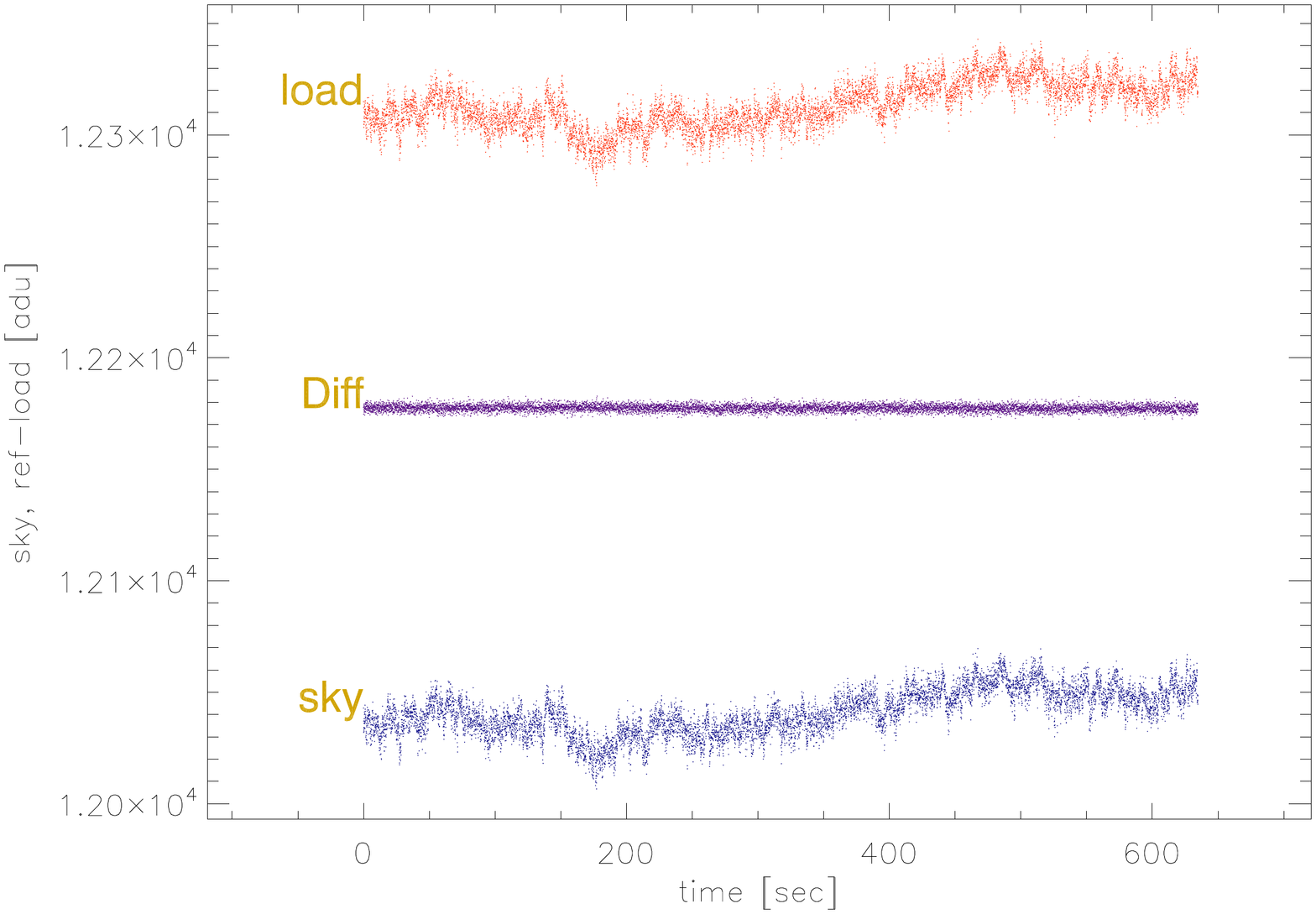}} & \rotatebox{90}{\includegraphics[height=5.5cm,width=5cm]{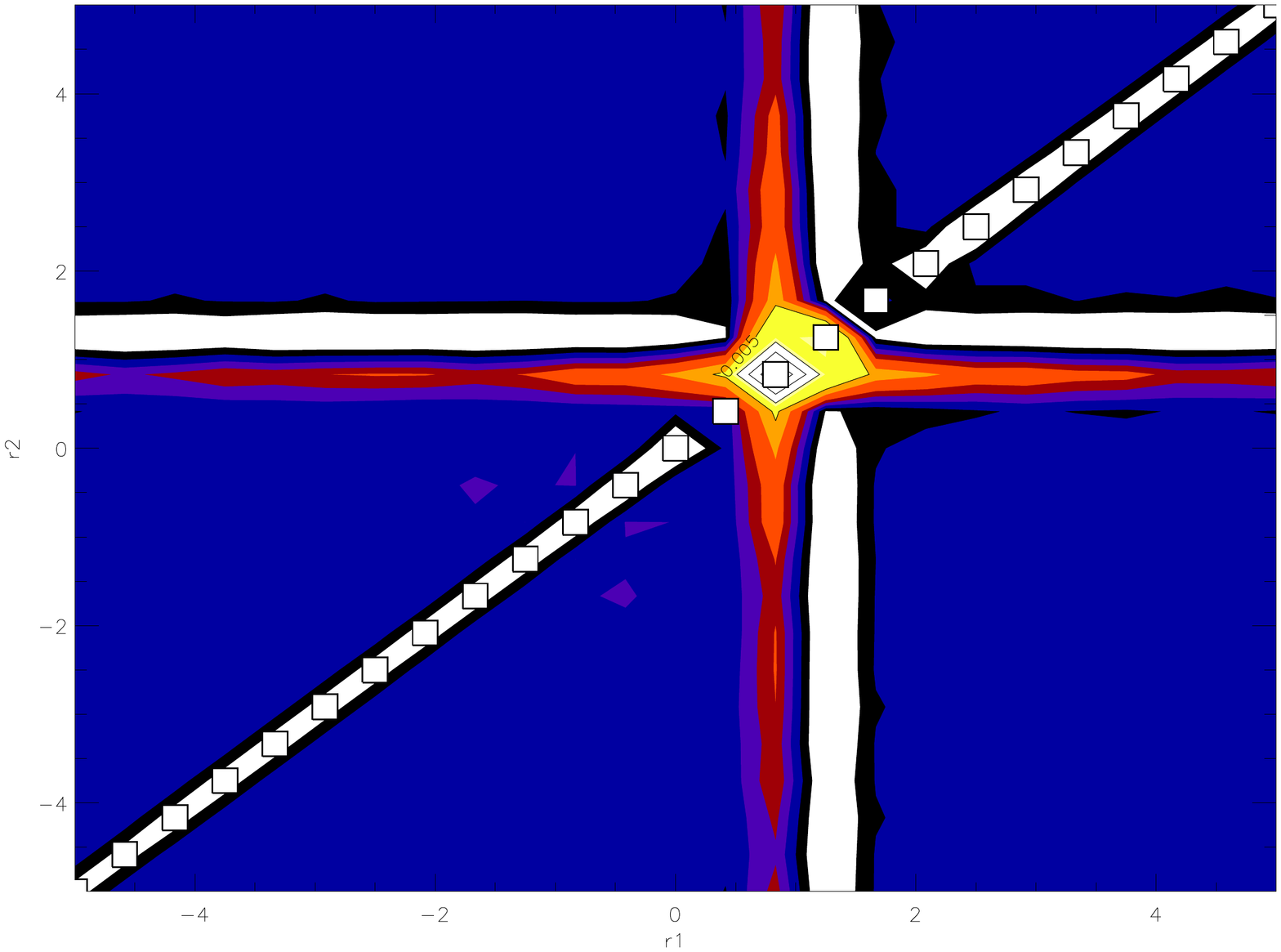}} & \rotatebox{90}{\includegraphics[height=5.5cm,width=5cm]{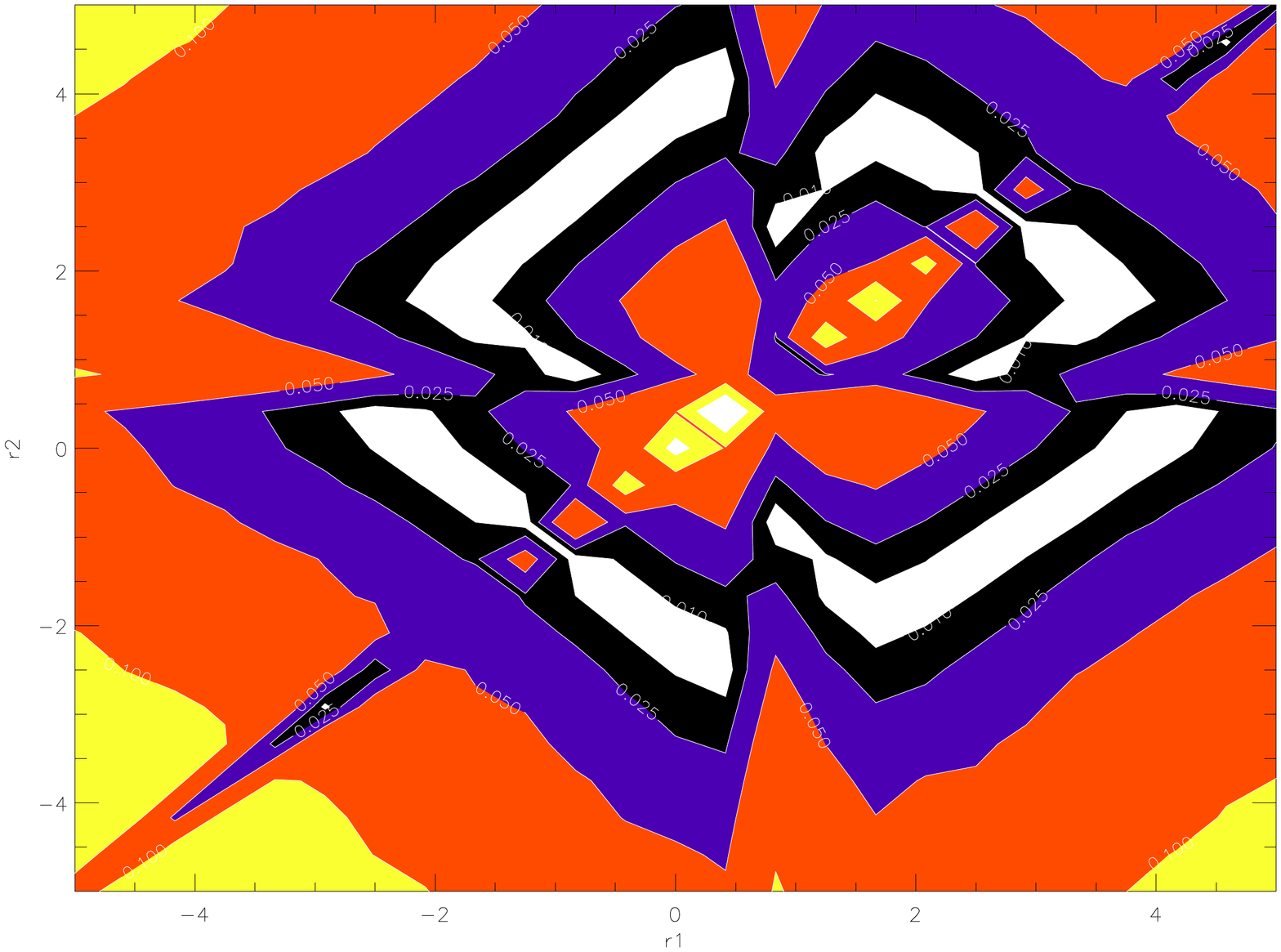}} \\
d) & e) & f) \\
\rotatebox{90}{\includegraphics[height=5.5cm,width=5cm]{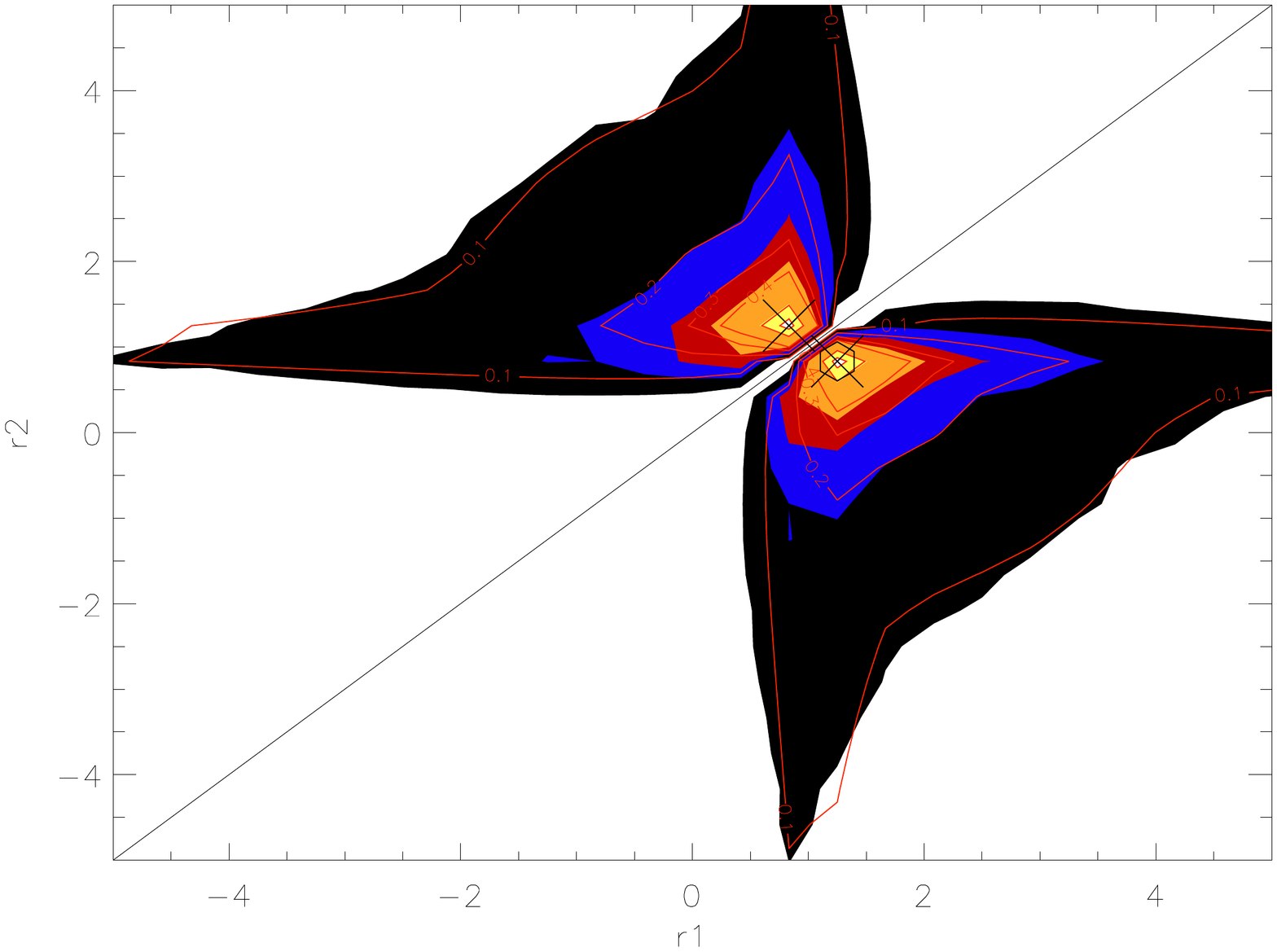}} & \rotatebox{90}{\includegraphics[height=5.5cm,width=5cm]{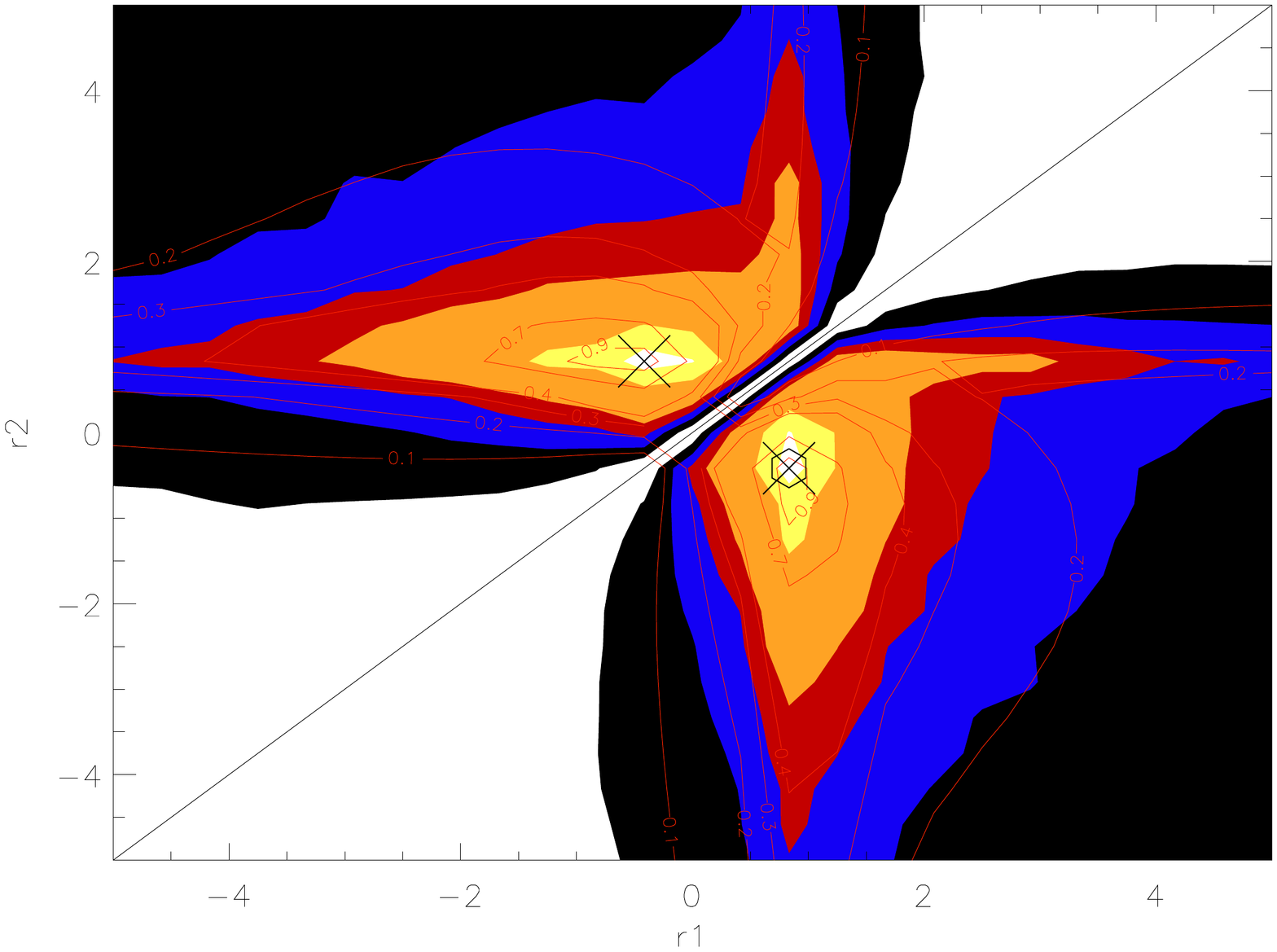}}& \rotatebox{90}{\includegraphics[height=5.5cm,width=5cm]{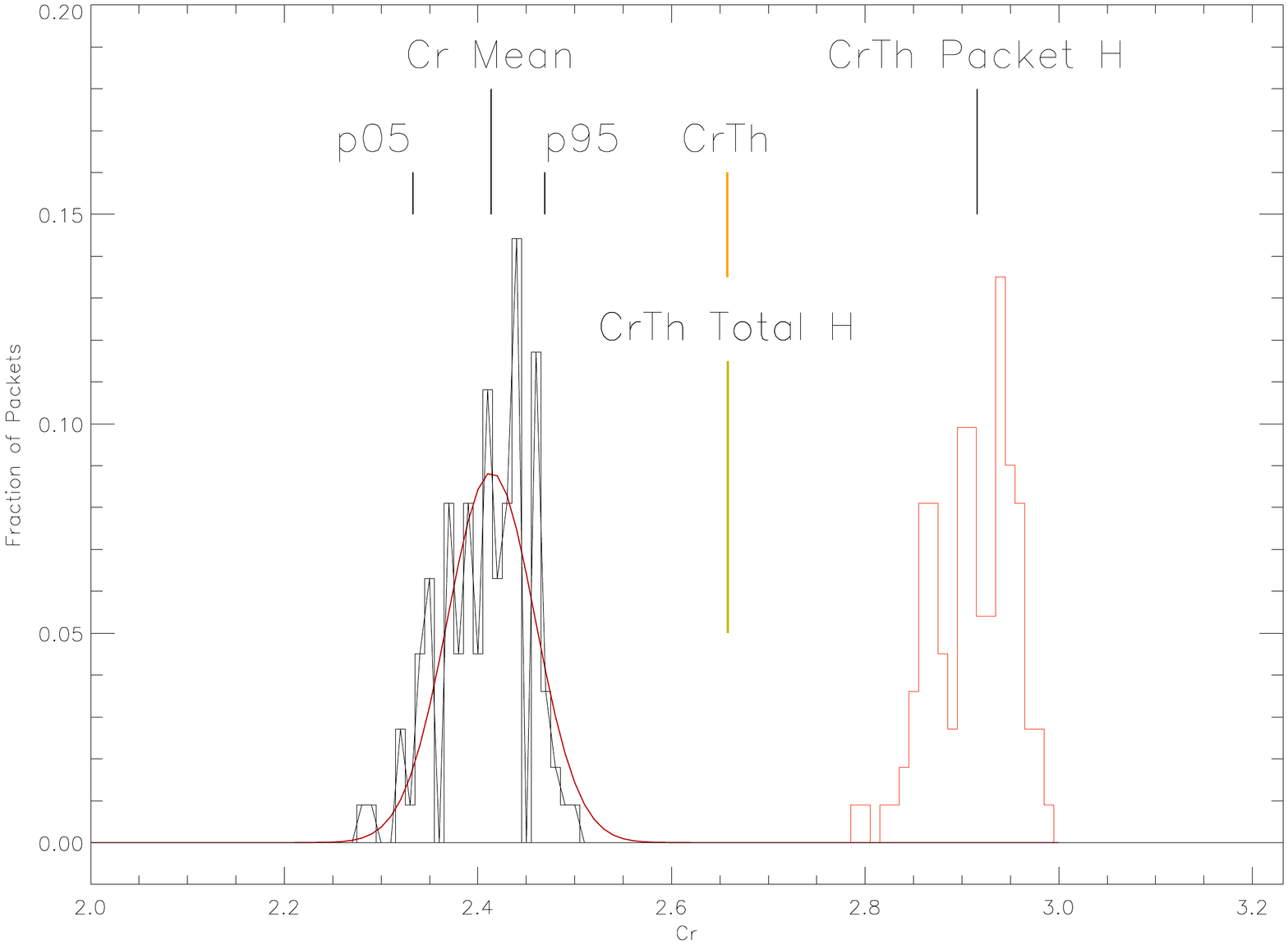}} \\
\end{tabular}
\end{center}
\caption[]{\label{fig:tests} 
Example of signal for the detector 2300 acquired during the tests and its optimization.
Frame a: are the \sky\ and \load\ and $\DeltaT$ samples in analog--to--digital units (ADU).
The differentiated signal has been shifted up, having null mean. Relevant statistics are 
reported in Tab.~\ref{tab:tests}a.
Frame b: 
represents the accuracy by which the analytical model is able to predict the entropy 
measured on the quantized signal as a function of $\GMF1$ and $\GMFT$ for $q=1$.
Contours of regions are for accuracies of 0.5\%, 1\%, 1.5\%, 2\%, 2.5\%. 
White boxes denote the regions where an accuracy worst than 3\% is obtained by the
simplest analytical model. Outside these regions the two models are completelly equivalent.
Frame c: 
represents the accuracy by which the most sophisticated analytical model presented in 
Appendix~\ref{app:normalized:entropy} is able to predict the $\Cr$ for $q=1$.
Contour lines are for accuracies of 5\%, 10\%, 15\%, 20\%, 25\%, and 30\%.
Frame d: 
compares $\Gammadiff$ for the analytical model with $\Gammadiff$ computed for the numerical model.
Colours are for $\Gammadiff$ analytical equal to 0.1, 0.2, 0.3, 0.4, 0.7, and 0.9 while contours are
for the same values of $\Gammadiff$ numerical. The black hexagon identifies the regions for the peaks 
of $\Gammadiff$ analytical, the crosses for the peaks of $\Gammadiff$ numerical.
Frame e: the same as of Frame d, but for $\Gammasky$.
Frame f: in black on the left, the histogram of $\Cr$ obtained on the 111 packets produced by the compression 
procedure with the optimized parameters of Tab.~\ref{tab:tests}b. The figure gives also the 5\% and 95\% 
percentiles (p05 and p95) and the mean, and in red a gaussian fit of the histogram. In addition the figure
shows in red at the right the $\Cr$ expected from the entropy directly measured on each packet, and the 
theoretical compression rate expected from the theroretical model (CrTh) and from the entropy measured on the 
whole quantized TOI.
 }
   \end{figure*}
}

\def\FIGcsl{ 
\begin{figure*}[h]
\begin{center}
\begin{tabular}{c}
 \includegraphics[height=20cm]{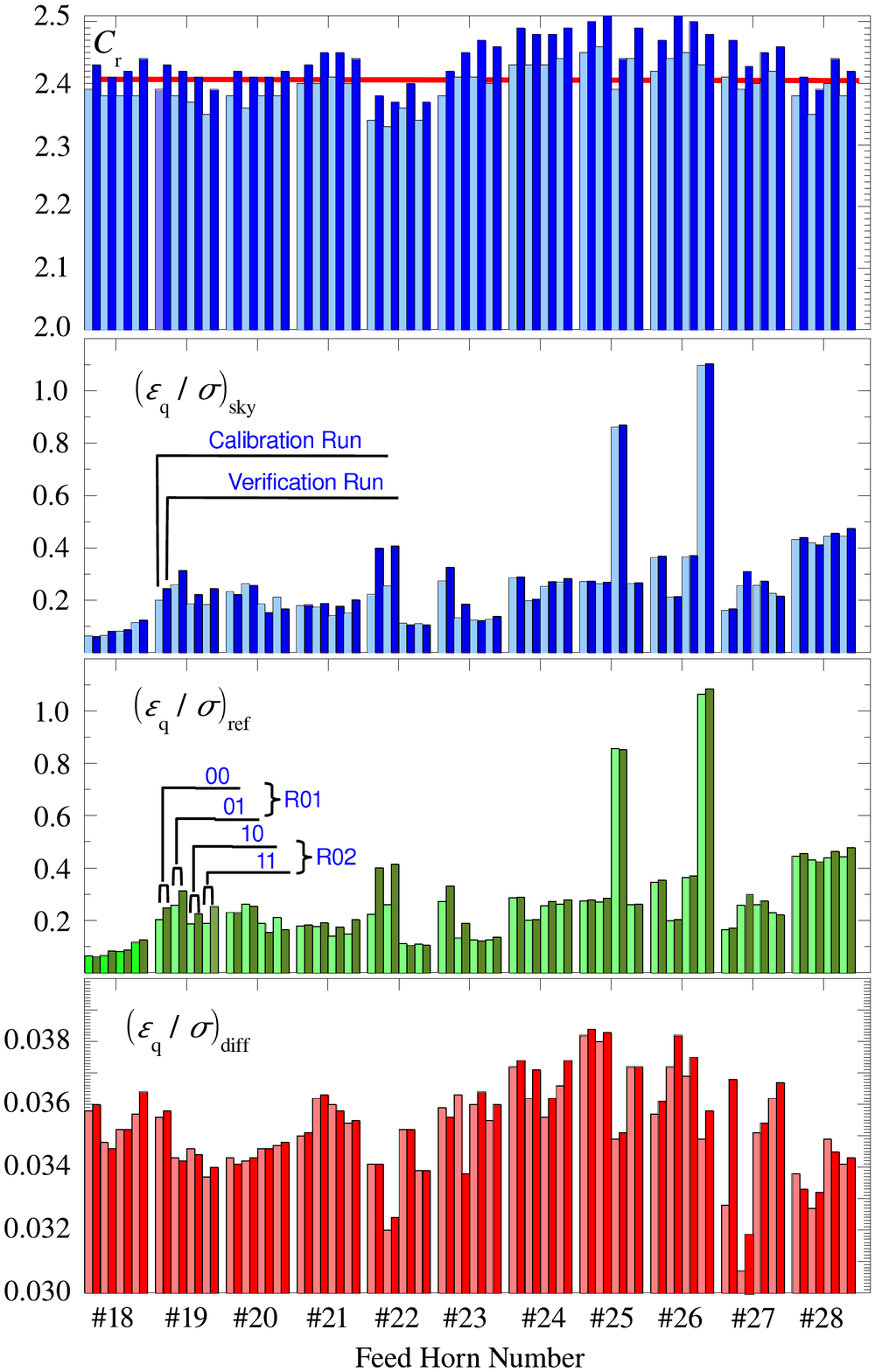}
\end{tabular}
\end{center}
 \vspace{2\baselineskip}
\caption{
\label{fig:csl}
Results for a typical session of REBA parameters tuning during the CSL test campaign.
From top to bottom the figure reports for each detector the mean $\Cr$, $\Qerrsky/\sigmasky$,
$\Qerrload/\sigmaload$, $\Qerrdiff/\sigma_{\mathrm{diff}}$ where $\sigma_{\mathrm{diff}}$ is the 
r.m.s. of the differentiated data. The red line in background of the top frame denotes the target
$\CrTgt = 2.4$. Values are represented by bars. Light--bars are the results from the calibration phase,
where raw data from the instrument are processed by \OCATWO.
Dark--bars are results from the verification phase, where processing is performed \onboard. 
The second frame from Top gives an example for detector 00 of Feed--Horn \#19.
Feed--horns are numbered according to the internal \Planck/LFI convetion 
assigning at \Planck/LFI the Feed--Horns numbers from \#18 to \#28. 
Detectors belonging to the same Feed--Horn are grouped together as shown in the third frame from top.
}
\end{figure*}
}

\def\FIGschematic{ 
\begin{figure}[t]
\begin{center}
\begin{tabular}{c}
 \ROTATEBOX{
 \includegraphics[height=10cm]{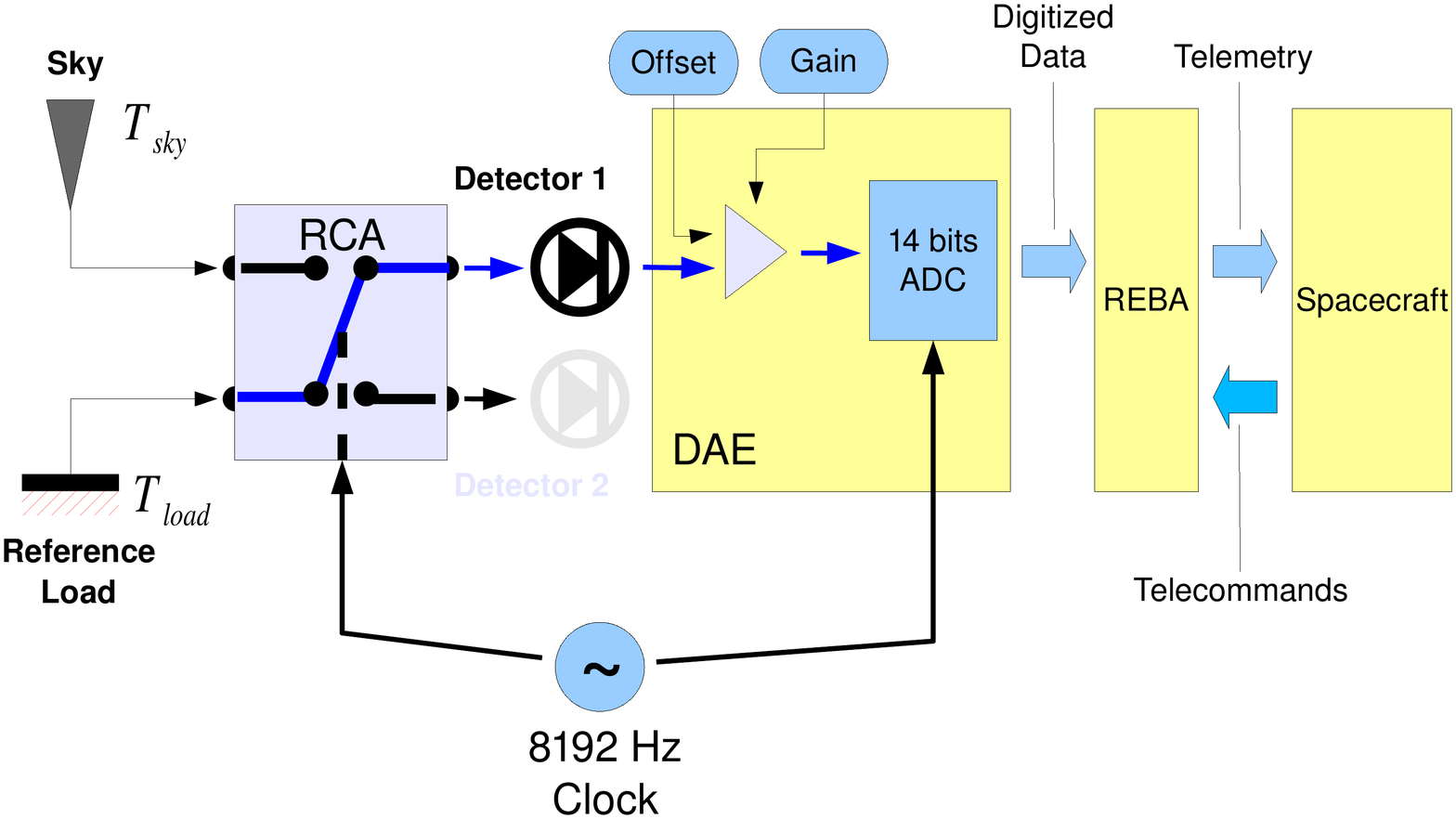}
 }
 \vspace{-2cm}
\end{tabular}
\end{center}
\caption{
\label{fig:schematic}
A schematic view of the main flow of scientific data for a single RCA of \Planck/LFI.
Each RCA has two detectors, but in this scheme only the first is represented
and schematized. For graphical purposes the scheme represents just 
the first detector while connected to the \load, while
Detector 2 would be connected to \sky. At a change of the Clock phase the two detectors will switch
their connections. The block arrows represents the flow of digitized data and telemetry toward the spacecraft and 
the flow of telecommands from the spacecraft.
}
\end{figure}
}

\def\FIGqoptimal{ 
\begin{figure}[t]
\begin{center}
\begin{tabular}{ccc}
 \hspace{-1.5cm}
 \rotatebox{90}{
 \includegraphics[width=5cm]{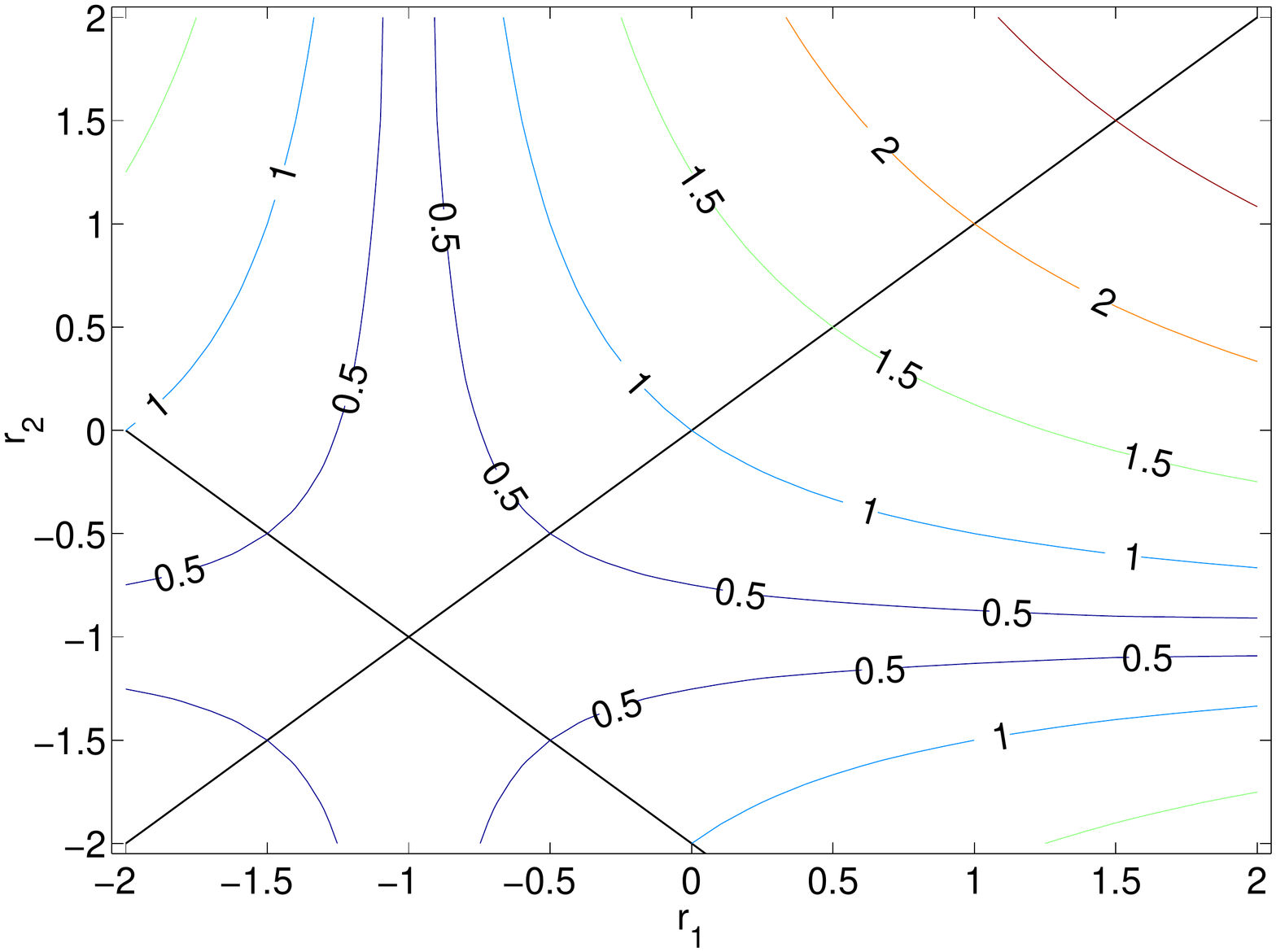}
 }&
 \hspace{-1cm}
 \rotatebox{90}{
 \includegraphics[width=5cm]{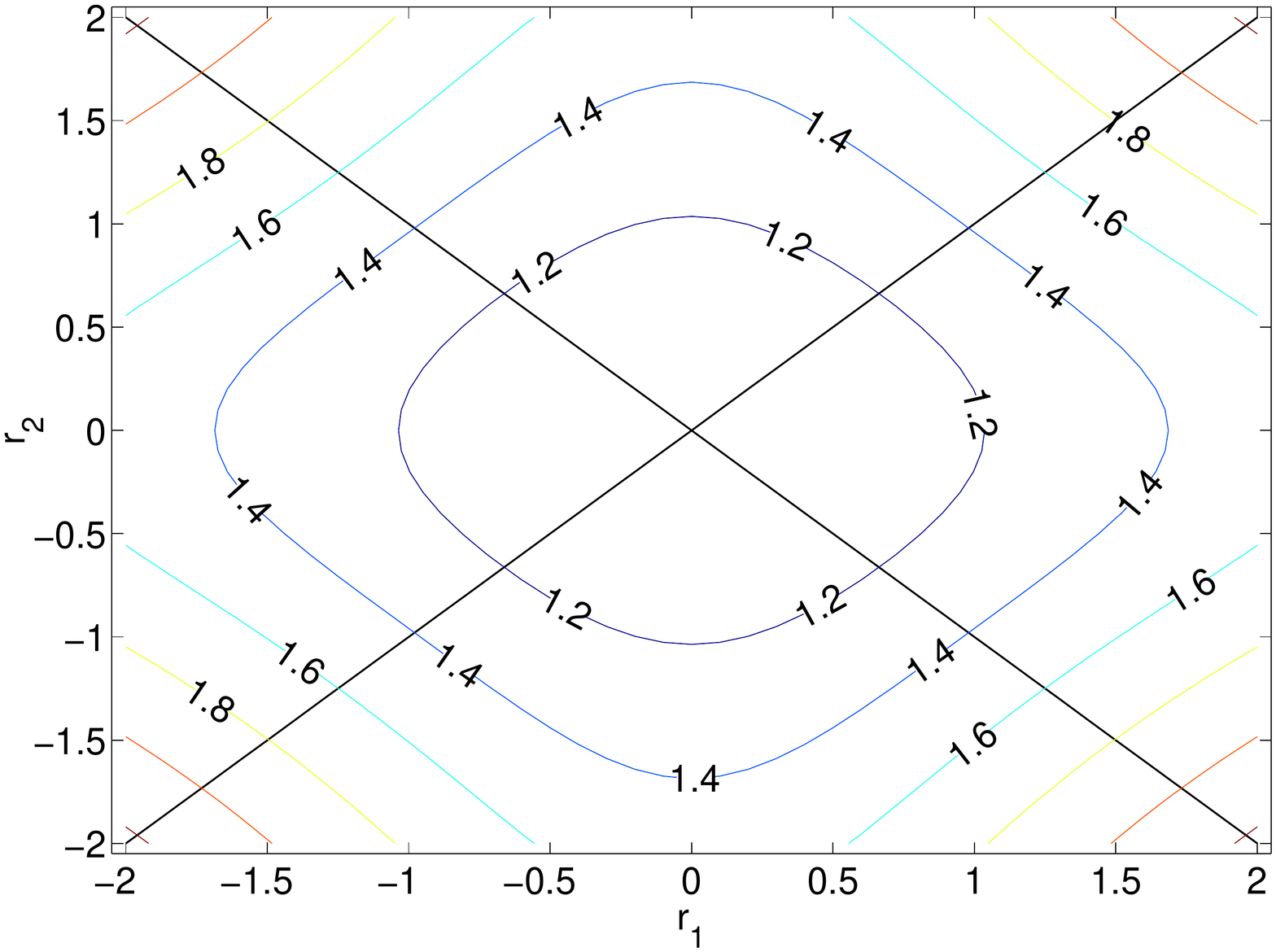}
 }&
 \hspace{-1cm}
 \rotatebox{90}{
 \includegraphics[width=5cm]{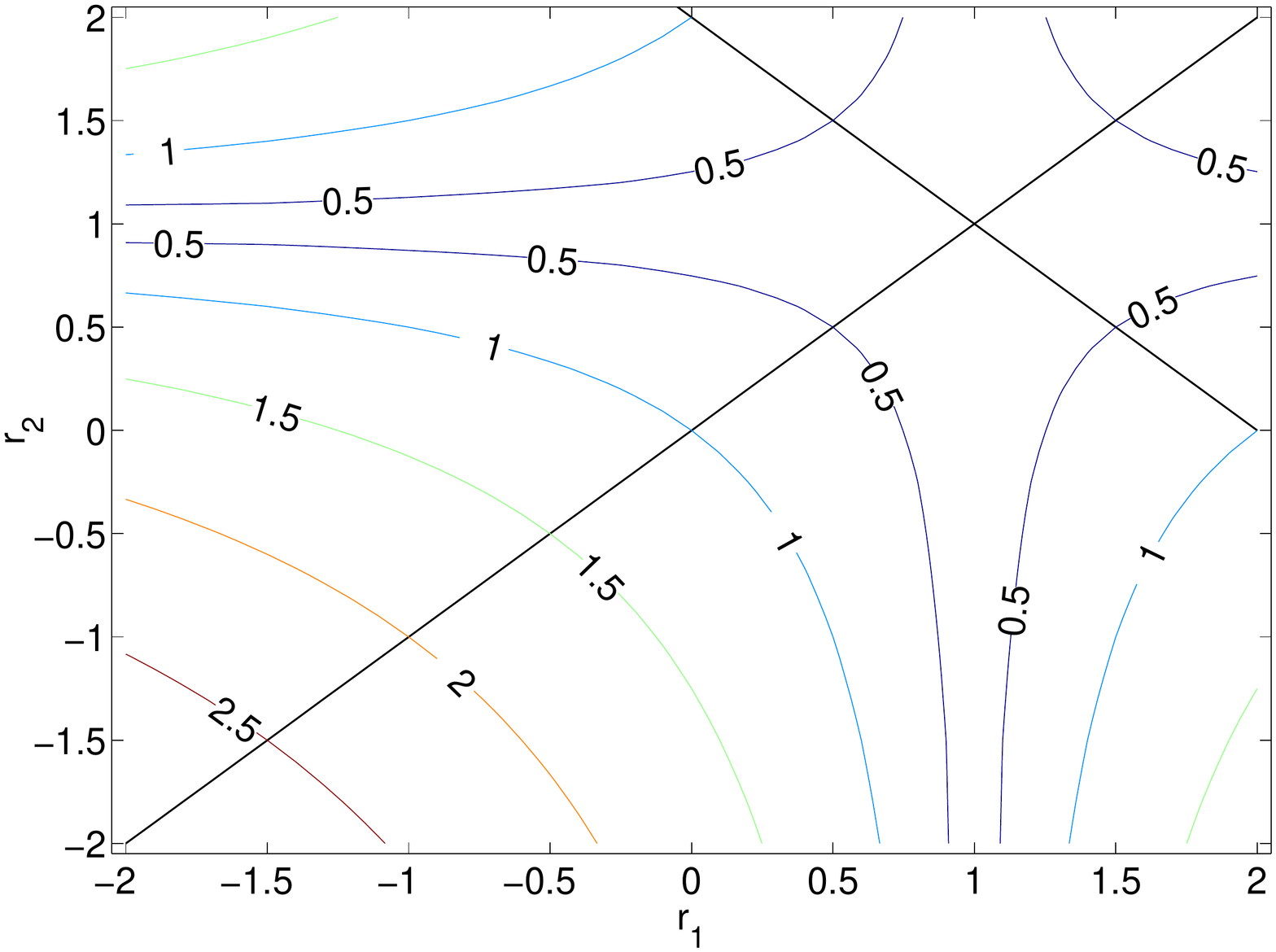}
 }\\
\end{tabular}
\end{center}
\caption{
\label{fig:qoptimal}
Iso--countour lines of the left side of Eq.~(2.52)
for $\corrsl=-1$ (left), 
$\corrsl=0$ (center) and $\corrsl=1$ (right).
In the first case the function is minimal in the $(-1,-1)$ point,
in the second in the $(0,0)$ point, while in the third
in the $(1,1)$ point.
}
\end{figure}
}